\documentclass[3p,twocolumn]{elsarticle}
\usepackage{bbm}
\usepackage{mathrsfs}
\usepackage{amsmath}
\usepackage{amsfonts}
\usepackage[colorlinks=true,linkcolor=blue,urlcolor=blue,citecolor=blue,anchorcolor=blue]{hyperref}
\usepackage{graphicx,epstopdf}
\usepackage{subfigure}
\usepackage{epsfig}
\usepackage{dcolumn}
\usepackage{bm}
\usepackage{color}
\usepackage{natbib}
\usepackage{amssymb}
\usepackage{xcolor}
\usepackage{braket}

\journal{Science Bulletin}

\biboptions{sort&compress}

\begin{document}
\title{Generating optical cat states via quantum interference of multi-path free-electron--photons interactions}

\author[add1,add2]{Feng-Xiao Sun}
\author[add1]{Yiqi Fang}
\author[add1,add2,add3,add4]{Qiongyi He\corref{cor1}}
\ead{qiongyihe@pku.edu.cn}
\author[add1,add2,add3,add5]{Yunquan Liu\corref{cor1}}
\ead{Yunquan.liu@pku.edu.cn}

\cortext[cor1]{Corresponding author}
\address[add1]{State Key Laboratory for Mesoscopic Physics, School of Physics, Frontiers Science Center for Nano-optoelectronics, $\&$ Collaborative Innovation Center of Quantum Matter, Peking University, Beijing 100871, China}
\address[add2]{Collaborative Innovation Center of Extreme Optics, Shanxi University, Taiyuan, Shanxi 030006, China}
\address[add3]{Peking University Yangtze Delta Institute of Optoelectronics, Nantong 226010, Jiangsu, China}
\address[add4]{Hefei National Laboratory, Hefei 230088, China}
\address[add5]{Beijing Academy of Quantum Information Sciences, Beijing 100193, China}

\begin{abstract}
The novel quantum effects induced by the free-electron--photons interaction have attracted increasing interest due to their potential applications in ultrafast quantum information processing. Here, we propose a scheme to generate optical cat states based on the quantum interference of multi-path free-electron--photons interactions that take place simultaneously with strong coupling strength. By performing a projection measurement on the electron, the state of light changes significantly from a coherent state into a non-Gaussian state with either Wigner negativity or squeezing property, both possess metrological power to achieve quantum advantage. More importantly, we show that the Wigner negativity oscillates with the coupling strength, and the optical cat states are successfully generated with high fidelity at all the oscillation peaks. This oscillation reveals the quantum interference effect of the multiple quantum pathways in the interaction of the electron with photons, by that various nonclassical states of light are promising to be fast prepared and manipulated. These findings inspire further exploration of emergent quantum phenomena and advanced quantum technologies with free electrons.  
\end{abstract}

\maketitle

\section{Introduction}
The substantial progress has been made in controllable creation of various quantum states of light~\cite{furusawa2015quantum}, providing a novel technology for implementing quantum communication~\cite{ralph1998teleportation,gisin2007quantum,lo2014secure,pirandola2015advances,armstrong2015multipartite,xu2020secure,liu2023eliminating}, quantum computing~\cite{slussarenko2019photonic,zhong2020quantum,chi2022programmable}, quantum imaging~\cite{moreau2019imaging}, and quantum metrology~\cite{giovannetti2011advances,aasi2013enhanced,barbieri2022optical}. Among them, non-Gaussian quantum optical states with negative Wigner function have been highlighted as indispensable quantum resources for achieving quantum advantages beyond Gaussian states~\cite{eisert2002distilling,mari2012positive}. Of particular interest is the creation of optical Schr\"{o}dinger cat states, defined as the superposition of two distinguishable coherent states~\cite{schrodinger1935gegenwartige}, for both testing the fundamentals of quantum mechanics~\cite{zavatta2004quantum} and developing quantum technologies such as fault-tolerant quantum computing~\cite{gilchrist2004schrodinger,lund2008fault}, quantum metrology against particle loss~\cite{joo2011quantum,lee2020optimal}, and quantum communications~\cite{sheng2022one,zhou2020measurement,pan2023free,kwek2021chip}. Over the last decades, many efforts have been made to generate optical cat states~\cite{ourjoumtsev2006generating,neergaard2006generation,vlastakis2013deterministically,joo2016implementation,le2018remote,hacker2019deterministic}. However, most of the reported results are still limited to the range of a few photons due to the current technology~\cite{sychev2017enlargement,le2018remote,hacker2019deterministic}. Very recently, new schemes for preparing weighted superposition of coherent states containing high-photon number have been proposed based on the intense laser-matter interactions for high-harmonic generation~\cite{lewenstein2021generation,rivera2021new,rivera2022strong,stammer2022high}, in which the created cat state takes the form of $|\phi\rangle\propto|\alpha+\delta\alpha\rangle-\xi|\alpha\rangle$ with $|\xi|<1$. Although not the ideal cat states (i.e. $|\alpha\rangle+e^{i\theta}|-\alpha\rangle$), it represents an important step forward in quantum optics and quantum information.

Modern developments in electron microscopy, especially the implementation of photon-induced near-field electron microscopy (PINEM)~\cite{barwick2009photon}, enable to control the quantum statistics of a large number of photons through free-electron--photons interactions~\cite{garcia2010multiphoton,park2010photon,piazza2015simultaneous}. With strong electron-light interactions~\cite{dahan2020resonant,wang2020coherent,kfir2020controlling,henke2021integrated,dahan2021imprinting}, the joint state of electrons and photons becomes an entangled state~\cite{kfir2019entanglements}, where the wavelength of the light can be set in the range of around $500-1600$nm~\cite{barwick2009photon,piazza2015simultaneous,dahan2021imprinting,henke2021integrated}. Consequently, one can extract and control the quantum statistics of photons by measuring the electron energy spectrum~\cite{di2019probing,ben2021shaping,kfir2021optical,ryabov2020attosecond,dahan2021imprinting,kurman2021spatiotemporal,zhang2021quantum}. Motivated by these inspiring achievements, it has become a worthwhile objective to explore free-electron--based methods to create optical Schr\"{o}dinger cat states with high fidelity, which hasn't been discovered yet.

\begin{figure}[t]
	\centering
	\includegraphics[width=0.5\textwidth]{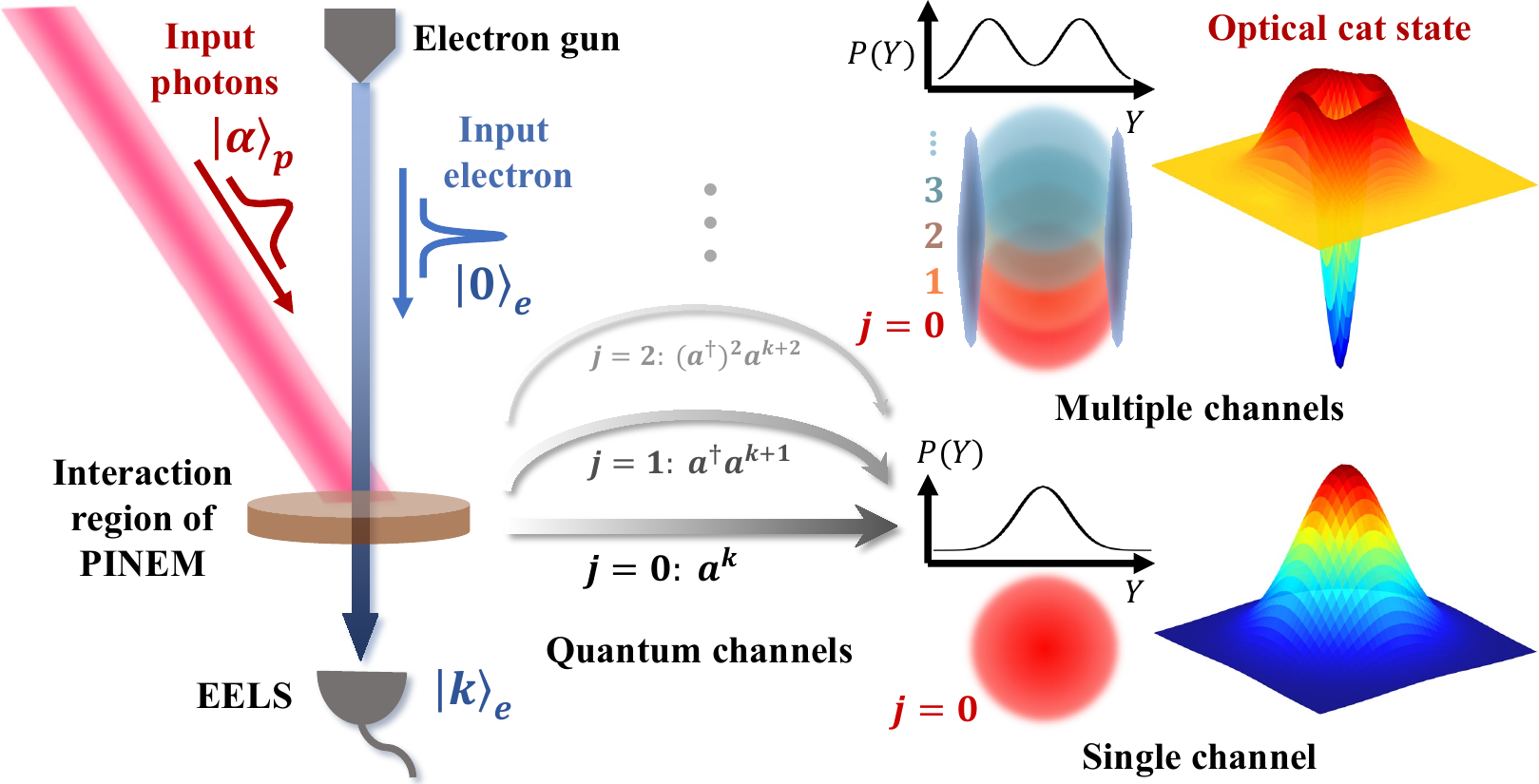}
	\caption{(Color online) The schematic diagram for generating optical cat states through the quantum interference of multi-path electron-photons interactions. Each free electron $|0\rangle_e$ interacts with the incident optical coherent state $|\alpha\rangle_p$ via PINEM, and then is measured by an EELS and postselected on a specific state $|k\rangle_e$. For a given projection $k\ge 0$, channel $\{j=0:a^k\}$ represents that the electron absorbs $k$ photons from the light. This makes dominant contribution when coupling strength is weak. As the photon-subtraction operation does not change coherent states, the produced state of light remains unchanged, indicated by the Gaussian Wigner function. While if the interaction gets stronger, more channels $\{j=1:a^\dagger a^{k+1}\}$,  $\{j=2:(a^\dagger)^2 a^{k+2}\}$, $\dots$ take part in and interfere with each other, resulting in a non-Gaussian state as characterized by the Wigner function with negative values. Two distinct outcomes are observed simultaneously in the probability distribution of phase-quadrature $Y$, which bears a resemblance to the optical cat state.}
	\label{fig:scheme}
\end{figure}

Here, we observe that the Wigner negativity of the conditional optical states oscillates with coupling strength in some region and provide a novel scheme for interpretation of this phenomenon based on the constructive and destructive effects of quantum interference of the multiple free-electron--photons interaction paths that take place simultaneously for each projection measurement on the electron spectrum. We prove that the quantum states at the peaks of oscillation are displaced odd cat states with fidelity $F\gtrsim0.99$. All those nonclassical states of light possess metrological power for the quantum precision measurement. Specifically, the input optical coherent state is coupled to a free electron via strong PINEM interaction, as illustrated in Fig.~\ref{fig:scheme}. Then due to the strong interaction, each electron either emits or absorbs multiple photons and consequently there exist multiple interaction channels that annihilate and create photons from the state of light. Those quantum channels take place simultaneously and interfere with each other due to the coherent nature of light. Then by performing a projective measurement on the electron with electron energy loss spectrometer (EELS), the state of light is collapsed into a superposition of conditional states induced by each quantum channel, and thus its Wigner negativity oscillates with the coupling strength due to the effect of quantum interference. When the coupling strength goes extremely strong, the oscillation becomes less pronounced and the size of optical cat states gets larger. These findings allow us to acquire a deep understanding of quantum phenomena produced by free-electron--photons interaction which would be useful for further exploration of advanced quantum technologies.  

\section{Proposal and methods}
The input light is prepared into a large-amplitude coherent state $|\alpha\rangle_p$, while the free electron is confined to the baseline energy $E_0$ and represented as an energy-ladder state $|0\rangle_e$ (Fig.~\ref{fig:scheme}). Here, a ladder state of free electron $|k\rangle_e$ corresponds to the energy of $E_0+k\hbar\omega$ by absorbing $k$ photons, where $\hbar\omega$ is the energy of a single photon. Please note that a non-relativistic free electron cannot emit or absorb photons, while in PINEM it has been proven that the Schr\"{o}dinger equation is still valid in the relativistic regime with taking the relativistic correction in the velocity of the electron~\cite{park2012relativistic,Li2022relativistic}. Thus, the relativistic correction in our work is only applied to the velocity of the electron, which is commonly used in PINEM theory~\cite{garcia2010multiphoton,park2010photon,di2019probing,kfir2019entanglements,ben2021shaping,dahan2021imprinting}.

Referred to as the theory of quantum PINEM (QPINEM), the electron-photon interaction is described by the scattering matrix~\cite{di2019probing,kfir2019entanglements,ben2021shaping,dahan2021imprinting} $S=\exp(g a^\dagger b-g^\ast a b^\dagger).$ Here, $g=e/\hbar\omega\int_{-\infty}^{\infty}dz'\exp(-i\omega z'/v)E_z(z')$ is the quantum coupling constant that describes the strength of the electron-photon interaction, with charge $e$ and velocity $v$ of the free electron and the electric field $E_z$ of light; $a$ ($a^{\dagger}$) is the optical annihilation (creation) operator; $b$ ($b^\dagger$) is the free-electron energy ladder operator describing losing (gaining) a single-photon energy quantum, $b|k\rangle_e=|k-1\rangle_e$ ($b^\dagger|k\rangle_e=|k+1\rangle_e$) and for more details see text SI in Supplementary Material (online). 

Then with the Baker-Campbell-Hausdorff formula~\cite{bonfiglioli2011topics}, we expand the scattering matrix with the Taylor series (see text SI online),
\begin{equation}
	S=e^{-\frac{|g|^2}{2}}\sum^\infty_{j,l=0}\frac{g^j(-g^\ast)^l}{j!l!}(a^\dagger b)^j(ab^\dagger)^l.
	\label{eq:scattering}
\end{equation}
This produces electron and photons into an entangled state, expressed as  
\begin{equation}
 |\psi\rangle_{p-e}=\mathcal{N}\sum_{j,l=0}^{\infty}\frac{(g)^j(-g^\ast\alpha)^l}{j!l!}(a^{\dagger})^{j}|\alpha\rangle_p|l-j\rangle_e,
 \label{p-e}
\end{equation}
where $\mathcal{N}$ is the normalized parameter. Note that we have used the fact that $a^{l}|\alpha\rangle_p=\alpha^{l}|\alpha\rangle_p$. 
Consequently, by detecting the electron with EELS and performing a projection $|k\rangle_e\langle k|$ on the electron spectrum, the conditional optical state $|\psi^{(k)}\rangle_p={}_e\langle k|\psi\rangle_{p-e}$ is obtained,
\begin{equation}
 |\psi^{(k)}\rangle_p=\mathcal{N}_p\sum_{j=\text{max}\{0,-k\}}^{\infty}\frac{(-\alpha|g|^2)^{j}}{j!(k+j)!}(a^{\dagger})^{j}|\alpha\rangle_p
 \label{eq:final}
\end{equation}
with normalization $\mathcal{N}_p$ and ${}_e\langle k|l-j\rangle_e=\delta_{k,l-j}$. Note that the coherent amplitude is in general complex, $\alpha=|\alpha| e^{i\phi}$, and the electron energy-ladder states are insensitive to the phase $\phi$. Hence, we can use the transition of $a\to a e^{-i\phi}$, $b\to b e^{i\phi}$ and focus on the case of $\alpha>0$. In fact, the phase $\phi$ just changes the direction of the generated optical state. The details can be found in text SI in the Supplemental Material (online).
\begin{figure*}[t]
	\centering
	\includegraphics[width=\textwidth]{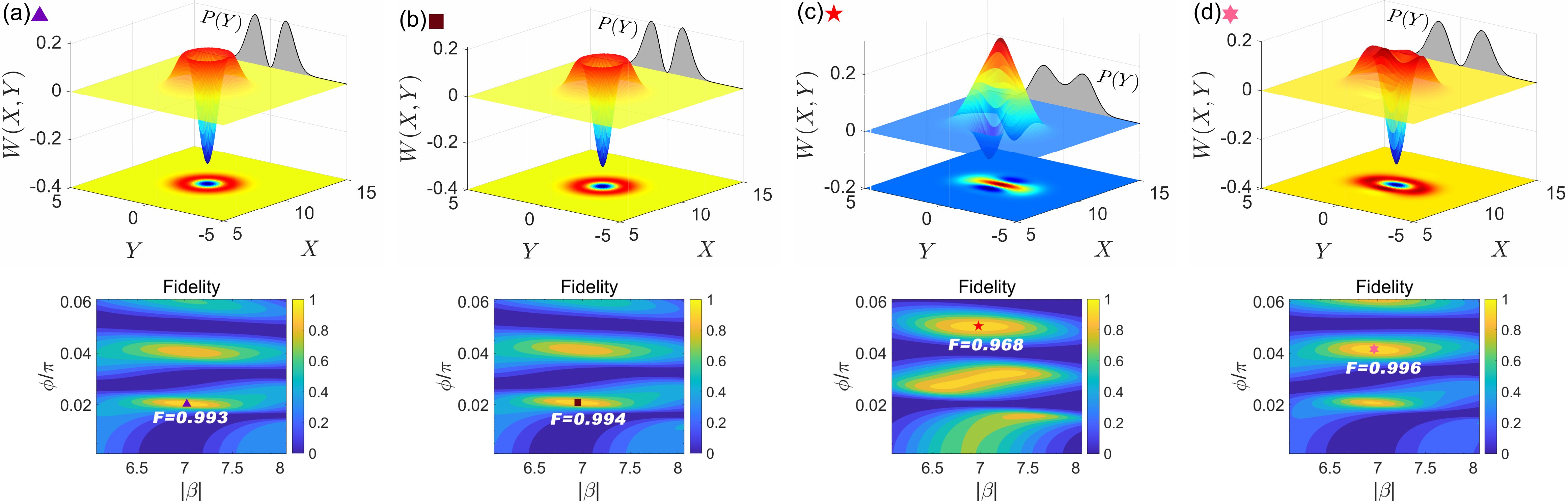}
	\caption{(Color online) Wigner function of the optical cat states acquired by setting (a) the interaction strength $|g|=0.17$ and post selection on the free electron $k=0$, (b) $|g|=0.275$ and $k=1$, (c) $|g|=0.95$ and $k=0$, (d) $|g|=0.95$ and $k=1$. The contour plots of Wigner functions are shown in the horizontal plane projection. And the probability distributions of the quadrature $Y$ are given by the vertical planes. The lower panel displays the fidelity comparing with an ideal cat state for each case, where the amplitudes $|\beta|$ and the phases $\phi$ of the cat states are marked by (a) triangle, (b) square, (c) star and (d) circle, respectively. Here the photon number of the initial coherent states is set as $|\alpha|^2=50$, and all quantum channels are considered}.
	\label{fig:nonclassical_state}
\end{figure*}

Now we will show that the optical state (\ref{eq:final}) becomes a cat state via the coherent superposition of multiple quantum-interaction channels denoted by $(a^\dagger)^j a^{k+j}$ for a fixed projective measurement $k$ on the electron (see `Quantum channels' in Fig.~\ref{fig:scheme}). To illustrate this, let's consider the initial coherent state of light containing $|\alpha|^2=50$ photons which is much larger than the achievable values of $k$ and $j$ (those with large coefficients given in Eq.~(\ref{eq:final})). For the weak interaction with $|g|\ll 1$, the coefficients in Eqs.~(\ref{p-e}) and (\ref{eq:final}) for higher-order interactions drop dramatically with the increase of $j$, and hence the electron interacts with photons mainly through the single channel of $\{j=0:a^{k}\}$ for $ k\ge 0$. As the photon-subtraction operation does not change the coherent state, the state of light maintains its original quantum statistics, as indicated by the Gaussian Wigner function in Fig.~\ref{fig:scheme}. For the case of $k<0$, the single channel $\{j=|k|:(a^\dagger)^{|k|}\}$ produces a $|k|$-photon-added optical state. It indeed changes the Wigner function of light from Gaussian to non-Gaussian, however, such change does not affect the coherent state significantly as $k$ is relatively small as compared to the number of photons $|\alpha|^2$. 

Then with enhancing the coupling strength $|g|$, more quantum channels take part in and interfere with each other, which changes the optical state substantially. This is evident in the appearance of negative values of the Wigner distribution function of the created optical states (Fig.~\ref{fig:scheme}). A given value of $k \ge 0$ means that the electron effectively absorbs $k$ photons, which corresponds to the quantum process either directly absorbing $k$ photons via the channel $\{j=0:a^{k}\}$, or first absorbing $k+j$ and then emitting $j$ photons through channels $\{j=1:a^\dagger a^{k+1}\}$,  $\{j=2:(a^\dagger)^2 a^{k+2}\}$, $\dots$. Note that the photon-subtraction and addition operations in each channel don't affect the coherent state significantly due to rather small values of $k$ and $j$ in comparison with $|\alpha|^2$. However, as shown in Fig.~\ref{fig:scheme}, the interference of them generates a non-Gaussian state with negative Wigner function. We provide an analytical interpretation together with the numerical calculation in SII (online) indicating that the generated non-Gaussian optical state can be regarded as a displaced cat state. In the following part we will show the high fidelity and the non-classical properties of the created cat states.

\section{Results and discussion}
In order to confirm the presence and evaluate the quality of the cat states, we have systematically ploted the Wigner functions~\cite{wigner1932quantum} and the fidelity comparing with an ideal cat state~\cite{jozsa1994fidelity} by continuously increasing the coupling strength $|g|$, which is shown in the Supplementary Video (online). Specially, we plot the Wigner function of the optical state produced by a projective measurement $k=0$ on the electron with $|g|=0.17$ in Fig.~\ref{fig:nonclassical_state}(a). The remarkable negative Wigner function is a clear indicator that the corresponding state exhibits strong quantum features. The projection on the vertical plane gives the probability distribution of the quadrature $Y=(a-a^\dagger)/\sqrt{2}i$. The observation of two distinct peaks in $Y$ hints an excellent resemblance to a cat state displaced along $X$-quadrature, 
$|\varphi_\text{cat}\rangle=\mathcal{N}_c(|\beta\rangle-|\beta^\ast\rangle)=\mathcal{N}_c(|\beta_x+i\beta_y\rangle-|\beta_x-i\beta_y\rangle)$. Here, $|\beta\rangle$ is coherent state with a complex amplitude $\beta=|\beta|\exp(i\phi)=\beta_x+i\beta_y$, and $\mathcal{N}_c$ is the normalized parameter (details are shown in text SII online). The fidelity between the created optical state $|\psi^{(k)}\rangle_p$ and the cat state with $|\beta|=7.021$, $\phi=0.020\pi$ is $F=\left|\langle\varphi_\text{cat}|\psi^{(k)}\rangle_p\right|^2=0.993$, as marked by a triangle in the bottom plot. 

We notice that there are $4$ quantum channels ($j=0,1,2,3$) with high probability that take place simultaneously (see the probability of channels in Fig.~S1 online), but each one does not affect the state of light much (the optical state created by each channel is `nearly' Gaussian with positive Wigner function, see Fig.~S2 online). This clearly evidences that it is the superposition of multi-interaction paths affects the state. If the postselection is set at $k=1$, a stronger coupling of $|g|=0.275$ is required to generate the cat state, as shown in Fig.~\ref{fig:nonclassical_state}(b). The high fidelity $0.994$ can be achieved with $|\beta|=6.951$, $\phi=0.021\pi$. It is obvious that the coefficients in Eq.~(\ref{eq:final}) drop more rapidly with large $|k|$. Thus to include multiple channels and obtain an optical cat state, a stronger interaction $|g|$ is required for larger $|k|$. 

It is generally considered that a stronger interaction will result in a higher-quality cat state. However, we find that the dependence of coupling strength shows a great deal of complexity. For the case of postselection on $k=0$, we see that two wider peaks do not become well separated for stronger coupling of $|g|=0.95$, reflecting a slightly lower fidelity comparing with the odd cat state [Fig.~\ref{fig:nonclassical_state}(c)]. While, if the postselection is performed on $k=1$ for the same coupling strength, two peaks are fully separated and a high-fidelity optical odd cat state of $F=0.996$ is obtained. These findings indicate that both the interaction strength $g$ and the projection choice of $k$ determine the interference effect and thus control the generated optical state.

\begin{figure}[t]
	\centering
	\includegraphics[width=0.4\textwidth]{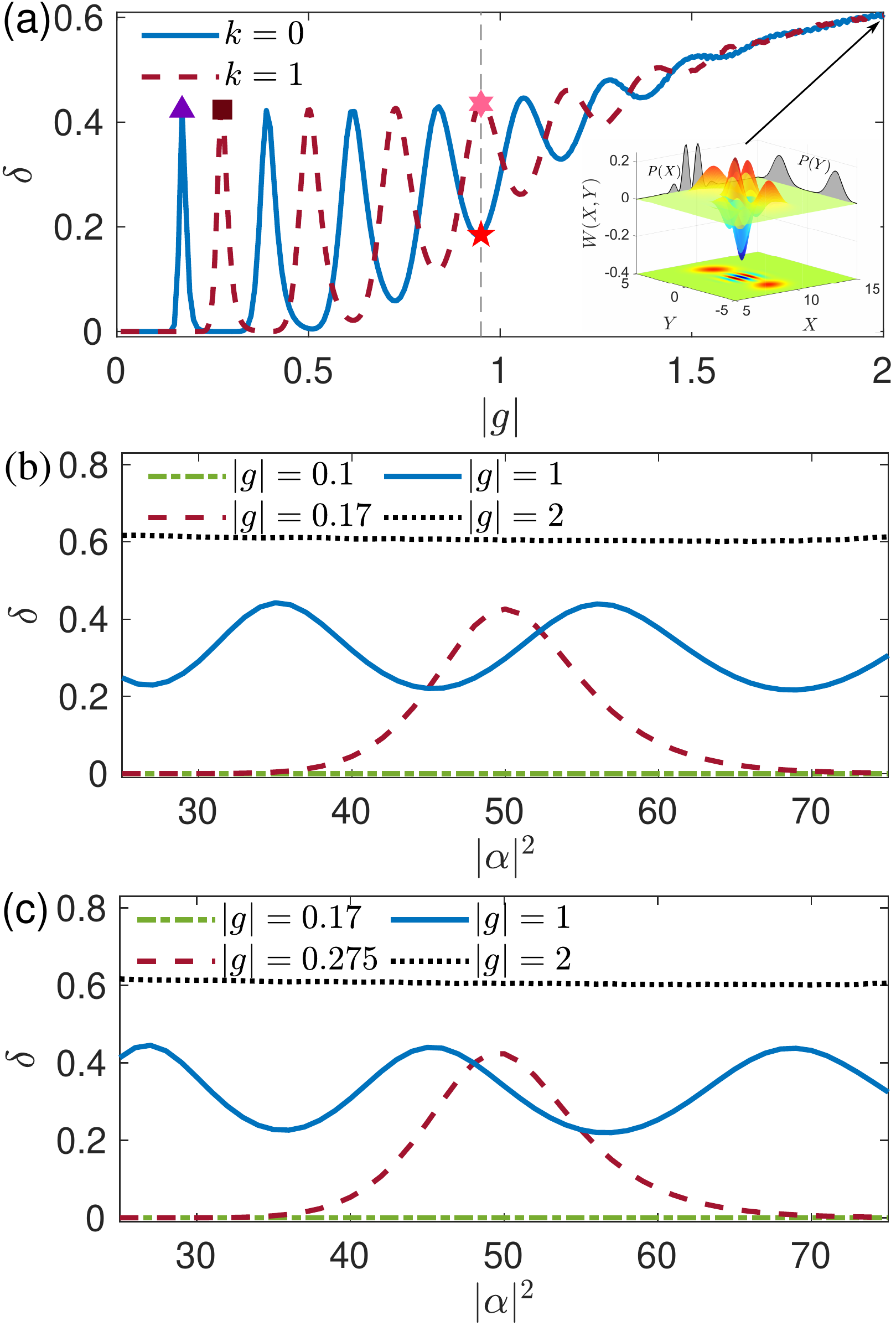}
	\caption{(Color online) The Wigner negativity $\delta$ of the created optical states changing with (a) the coupling strength $|g|$ and (b,c) the average photon number $|\alpha|^2$. (a) The postselected energy of free electron is set as $k=0$ (solid) and $k=1$ (dashed), $|\alpha|^2=50$. The Wigner negativities of the generated optical states in Fig.~\ref{fig:nonclassical_state} from (a) to (d) are marked with the triangle, square, star and circle, respectively. The inset of (a) shows the Wigner function, the probability distributions $P(X)$ and $P(Y)$ of a large optical cat state generated with $|g|=2$ and $k=0$. The coupling strength is (b) $|g|=0.01$ (dashdotted), $|g|=0.17$ (dashed), $|g|=1$ (solid), $|g|=2$ (dotted), and (c) $|g|=0.17$ (dashdotted), $|g|=0.275$ (dashed), $|g|=1$ (solid) and $|g|=2$ (dotted). And the postselected energy is set as (b) $k=0$ and (c) $k=1$.}
	\label{fig:WN}
\end{figure}

A further highlight of our scheme is the observation of the Wigner negativity of the conditional optical state oscillates with the coupling strength due to the constructive and destructive interference of the multiple channels, as shown in Fig.~\ref{fig:WN}. The Wigner negativity, defined by the volume of the negative part of the Wigner function $\delta=\int |W(\alpha,\alpha^\ast)|d^2 \alpha-1$~\cite{kenfack2004negativity}, acts as an important signature of cat states~\cite{shen2015nonlinear,sun2019schrodinger,teh2020dynamics,sun2021remote}. As shown in Fig.~\ref{fig:WN}(a) for postselections $k=0$ (solid) and $k=1$ (dashed), when $|g|\ll 1$, or more specifically, $|g|\lesssim0.1$, the generated optical states remain positive in their Wigner functions and thus $\delta=0$. As explained above, in such cases, only a single quantum channel makes contribution, which does not affect the optical states much to show negativity. With the increase of the coupling strength, the effects of other quantum channels accumulate gradually, and the interference of all involved channels leads to the oscillation of Wigner negativity. The triangle, square, star, and circle mark the Wigner negativity of the create optical cat states shown in Fig.~\ref{fig:nonclassical_state} from (a) to (d), respectively.

We find that the optical states at the oscillation peaks can be regarded as displaced odd cat states with fidelity higher than $0.99$, where the constructive interference takes place. Considering that for different postselection $k$, the peaks appear at different coupling strengths $|g|$, which means that the optical cat states can be controllably prepared by adjusting $|g|$ and $k$. The detailed analysis on the contribution of the individual and the interference of quantum channels are provided in the text SIII and the Supplementary Video (online). In brief, the quantum channels are divided into two sets with opposite phases, i.e. channels of even $j$ with positive coefficients and channels of odd $j$ with negative coefficients in Eq.~(\ref{eq:final}). We find that when the two sets of channels contribute equally, the Wigner negativity reaches its maxima due to the constructive interference.

By further increasing the coupling to $|g|>1.5$, the oscillation of Wigner negativity becomes less pronounced. It is because that with the coupling strength becomes stronger, a larger number of quantum channels are involved and the interference saturates (see Supplemental Material online). In this regime, the Wigner negativity exhibits no oscillation and becomes larger with increasing $g$ and less sensitive to the choice of $k$, as shown by Fig.~\ref{fig:WN}(a). The inset plot shows the created large-size optical cat state with strong coupling $|g|=2$. The notable distance between the two peaks in $Y$ quadrature implies the large size of the cat state~\cite{sychev2017enlargement}. At the same time, the large Wigner negativity is an indicator of the quantum nature of the created state, ruling out the possibility of the state being in a classical mixture of two coherent states.

Besides of $g$ and $k$, the mean photon number $|\alpha|^2$ of the initial coherent state also affects the Wigner negativity, as shown in Figs.~\ref{fig:WN}(b) and (c). With the postselection $k=0$, Fig.~\ref{fig:WN}(b) shows that for a very weak coupling $|g|=0.01$ (dashdotted), any value of amplitude is unable to generate Wigner negativity since there is no entanglement between the photons and the free electron. By increasing the coupling to $|g|=0.17$ (dashed), Wigner-negative states can be produced within appropriate range of mean photon number. In this case, multiple quantum channels are taken part in and interfere with each other. Since $|\alpha|^2$ is also included in the superposition coefficients of the output optical states given in Eq.~(\ref{eq:final}), the oscillation of Wigner negativity is expected, which is more obvious for a stronger coupling $|g|=1$ (solid). Further increasing the coupling to $|g|=2$ (dotted), the oscillation of Wigner negativity vanishes and a high Wigner negativity is always obtained, which is consistent with Fig.~\ref{fig:WN}(a). As for the case of $k=1$ in Fig.~\ref{fig:WN}(c), the Wigner negativity vanishes for any value of amplitude with the coupling $|g|=0.17$ (dashdotted). It is understandable that the coefficients in Eq.~(\ref{eq:final}) drop more rapidly with large $|k|$, and thus a stronger coupling $|g|$ is required to include multiple channels and observe a non-zero Wigner negativity. When the coupling is increased to $|g|=0.275$ (dashed), Wigner-negative states can be observed with appropriate $|\alpha|^2$. The pronounced oscillation of Wigner negativity is also obtained with $|g|=1$ (solid), and the saturation with $|g|=2$ (dotted). This is consistent with Figs.~\ref{fig:WN}(a) and (b).
\begin{figure}[t]
	\centering
	\includegraphics[width=0.5\textwidth]{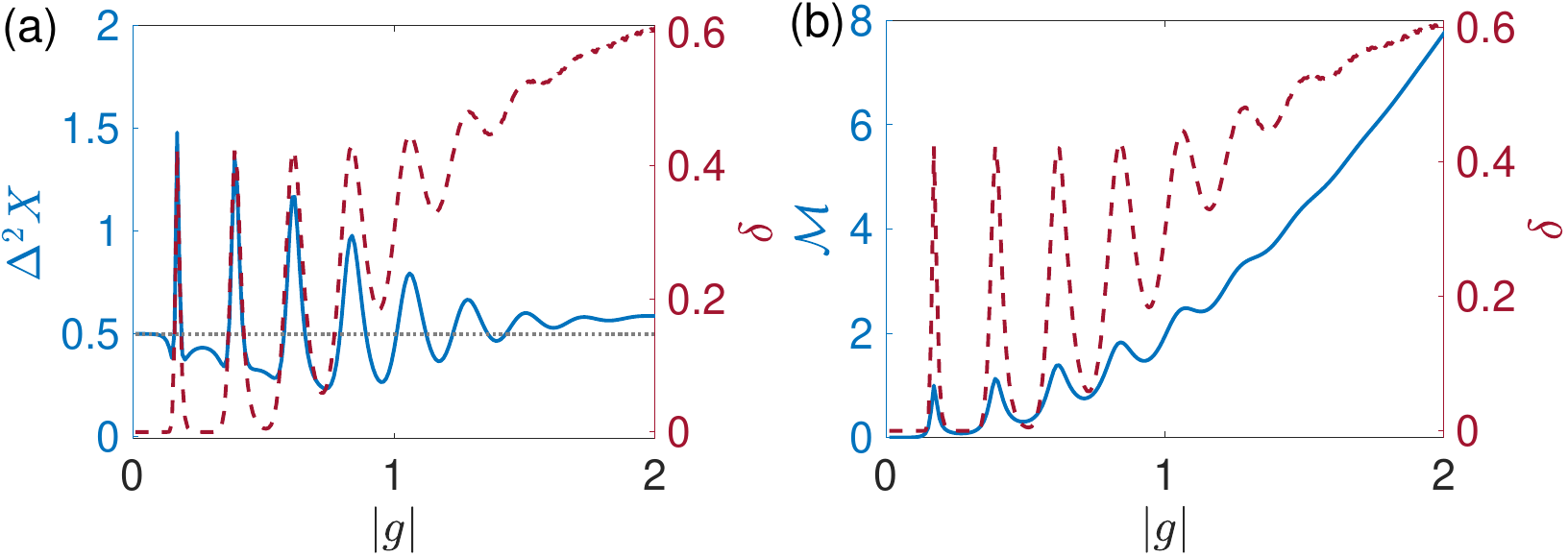}
	\caption{(Color online) The dependence of (a) the variance of amplitude $\Delta^2X$ and (b) the metrological power $\mathcal{M}$ on the coupling strength $|g|$, which is compared with the Wigner negativity $\delta$ (dashed). The parameters are set as $k=0$ and $|\alpha|^2=50$.}
	\label{fig:MP}
\end{figure}

To reveal its application in quantum metrology, we plot the variance $\Delta^2 X$ in Fig.~\ref{fig:MP}(a) and the metrological power $\mathcal{M}$ in Fig.~\ref{fig:MP}(b). It is found that the squeezed quadrature of $\Delta^2 X<0.5$ is produced at the oscillation valleys of $\delta$ (dashed). Please note that those state are not squeezed coherent states (Gaussian states) since they show negative values in their Wigner distribution. In fact, those states are more like displaced even cat states which we will further study somewhere else. Here we focus on the optical odd cat states created at peaks of $\delta$ which are more powerful in precision measurement, as indicated in Fig.~\ref{fig:MP}(b). The metrological power is defined by $\mathcal{M}=\text{max}[F_X(\rho)-2,0]/4$~\cite{kwon2019nonclassicality,ge2020operational,ge2020evaluating}, where $F_X(\rho)$ is the quantum Fisher information optimized over quadratures. We can see that the created odd cat states at peaks of $\delta$ with larger amount of Wigner negativity have higher metrological power. And $\mathcal{M}$ further increases when large-size cat states can be observed with $|g|>1$. As the strong coupling of $|g|\sim 1$ has been reported~\cite{adiv2022observation}, it is promising to observe the optical cat states and the oscillated Wigner negativity in PINEM experiments. 

A detailed discussion on experimental feasibility is given in text SIV (online), including the methods to achieve strong couplings, the success probability of the postselection, and the effects of experimental imperfections. In brief, as the coupling strength takes the form of $g=e/\hbar\omega\int_{-\infty}^{\infty}dz'\exp(-i\omega z'/v)E_z(z')$, the most commonly used method to modify $g$ is to change the strength and the distribution of $E_z$, which represents the electric-field component of a single photon in the near-field of PINEM. By using designed photonic nanostructures such as micro-cavities~\cite{kfir2019entanglements} and energy-momentum phase-matched structures~\cite{dahan2020resonant}, the near-field can be controlled and enhanced in a large range. As for the inevitable errors when controlling the coupling strength, we have found that the robustness against the fluctuation $\Delta g$ can be enhanced by achieving strong coupling strength $g$ (see Fig.~S8 online). In addition, the free-electron-photons interaction length in PINEM is typically $L\leq500\mu$m~\cite{dahan2020resonant}, which indicates that the timescale of the interaction is usually $t\leq 1000$fs. Hence, a large number of optical cat states are promising to be obtained within a finite timescale, which facilitates our method in various quantum information processing.

\section{Conclusion}
In summary, we have proposed a novel method to generate high-fidelity optical cat states through the QPINEM interaction. The cat states result from the multi-path interference effect related to the strong free-electron--photons interaction. By performing postselection on the electron that is entangled with photons, the quantum statistics of photons will be changed and the optical odd cat states with high fidelity are generated due to the constructive interference of the multi-path electron-photon interactions. The Wigner negativity, an important signature of the cat state, oscillates with the coupling strength, confirming the quantum interference effect. We find that the Wigner negativity of the generated states achieves maximum at all the oscillation peaks, and two Gaussian peaks in $Y$ quadrature are well separated, clearly reflecting the presence of high-quality optical cat states. Comparing with an ideal displaced odd cat state, we verify that high fidelity up to $0.99$ has been achieved. Continuously increase the interaction strength, the interference saturates and the large size of cat state can be obtained. The conceptual ideas that we presented can be observed with the prospective application for modern quantum technologies, considering that the strong coupling $|g|\sim 1$ of the free-electron--photons interaction has been reported~\cite{adiv2022observation}. In the light of the importance of optical cat states in quantum computation and precision measurements, it takes a step toward the development of essential quantum resources based on the interaction of free electron with ultrafast optics. \\

\noindent\textbf{Conflict of interest}\\

\indent The authors declare that they have no conflict of interest.\\

\noindent\textbf{Acknowledgments}\\

\indent This work is supported by the State Key R\&D Program (No.~2022YFA1604301), the National Natural Science Foundation of China (Grants No.~92050201, 92250306, 11975026, 12125402, 12147148),  the Key R\&D Program of Guangzhou Province (Grant No.~2018B030329001), and the Beijing Natural Science Foundation (Grant No.~Z190005). F.-X. S. acknowledges the China Postdoctoral Science Foundation (Grant No.~2020M680186). Q. H. acknowledges the Innovation Program for Quantum Science and Technology (No. 2021ZD0301500).\\

\noindent\textbf{Author contributions} \\

\indent Feng-Xiao Sun, Qiongyi He and Yunquan Liu conceptualized and designed the study. Feng-Xiao Sun and Yiqi Fang carried out the calculations and analyzed the results. All authors contributed to discussions and writing of the manuscript.

\appendix
\section{Supplementary Material}
Supplementary materials to this article can be found online.


\begin{thebibliography}{}

\bibitem[Furusawa, 2015]{furusawa2015quantum}
Furusawa A.
\newblock Quantum states of light.
\newblock Springer, Tokyo; 2015.

\bibitem[Ralph and Lam, 1998]{ralph1998teleportation}
Ralph TC, Lam PK.
\newblock Teleportation with bright squeezed light.
\newblock Phys Rev Lett 1998;81:5668.

\bibitem[Gisin and Thew, 2007]{gisin2007quantum}
Gisin N, Thew R.
\newblock Quantum communication.
\newblock Nat Photon 2007;1:165-171.

\bibitem[Lo et~al., 2014]{lo2014secure}
Lo HK, Curty M, Tamaki K.
\newblock Secure quantum key distribution.
\newblock Nat Photon 2014;8:595-604.

\bibitem[Pirandola et~al., 2015]{pirandola2015advances}
Pirandola S, Eisert J, Weedbrook C, et~al.
\newblock Advances in quantum teleportation.
\newblock Nat Photon 2015;9:641-652.

\bibitem[Armstrong et~al., 2015]{armstrong2015multipartite}
Armstrong S, Wang M, Teh RY, et~al.
\newblock Multipartite Einstein--Podolsky--Rosen steering and genuine
  tripartite entanglement with optical networks.
\newblock Nat Phys 2015;11:167-172.

\bibitem[Xu et~al., 2020]{xu2020secure}
Xu F, Ma X, Zhang Q, et~al.
\newblock Secure quantum key distribution with realistic devices.
\newblock Rev Mod Phys 2020;92:025002.

\bibitem[Liu et~al, 2023]{liu2023eliminating}
Liu RZ, Qiao YK, Zhong HS, et~al.
\newblock Eliminating temporal correlation in quantum-dot entangled photon source by quantum interference.
\newblock Sci Bull 2023;68:807-812.

\bibitem[Slussarenko and Pryde, 2019]{slussarenko2019photonic}
Slussarenko S, Pryde GJ.
\newblock Photonic quantum information processing: A concise review.
\newblock Appl Phys Rev 2019;6:041303.

\bibitem[Zhong et~al., 2020]{zhong2020quantum}
Zhong HS, Wang H, Deng YH, et~al.
\newblock Quantum computational advantage using photons.
\newblock Science 2020;370:1460-1463.

\bibitem[Chi et~al., 2022]{chi2022programmable}
Chi Y, Huang J, Zhang Z, et~al.
\newblock A programmable qudit-based quantum processor.
\newblock Nat Commun 2022;13:1166.

\bibitem[Moreau et~al., 2019]{moreau2019imaging}
Moreau PA, Toninelli E, Gregory T, et~al.
\newblock Imaging with quantum states of light.
\newblock Nat Rev Phys 2019;1:367-380.

\bibitem[Giovannetti et~al., 2011]{giovannetti2011advances}
Giovannetti V, Lloyd S, Maccone L.
\newblock Advances in quantum metrology.
\newblock Nat Photon 2011;5:222-229.

\bibitem[Aasi et~al., 2013]{aasi2013enhanced}
Aasi J, Abadie J, Abbott B, et~al.
\newblock Enhanced sensitivity of the LIGO gravitational wave detector by using
  squeezed states of light.
\newblock Nat Photon 2013;7:613-619.

\bibitem[Barbieri, 2022]{barbieri2022optical}
Barbieri M.
\newblock Optical quantum metrology.
\newblock PRX Quantum 2022;3:010202.

\bibitem[Eisert et~al., 2002]{eisert2002distilling}
Eisert J, Scheel S, Plenio MB.
\newblock Distilling Gaussian states with Gaussian operations is impossible.
\newblock Phys Rev Lett 2002;89:137903.

\bibitem[Mari and Eisert, 2012]{mari2012positive}
Mari A, Eisert J.
\newblock Positive Wigner functions render classical simulation of quantum
  computation efficient.
\newblock Phys Rev Lett 2012;109:230503.

\bibitem[Schr{\"o}dinger, 1935]{schrodinger1935gegenwartige}
Schr{\"o}dinger E.
\newblock Die gegenw{\"a}rtige situation in der quantenmechanik.
\newblock Naturwissenschaften 1935;23:823.

\bibitem[Zavatta et~al., 2004]{zavatta2004quantum}
Zavatta A, Viciani S, Bellini M.
\newblock Quantum-to-classical transition with single-photon-added coherent
  states of light.
\newblock Science 2004;306:660-662.

\bibitem[Gilchrist et~al., 2004]{gilchrist2004schrodinger}
Gilchrist A, Nemoto K, Munro W.~J, et~al.
\newblock Schr{\"o}dinger cats and their power for quantum information
  processing.
\newblock J Opt B: Quantum Semiclass Opt 2004;6:S828.

\bibitem[Lund et~al., 2008]{lund2008fault}
Lund AP, Ralph TC, Haselgrove HL.
\newblock Fault-tolerant linear optical quantum computing with small-amplitude
  coherent states.
\newblock Phys Rev Lett 2008;100:030503.

\bibitem[Joo et~al., 2011]{joo2011quantum}
Joo J, Munro WJ, Spiller TP.
\newblock Quantum metrology with entangled coherent states.
\newblock Phys Rev Lett 2011;107:083601.

\bibitem[Lee et~al., 2020]{lee2020optimal}
Lee SW, Lee SY, Kim J.
\newblock Optimal quantum phase estimation with generalized multi-component
  Schr{\"o}dinger cat states.
\newblock J Opt Soc Am B 2020;37:2423-2429.

\bibitem[Sheng et~al., 2022]{sheng2022one}
Sheng YB, Zhou L, Long GL.
\newblock One-step quantum secure direct communication.
\newblock Sci Bull 2022;67:367-374.

\bibitem[Zhou et~al., 2020]{zhou2020measurement}
Zhou ZR, Sheng YB, Niu PH, et~al.
\newblock Measurement-device-independent quantum secure direct communication.
\newblock Sci China Phys Mech Astron 2020;63:230362.

\bibitem[Pan et~al., 2023]{pan2023free}
Pan D, Song XT, Long GL.
\newblock Free-space quantum secure direct communication: basics, progress, and outlook.
\newblock Adv Devices Instrum 2023;4:0004.

\bibitem[Kwek et~al., 2021]{kwek2021chip}
Kwek LC, Cao L, Luo W, et~al.
\newblock Chip-based quantum key distribution.
\newblock AAPPS Bull 2021;31:15.

\bibitem[Ourjoumtsev et~al., 2006]{ourjoumtsev2006generating}
Ourjoumtsev A, Tualle-Brouri R, Laurat J, et~al.
\newblock Generating optical Schr{\"o}dinger kittens for quantum information
  processing.
\newblock Science 2006;312:83-86.

\bibitem[Neergaard-Nielsen et~al., 2006]{neergaard2006generation}
Neergaard-Nielsen JS, Nielsen BM, Hettich C, et~al.
\newblock Generation of a superposition of odd photon number states for quantum
  information networks.
\newblock Phys Rev Lett 2006;97:083604.

\bibitem[Vlastakis et~al., 2013]{vlastakis2013deterministically}
Vlastakis B, Kirchmair G, Leghtas Z, et~al.
\newblock Deterministically encoding quantum information using 100-photon
  Schr{\"o}dinger cat states.
\newblock Science 2013;342:607-610.

\bibitem[Joo et~al., 2016]{joo2016implementation}
Joo J, Lee SY, Kim J.
\newblock Implementation of traveling odd Schr{\"o}dinger cat states in
  circuit-QED.
\newblock Photonics, 2016;3:57.

\bibitem[Le~Jeannic et~al., 2018]{le2018remote}
Le~Jeannic H, Cavaill{\`e}s A, Raskop J, et~al.
\newblock Remote preparation of continuous-variable qubits using loss-tolerant
  hybrid entanglement of light.
\newblock Optica 2018;5:1012-1015.

\bibitem[Hacker et~al., 2019]{hacker2019deterministic}
Hacker B, Welte S, Daiss S, et~al.
\newblock Deterministic creation of entangled atom--light Schr{\"o}dinger-cat
  states.
\newblock Nat Photon 2019;13:110-115.

\bibitem[Sychev et~al., 2017]{sychev2017enlargement}
Sychev DV, Ulanov AE, Pushkina AA, et~al.
\newblock Enlargement of optical Schr{\"o}dinger's cat states.
\newblock Nat Photon 2017;11:379-382.

\bibitem[Lewenstein et~al., 2021]{lewenstein2021generation}
Lewenstein M, Ciappina M, Pisanty E, et~al.
\newblock Generation of optical Schr{\"o}dinger cat states in intense
  laser--matter interactions.
\newblock Nat Phys 2021;17:1104-1108.

\bibitem[Rivera-Dean et~al., 2021]{rivera2021new}
Rivera-Dean J, Stammer P, Pisanty E, et~al.
\newblock New schemes for creating large optical Schr{\"o}dinger cat states
  using strong laser fields.
\newblock J Comput Electron 2021;20:2111-2123.

\bibitem[Rivera-Dean et~al., 2022]{rivera2022strong}
Rivera-Dean J, Lamprou T, Pisanty E, et~al.
\newblock Strong laser fields and their power to generate controllable
  high-photon-number coherent-state superpositions.
\newblock Phys Rev A 2022;105:033714.

\bibitem[Stammer et~al., 2022]{stammer2022high}
Stammer P, Rivera-Dean J, Lamprou T, et~al.
\newblock High photon number entangled states and coherent state superposition
  from the extreme ultraviolet to the far infrared.
\newblock Phys Rev Lett 2022;128:123603.

\bibitem[Barwick et~al., 2009]{barwick2009photon}
Barwick B, Flannigan DJ, Zewail AH.
\newblock Photon-induced near-field electron microscopy.
\newblock Nature 2009;462:902-906.

\bibitem[Garc{\'\i}a~de Abajo et~al., 2010]{garcia2010multiphoton}
Garc{\'\i}a~de Abajo FJ, Asenjo-Garcia A, Kociak M.
\newblock Multiphoton absorption and emission by interaction of swift electrons
  with evanescent light fields.
\newblock Nano Lett 2010;10:1859-1863.

\bibitem[Park et~al., 2010]{park2010photon}
Park ST, Lin M, Zewail AH.
\newblock Photon-induced near-field electron microscopy (pinem): theoretical
  and experimental.
\newblock New J Phys 2010;12:123028.

\bibitem[Piazza et~al., 2015]{piazza2015simultaneous}
Piazza L, Lummen T, Quinonez E, et~al.
\newblock Simultaneous observation of the quantization and the interference
  pattern of a plasmonic near-field.
\newblock Nat Commun 2015;6:6407.

\bibitem[Dahan et~al., 2020]{dahan2020resonant}
Dahan R, Nehemia S, Shentcis M, et~al.
\newblock Resonant phase-matching between a light wave and a free-electron
  wavefunction.
\newblock Nat Phys 2020;16:1123-1131.

\bibitem[Wang et~al., 2020]{wang2020coherent}
Wang K, Dahan R, Shentcis M, et~al.
\newblock Coherent interaction between free electrons and a photonic cavity.
\newblock Nature 2020;582:50-54.

\bibitem[Kfir et~al., 2020]{kfir2020controlling}
Kfir O, Louren{\c{c}}o-Martins H, Storeck G, et~al.
\newblock Controlling free electrons with optical whispering-gallery modes.
\newblock Nature 2020;582:46-49.

\bibitem[Henke et~al., 2021]{henke2021integrated}
Henke JW, Raja AS, Feist A, et~al.
\newblock Integrated photonics enables continuous-beam electron phase
  modulation.
\newblock Nature 2021;600:653-658.

\bibitem[Dahan et~al., 2021]{dahan2021imprinting}
Dahan R, Gorlach A, Haeusler U, et~al.
\newblock Imprinting the quantum statistics of photons on free electrons.
\newblock Science 2021;373:eabj7128.

\bibitem[Kfir, 2019]{kfir2019entanglements}
Kfir O.
\newblock Entanglements of electrons and cavity photons in the strong-coupling
  regime.
\newblock Phys Rev Lett 2019;123:103602.

\bibitem[Di~Giulio et~al., 2019]{di2019probing}
Di~Giulio V, Kociak M, de~Abajo FJG.
\newblock Probing quantum optical excitations with fast electrons.
\newblock Optica 2019;6:1524-1534.

\bibitem[Ben~Hayun et~al., 2021]{ben2021shaping}
Ben~Hayun A, Reinhardt O, Nemirovsky J, et~al.
\newblock Shaping quantum photonic states using free electrons.
\newblock Sci Adv 2021;7:eabe4270.

\bibitem[Kfir et~al., 2021]{kfir2021optical}
Kfir O, Di~Giulio V, de~Abajo FJG, et~al.
\newblock Optical coherence transfer mediated by free electrons.
\newblock Sci Adv 2021;7:eabf6380.

\bibitem[Ryabov et~al., 2020]{ryabov2020attosecond}
Ryabov A, Thurner JW, Nabben D, et~al.
\newblock Attosecond metrology in a continuous-beam transmission electron
  microscope.
\newblock Sci Adv 2020;6:eabb1393.

\bibitem[Kurman et~al., 2021]{kurman2021spatiotemporal}
Kurman Y, Dahan R, Sheinfux HH, et~al.
\newblock Spatiotemporal imaging of 2D polariton wave packet dynamics using
  free electrons.
\newblock Science 2021;372:1181-1186.

\bibitem[Zhang et~al., 2021]{zhang2021quantum}
Zhang B, Ran D, Ianconescu R, et~al.
\newblock Quantum state interrogation using a preshaped free electron wavefunction.
\newblock Phys Rev Res 2022;4:033071.

\bibitem[Park et~al., 2012]{park2012relativistic}
Park ST, Zewail AH.
\newblock Relativistic effects in photon-induced near field electron microscopy.
\newblock J Phys Chem A 2012;116:11128-11133.

\bibitem[Li et~al., 2022]{Li2022relativistic}
Li J, Liu Y-Q.
\newblock Relativistic free electrons based quantum physics.
\newblock Acta Phys Sin 2022;71:233302.

\bibitem[Bonfiglioli and Fulci, 2011]{bonfiglioli2011topics}
Bonfiglioli A, Fulci R.
\newblock Topics in noncommutative algebra: the theorem of Campbell,
  Baker, Hausdorff and Dynkin.
\newblock Springer, Berlin, 2011.

\bibitem[Wigner, 1932]{wigner1932quantum}
Wigner E.
\newblock On the quantum correction for thermodynamic equilibrium.
\newblock Phys Rev 1932;40:749.

\bibitem[Jozsa, 1994]{jozsa1994fidelity}
Jozsa R.
\newblock Fidelity for mixed quantum states.
\newblock J Mod Opt 1994;41:2315-2323.

\bibitem[Kenfack and {\.Z}yczkowski, 2004]{kenfack2004negativity}
Kenfack A, {\.Z}yczkowski K.
\newblock Negativity of the Wigner function as an indicator of
  non-classicality.
\newblock J Opt B: Quantum Semiclass Opt 2004;6:396.

\bibitem[Shen et~al., 2015]{shen2015nonlinear}
Shen Y, Assad SM, Grosse NB, et~al.
\newblock Nonlinear entanglement and its application to generating cat states.
\newblock Phys Rev Lett 2015;114:100403.

\bibitem[Sun et~al., 2019]{sun2019schrodinger}
Sun FX, He Q, Gong Q, et~al.
\newblock Schr{\"o}dinger cat states and steady states in subharmonic
  generation with kerr nonlinearities.
\newblock Phys Rev A 2019;100:033827.

\bibitem[Teh et~al., 2020]{teh2020dynamics}
Teh RY, Sun FX, Polkinghorne RES, et~al.
\newblock Dynamics of transient cat states in degenerate parametric oscillation
  with and without nonlinear Kerr interactions.
\newblock Phys Rev A 2020;101:043807.

\bibitem[Sun et~al., 2021]{sun2021remote}
Sun FX, Zheng SS, Xiao Y, et~al.
\newblock Remote generation of magnon Schr{\"o}dinger cat state via
  magnon-photon entanglement.
\newblock Phys Rev Lett 2021;127:087203.

\bibitem[Kwon et~al., 2019]{kwon2019nonclassicality}
Kwon H, Tan KC, Volkoff T, et~al.
\newblock Nonclassicality as a quantifiable resource for quantum metrology.
\newblock Phys Rev Lett 2019;122:040503.

\bibitem[Ge et~al., 2020]{ge2020operational}
Ge W, Jacobs K, Asiri S, et~al.
\newblock Operational resource theory of nonclassicality via quantum metrology.
\newblock Phys Rev Res 2020;2:023400.

\bibitem[Ge and Zubairy, 2020]{ge2020evaluating}
Ge W, Zubairy MS.
\newblock Evaluating single-mode nonclassicality.
\newblock Phys Rev A 2020;102:043703.

\bibitem[Adiv et~al., 2022]{adiv2022observation}
Adiv Y, Hu H, Tsesses S, et~al.
\newblock Observation of 2D Cherenkov radiation.
\newblock Phys Rev X 2023;13:011002.

\end{thebibliography}
\end{document}


\title{Supplemental Material on ``Generating optical cat states via quantum interference of multi-path free-electron--photons interactions''}

\author[add1,add2]{Feng-Xiao Sun}
\author[add1]{Yiqi Fang}
\author[add1,add2,add3,add4]{Qiongyi He\corref{cor1}}
\ead{qiongyihe@pku.edu.cn}
\author[add1,add2,add3,add5]{Yunquan Liu\corref{cor1}}
\ead{Yunquan.liu@pku.edu.cn}

\cortext[cor1]{Corresponding author}
\address[add1]{State Key Laboratory for Mesoscopic Physics, School of Physics, Frontiers Science Center for Nano-optoelectronics, $\&$ Collaborative Innovation Center of Quantum Matter, Peking University, Beijing 100871, China}
\address[add2]{Collaborative Innovation Center of Extreme Optics, Shanxi University, Taiyuan, Shanxi 030006, China}
\address[add3]{Peking University Yangtze Delta Institute of Optoelectronics, Nantong 226010, Jiangsu, China}
\address[add4]{Hefei National Laboratory, Hefei 230088, China}
\address[add5]{Beijing Academy of Quantum Information Sciences, Beijing 100193, China}

\begin{abstract}
In this Supplemental Material, we provide a detailed description of the theoretical analysis on the free-electrons--photons interactions in PINEM, the corresponding ideal cat states, the interference of the multi-channel interactions, and the discussion on the experimental realization.
\end{abstract}

\maketitle

\section{Analysis on the free-electron--photon interaction of PINEM}

To investigate the optical cat states generated by the free-electron--photon interactions of photon-induced near-field microscopic (PINEM), we will firstly review the main results of the quantum-optical generalization of PINEM (QPINEM)  theory~\cite{di2019probing,kfir2019entanglements,henke2021integrated,ben2021shaping,dahan2021imprinting}.
For simplicity, we assume that only one electron $e$ passes through the PINEM at a time, whose velocity $v$ is along the $z$ direction. Then the Hamiltonian of PINEM takes the form of,
\begin{equation}
	H=H_e+H_p+H_{\text{int}}=-i\hbar v\frac{\partial}{\partial z}+\hbar\omega a^{\dagger}a+evA_z,
	\label{eq:initial_hamiltonian}
\end{equation}
where $a$ ($a^{\dagger}$) is the annihilation (creation) operator of the optical mode, $\omega$ is the optical frequency, and $A_z$ is the $z$ projection of the vector potential of light. Here $H_e=-i\hbar v{\partial}/{\partial z}$ is the free Hamiltonian of the electron derived from the Dirac equation~\cite{di2019probing}, $H_p=\hbar\omega a^{\dagger}a$ is the free Hamiltonian of the optical mode, and $H_{\text{int}}=evA_z$ is the free-electron--photon interaction Hamiltonian. An assumption has been taken that there is only a single optical mode $a$ that interacts with the electron. 

The vector potential $A_z$ can be further quantized as,
\begin{equation}
  A_z=\sqrt{\frac{\hbar}{2 \epsilon \omega}}\left(u_z a+u^{\ast}_z a^{\dagger}\right).
\end{equation}
Here $\epsilon$ is the optical permittivity, and $u_z$ is the vector mode function. Thus, the interaction Hamiltonian can be characterized as,
\begin{equation}
	H_{\text{int}}=\sqrt{\frac{\hbar}{2 \epsilon \omega}}ev(u_z a+u_{z}^{\ast}a^{\dagger}).
	\label{eq:hamiltonian}
\end{equation}
By taking the ladder operators of electron into account, $b=\exp(-i\omega z/v)$, $b^{\dagger}=\exp(i\omega z/v)$, which describe the translation in momentum (and equivalently energy) of the electron, the interaction Hamiltonian $H_{\text{int}}$ can be further simplified as~\cite{kfir2019entanglements,henke2021integrated}, 
\begin{equation}
H_{\text{int}}=\hbar\xi(ba^\dagger-b^\dagger a),
\end{equation} 
where $\xi=\sqrt{\frac{e^2v^2}{2\epsilon\hbar\omega}}$ is the coupling strength.

In the interaction picture, the interaction Hamiltonian $H^{I}_{\text{int}}=\exp({iH_0t}/{\hbar})H_{\text{int}}\exp(-{iH_0t}/{\hbar})$ with $H_0=H_e+H_p$. Note that $a_{I}(t)=\exp({iH_0t}/{\hbar})a\exp(-{iH_0t}/{\hbar})=a e^{-i\omega t}$, $b_{I}(t)=\exp({iH_0t}/{\hbar})b\exp(-{iH_0t}/{\hbar})=e^{vt\frac{\partial}{\partial z}}e^{-i\omega z/v}e^{-vt\frac{\partial}{\partial z}}=e^{-i\omega (z+vt)/v}=b e^{-i\omega t}$. Thus, the interaction Hamiltonian remains time-independent, $H^{I}_{\text{int}}=\hbar\xi[b_{I}(t)a_{I}^\dagger(t)-b_{I}^\dagger(t) a_{I}(t)]=\hbar\xi(ba^\dagger-b^\dagger a)$. Considering an arbitrary initial state of photons and electron $|\psi_i\rangle_{p-e}$, the state after the PINEM interaction can be described as $|\psi\rangle_{p-e}=S|\psi_i\rangle_{p-e}$, where the scattering matrix $S$ can be calculated by,
\begin{equation}
	S=\mathcal{T}\exp\left[-\frac{i}{\hbar}\int_{-\infty}^{\infty}dt H^{I}_{\text{int}}(t')\right]= \exp(ga^{\dagger}b-g^{\ast}ab^{\dagger}).
	\label{eq:scattering}
\end{equation}
Here $\mathcal{T}$ is the time-ordering operator, and the quantum coupling constant $g=-i\xi\tau$ with $\tau$ being the effective interaction duration. In the QPINEM theory~\cite{di2019probing,kfir2019entanglements,ben2021shaping,henke2021integrated,dahan2021imprinting}, the quantum coupling constant $g$ is usually denoted as $g_q$ or $g_{Qu}$, where we just mark it as $g$ for simplicity. And in the above discussion, we focus on the case where only single-mode electromagnetic fields are included. Whereas, it can be generalized for multi-mode cases and the effective electron-light coupling strength per photon takes the following formation of~\cite{ben2021shaping,henke2021integrated,dahan2021imprinting},
\begin{equation}
	g=\frac{e}{\hbar\omega}\int_{-\infty}^{\infty}dz' e^{-i\omega z'/v}E_{z}(z'),
	\label{eq:coupling}
\end{equation}
with the electric field per photon $E_z=-\partial A_z/\partial t$.

By using the Baker-Campbell-Hausdorff formula~\cite{bonfiglioli2011topics}, 
\begin{equation}
	\exp X\exp Y=\exp\left\{X+Y+\frac{1}{2}[X,Y]+\frac{1}{12}[X,[X,Y]]-\frac{1}{12}[Y,[X,Y]]+\dots\right\},
	\label{eq:BCH}
\end{equation}
and the commutation relation of $[ga^{\dagger}b,-g^{\ast}ab^{\dagger}]=|g|^2$, the scattering matrix can be rewritten as,
\begin{align}
	S&=exp(-\frac{1}{2}|g|^2)\exp(ga^{\dagger}b)\exp(-g^{\ast}ab^{\dagger})\nonumber\\
	&=\exp(-\frac{|g|^2}{2})\sum_{j,l}\frac{g^{j}(-g^{\ast})^{l}}{j!l!}(a^{\dagger}b)^{j}(ab^{\dagger})^{l},
	\label{eq:sm}
\end{align}
which is exactly Eq.~(2) of the main text.  Here it is noticed that the ladder operators of electron satisfy $bb^{\dagger}=\exp(-i\omega z/v)\exp(i\omega z/v)=1$, $b^{\dagger}b=\exp(i\omega z/v)\exp(-i\omega z/v)=1$, which results in $[b,b^{\dagger}]=0$.

In the following, we focus on the experimentally feasible case where the input optical state is a coherent state $|\alpha\rangle_p$ and the electron state is $|0\rangle_e$ with the baseline energy $E_0$. The electron states are marked with the basis of energy-ladder states $|k\rangle_e$, which corresponds to an electron with energy $E_0+k\hbar\omega$. 
By applying the scattering matrix (\ref{eq:sm}) to the initial state $|\psi_i\rangle_{p-e}=|\alpha\rangle_p|0\rangle_e$, the final state is obtained,
\begin{equation}
	|\psi\rangle_{p-e}=S|\psi_i\rangle_{p-e}=\mathcal{N}\sum_{j,l=0}^{\infty}\frac{(g)^j(-g^\ast\alpha)^l}{j!l!}(a^{\dagger})^{j}|\alpha\rangle_p|l-j\rangle_e.
	\label{eq:final}
\end{equation}
Here $\mathcal{N}$ is the normalized parameter, and the fact that $a^l|\alpha\rangle_p=\alpha^l|\alpha\rangle_p$ has been used. Then by performing project measurement on the electrons and postselecting a particular state $|k\rangle_e$, the conditional optical state takes the form,
\begin{align}
	|\psi^{(k)}\rangle_p&={}_e\langle k|\psi\rangle_{p-e}\nonumber\\
	&=\mathcal{N}\sum_{j,l=0}^{\infty}\frac{(g)^j(-g^\ast\alpha)^l}{j!l!}(a^{\dagger})^{j}|\alpha\rangle_p{}_e\langle k|l-j\rangle_e\nonumber\\
	&=\mathcal{N}\sum_{j,l=0}^{\infty}\frac{(g)^j(-g^\ast\alpha)^l}{j!l!}(a^{\dagger})^{j}|\alpha\rangle_p\delta_{k,l-j}\nonumber\\
	&=\mathcal{N}_p\sum_{j=\text{max}\{0,-k\}}^{\infty}\frac{(-\alpha|g|^2)^{j}}{j!(k+j)!}(a^{\dagger})^{j}|\alpha\rangle_p.
	\label{eq:photonic}
\end{align}
We then define $C_j^{(k)}=\mathcal{N}_p(-\alpha|g|^2)^{j}/j!(j+k)!$ as the superposition coefficient of the optical state $|\phi_j\rangle_p=(a^{\dagger})^{j}|\alpha\rangle_p$ created by the individual quantum channel $(a^{\dagger})^{j}a^{j+k}$. According to its formation, $C_j^{(k)}$ will drop dramatically with the increase of $j$. Hence, in the case where the coupling strength is not too large, there will be only a small number of quantum channels that contribute to the generation of optical states. 
\begin{figure}[tb]
	\centering
	\includegraphics[width=0.3\textwidth]{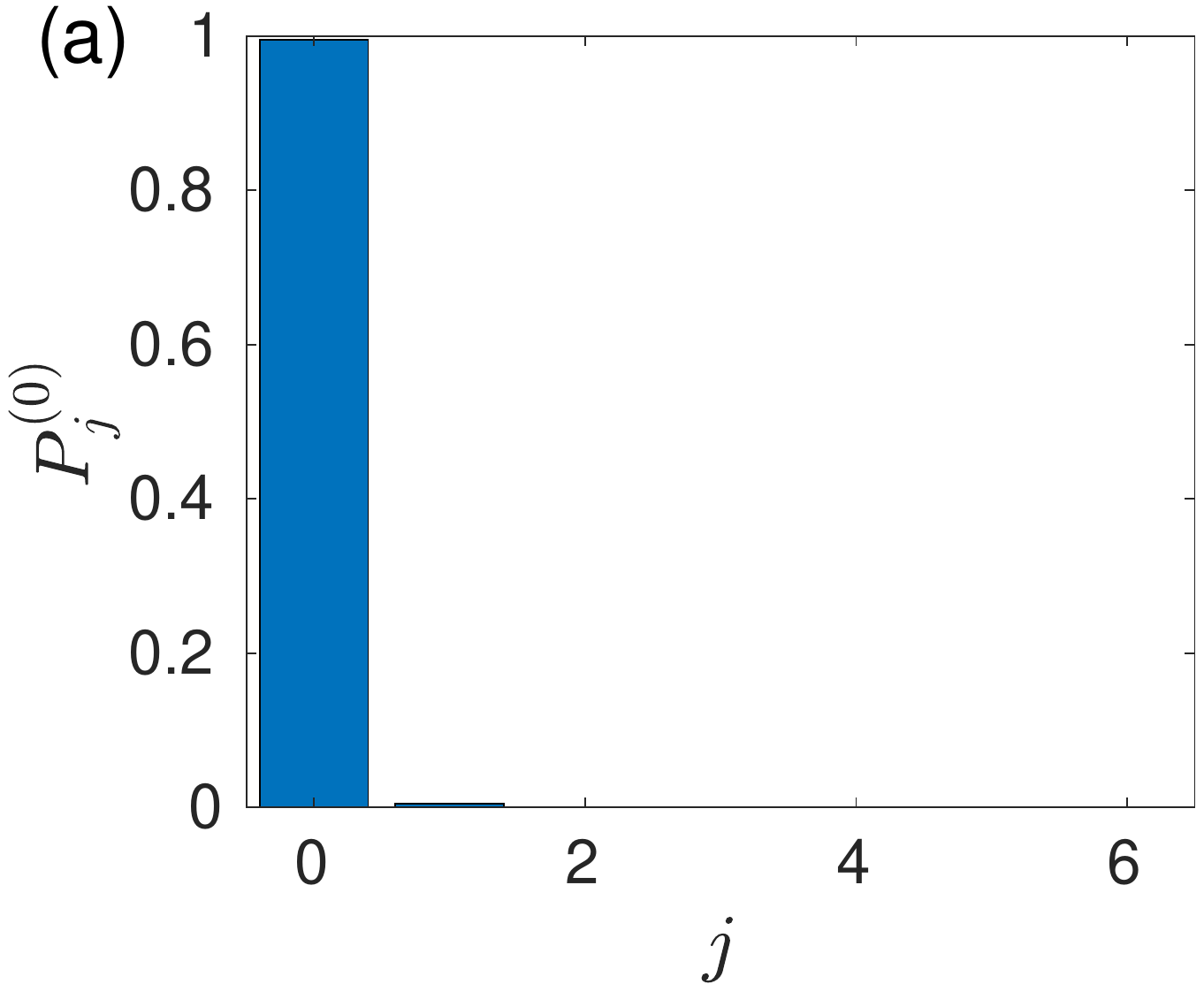}\hspace{1em}
	\includegraphics[width=0.3\textwidth]{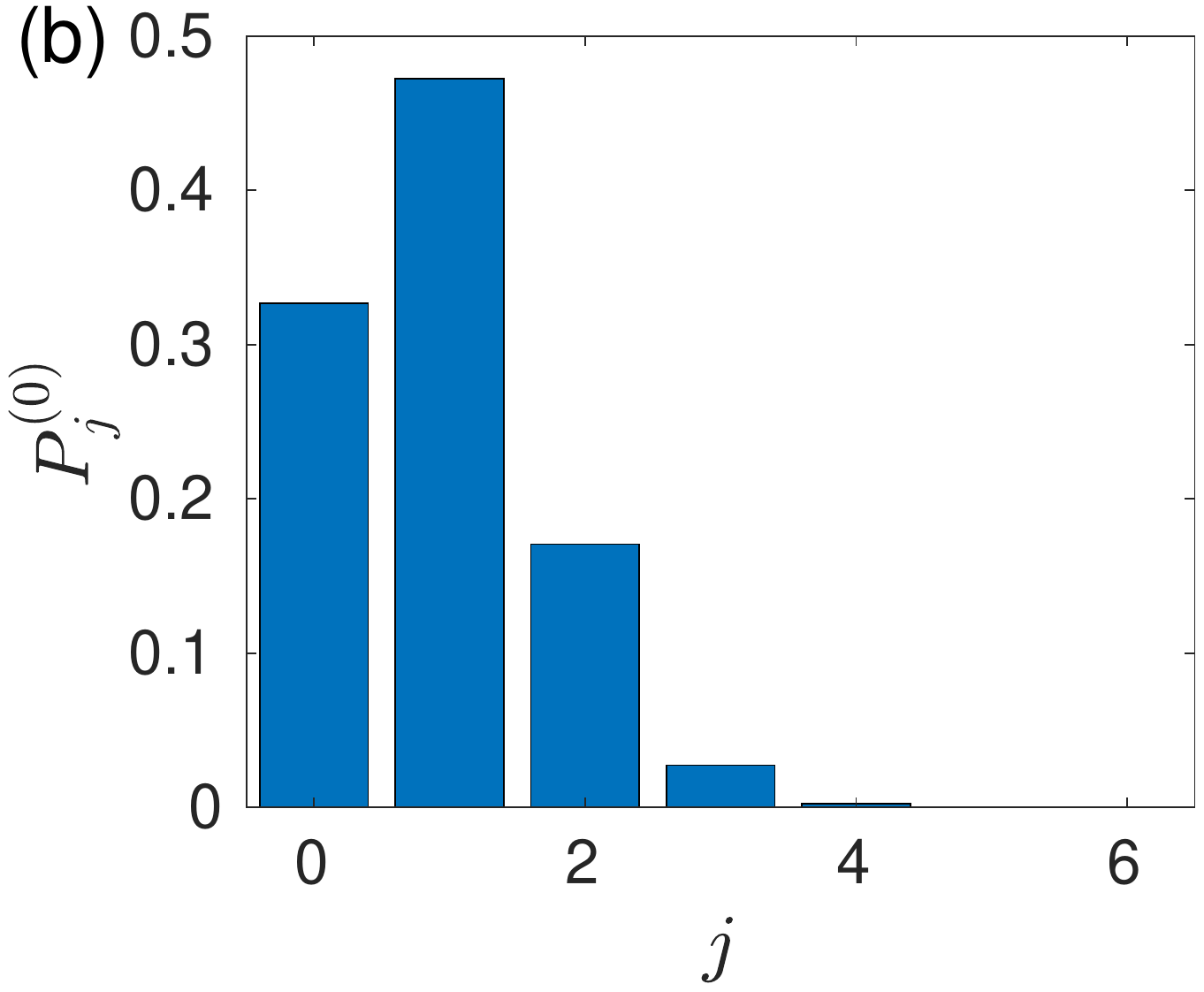}
	\caption{The weight of each quantum channel in forming the optical state with (a) $|g|=0.01$ and (b) $|g|=0.17$. The postselection on the electron is chosen as $k=0$, and the photon number of the input coherent state is $|\alpha|^2=50$.}
	\label{fig:weight}
\end{figure}
\begin{figure}[tb]
	\centering
	\includegraphics[width=0.24\textwidth]{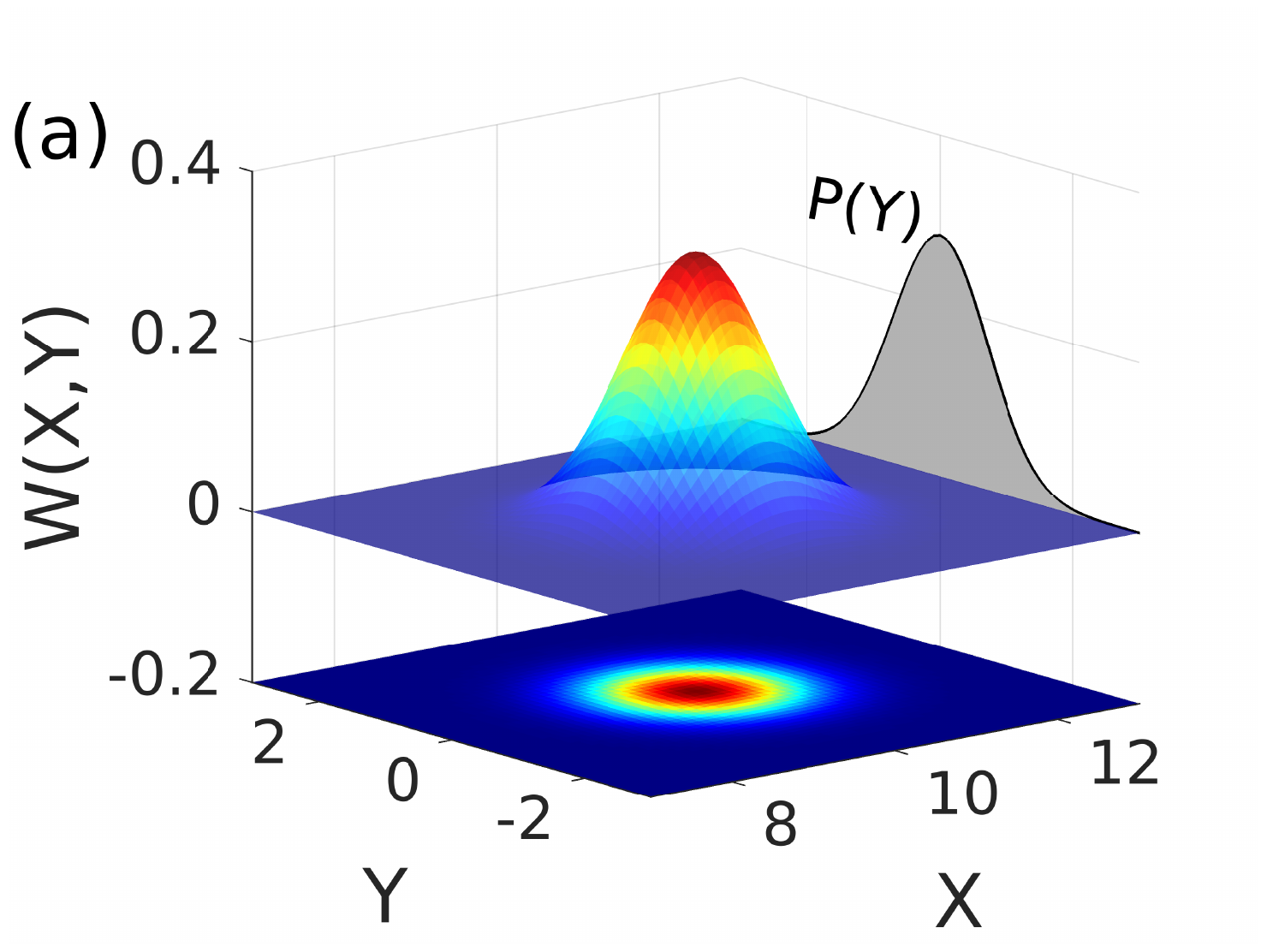}
	\includegraphics[width=0.24\textwidth]{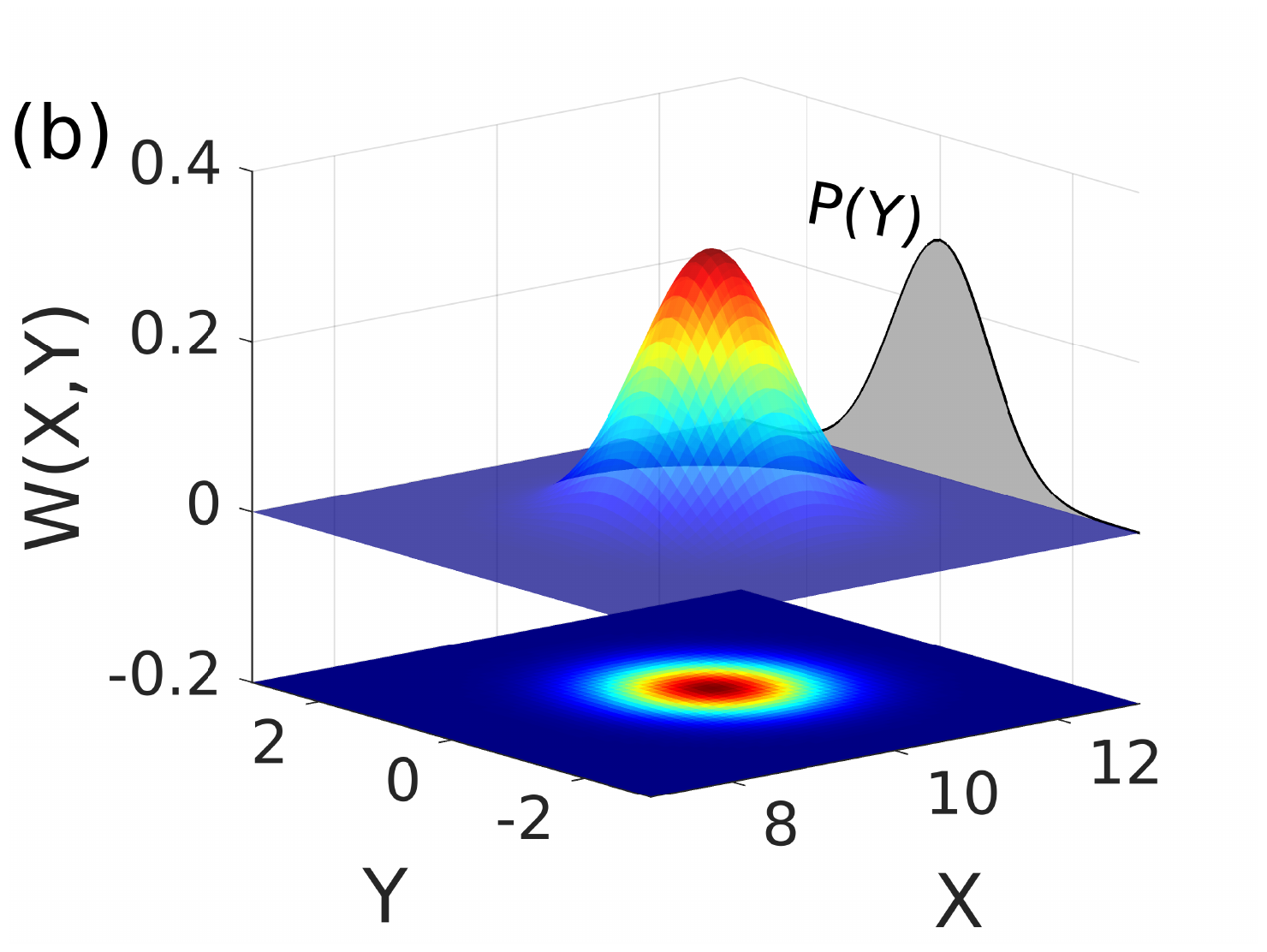}
	\includegraphics[width=0.24\textwidth]{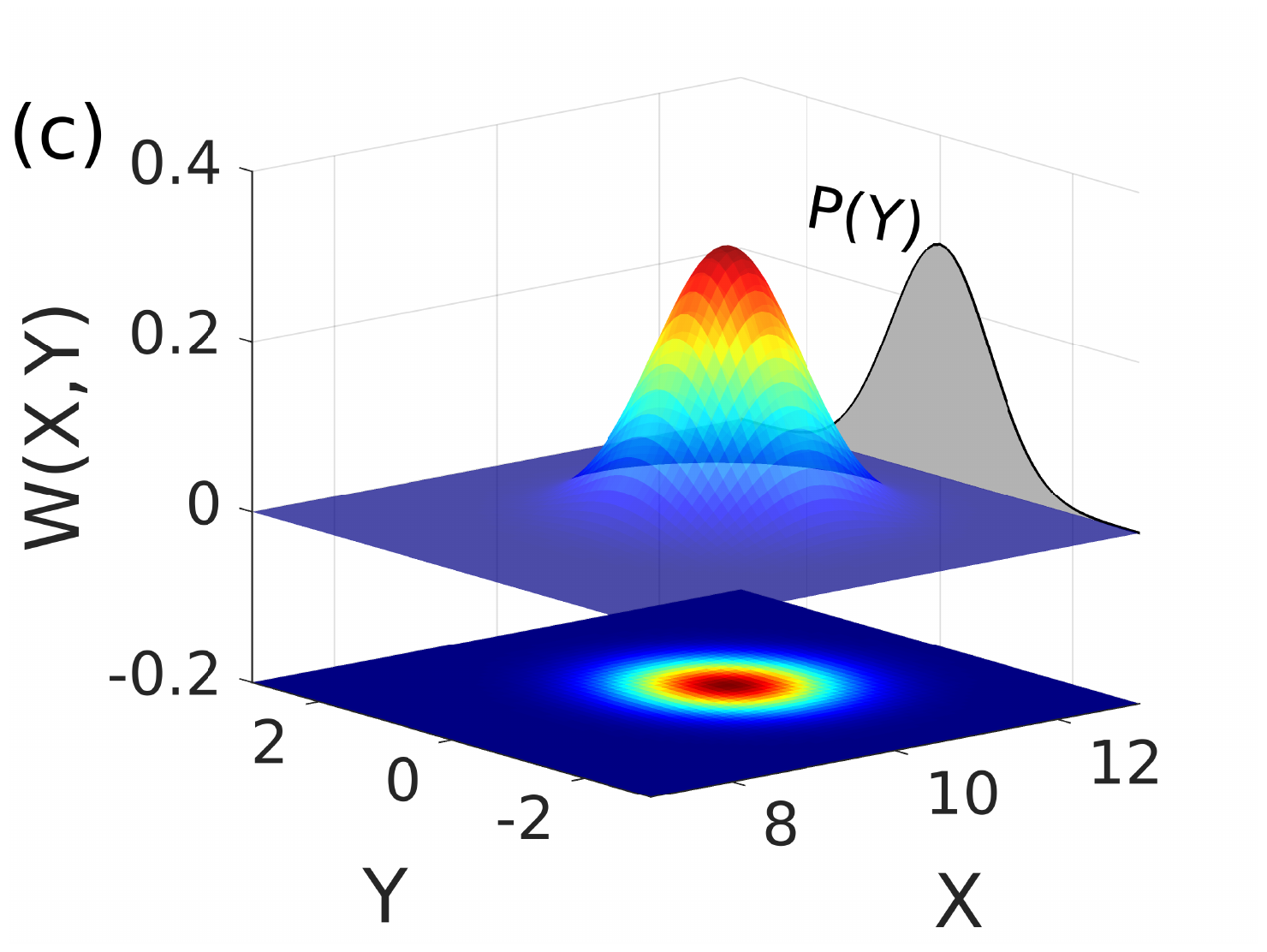}
	\includegraphics[width=0.24\textwidth]{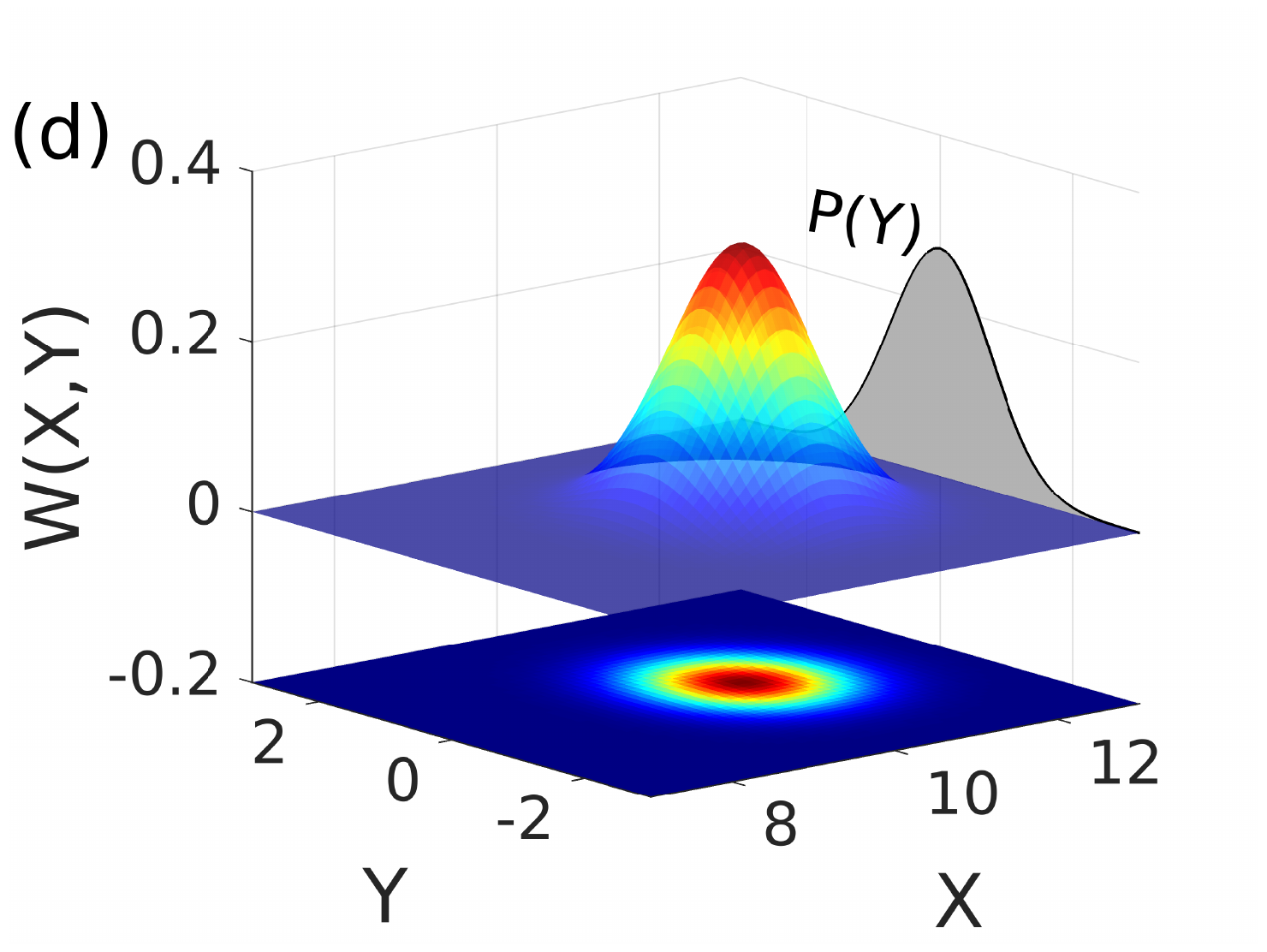}
	\caption{Wigner functions for the optical states $|\phi_j\rangle_p$ generated by each quantum channel $(a^{\dagger})^{j}a^{j+k}$ individually. Here the quantum channels are chosen as (a) $\{j=0:I\}$, (b) $\{j=1:a^{\dagger} a\}$, (c) $\{j=2:(a^{\dagger})^2 a^2\}$ and (d) $\{j=3:(a^{\dagger})^3 a^3\}$. The postselection on the electron is set as $k=0$, and the initial photon number of optical state is $|\alpha|^2=50$.}
	\label{fig:channel}
\end{figure}

In order to study the interference effect of the multiple interaction channels, we define the weight of each quantum channel. Please note that the generated optical state $|\phi_j\rangle_p$ in each channel has not been normalized. Thus the weight of interaction channel should be evaluated by the ratio $P_{j}^{(k)}=|\tilde{C}_j^{(k)}|/\sum_j|\tilde{C}_j^{(k)}|$ with $\tilde{C}_j^{(k)}=\sqrt{{}_p\langle\phi_j|\phi_j\rangle_p}C_j^{(k)}$. Specially, for the case of $|\alpha|^2\gg 1$, we have $\sqrt{{}_p\langle\phi_j|\phi_j\rangle_p}\approx (\alpha^{\ast})^j$ and thus $\tilde{C}_j^{(k)}\approx \mathcal{N}_p(-|\alpha|^2|g|^2)^{j}/j!(j+k)!$. The results of $P_{j}^{(k)}$ for the case of weak PINEM interaction $|g|=0.01$ and the postselection on electron energy $k=0$ are shown in Fig.~\ref{fig:weight}(a), where only a single quantum channel of $j=0$ occupies a large probability. The weights of other quantum channels are negligible compared to quantum channel $j=0$. Since the quantum channel is $\{j=0:I\}$ with identity operator $I$, the generated optical state will remain unchanged as a coherent state, which is positive in Wigner function. Furthermore, for the case of  $k>0$, the single channel is $\{j=0:a^{k}\}$. As the photon-subtraction operation doesn't change the coherent state, its original quantum statistics also maintains. And for the case of $k<0$, the single channel $\{j=|k|:(a^\dagger)^{|k|}\}$ produces a $|k|$-photon-added optical state. It indeed changes the Wigner function of initial coherent state from Gaussian to non-Gaussian due to the photon-addition operation. However, such change doesn't affect the coherent state significantly as $k$ is relatively small as compared to the number of photons $|\alpha|^2$. 

By increasing the PINEM interaction $|g|$, more quantum channels will be involved and interfere with each other. For the case where an optical cat state is observed with $|g|=0.17$ and $k=0$ shown in Fig.~2(a) of the main text, the weights of quantum channels $P_{j}^{(k)}$ are displayed in Fig.~\ref{fig:weight}(b). It is found that there are $4$ channels ($j=0,1,2,3$) whose weights are not ignorable. And the output optical state of each channel is shown in Fig.~\ref{fig:channel}, where all the output optical states remain positive in Wigner functions. In this case, the average photon number in the initial optical states $|\alpha|^2=50$ is much large than the values of $k$ and $j$. Although there are photon-addition operations contributed in the channels which change the Wigner function of initial coherent state from Gaussian to non-Gaussian, the change is not large enough to observe Wigner negativity. Thus, the generation of optical cat states with negative Wigner function shown in Fig.~2(a) of the main text can only result from the interference between the interaction channels.
\begin{figure}[tb]
	\centering
	\includegraphics[width=0.3\textwidth]{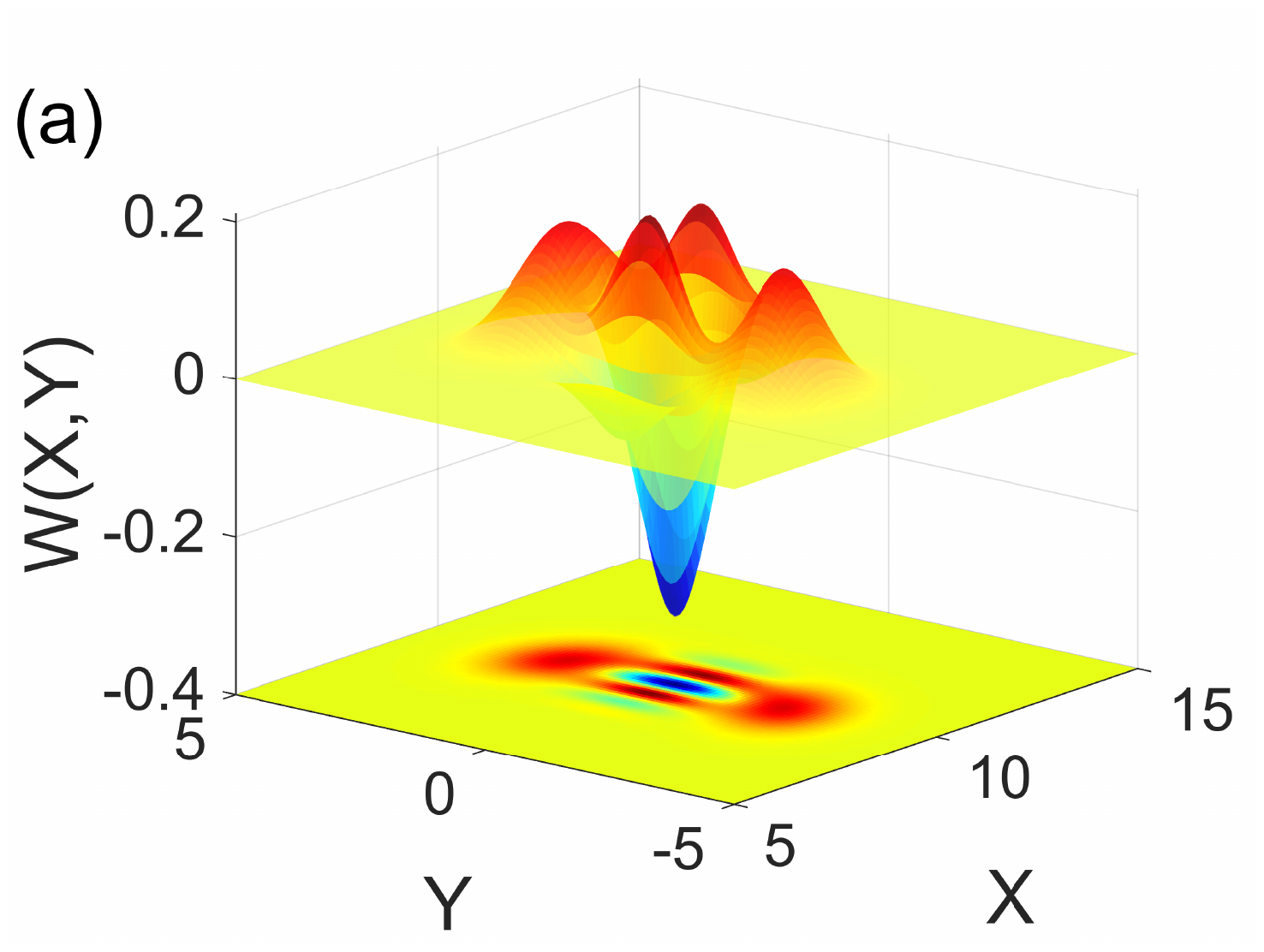}
	\includegraphics[width=0.3\textwidth]{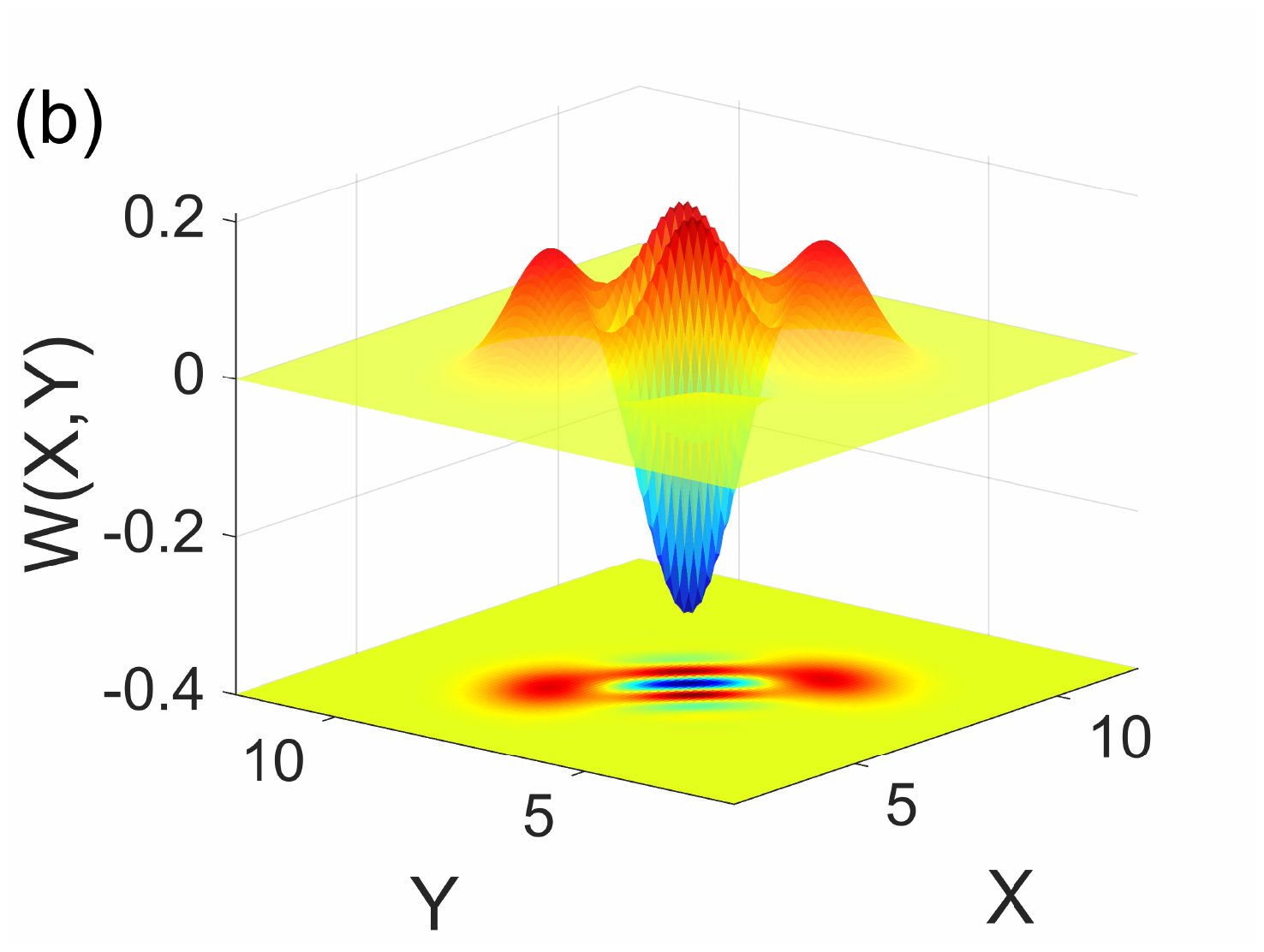}
	\includegraphics[width=0.3\textwidth]{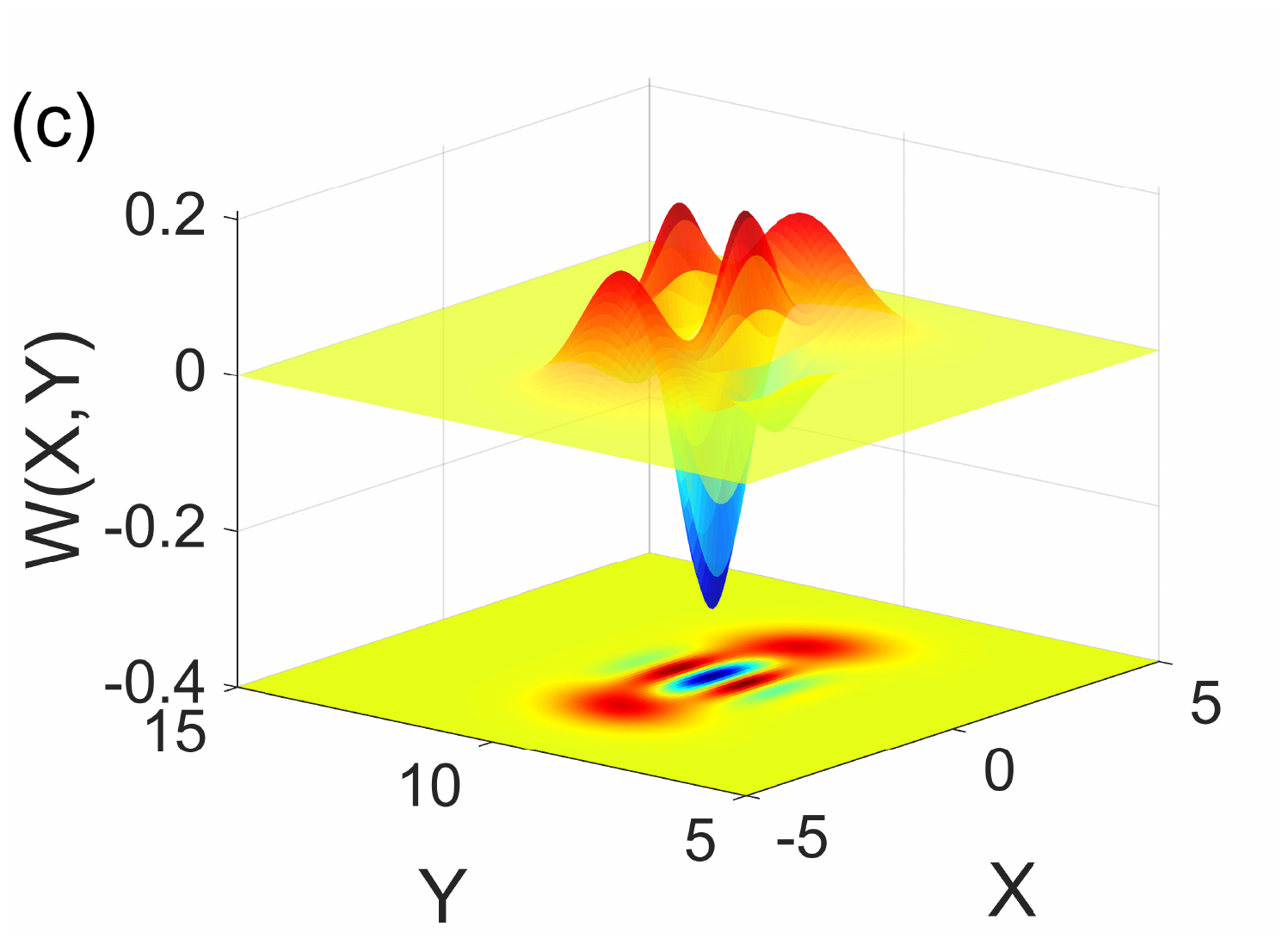}\\
	\includegraphics[width=0.3\textwidth]{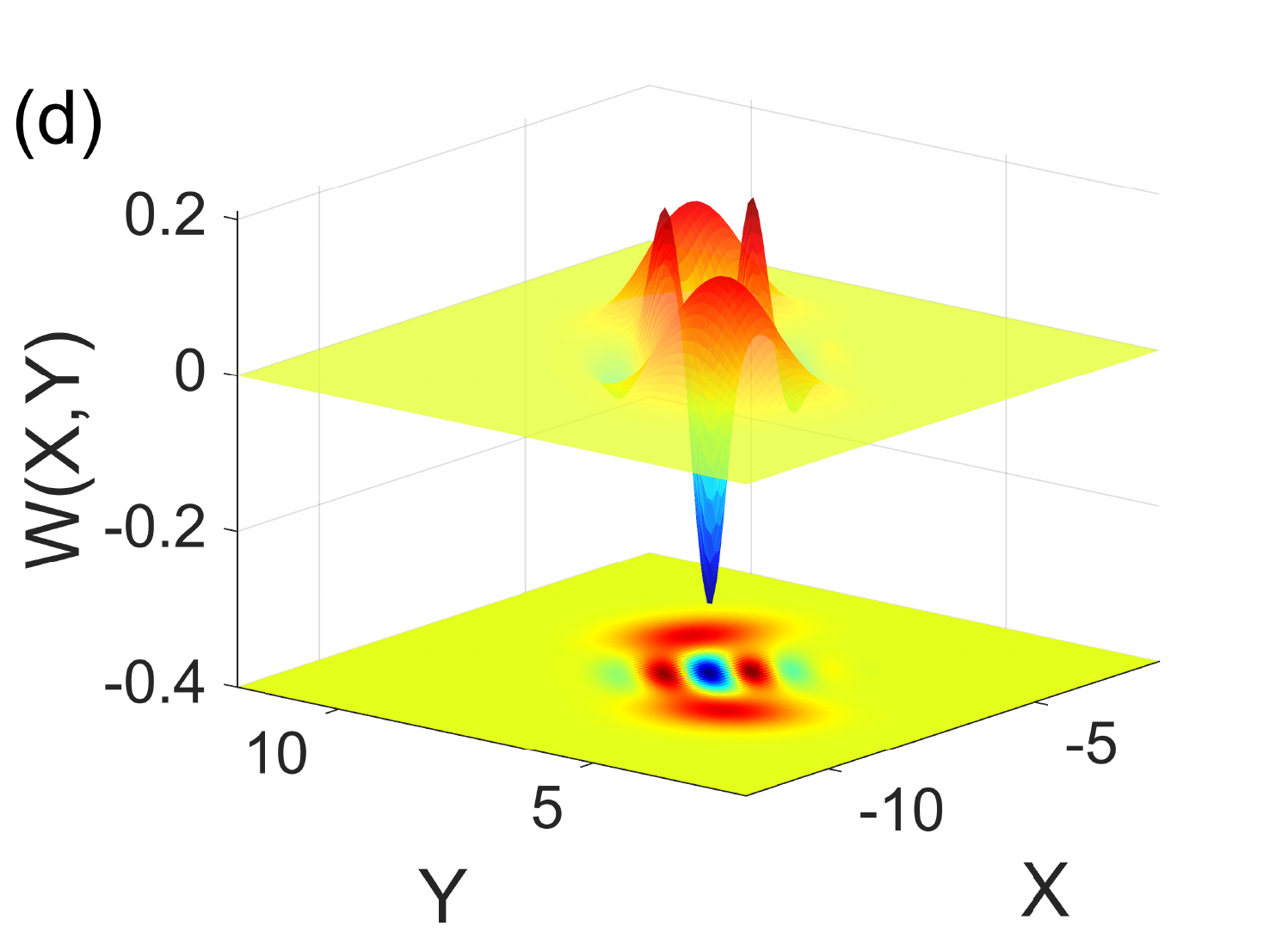}
	\includegraphics[width=0.3\textwidth]{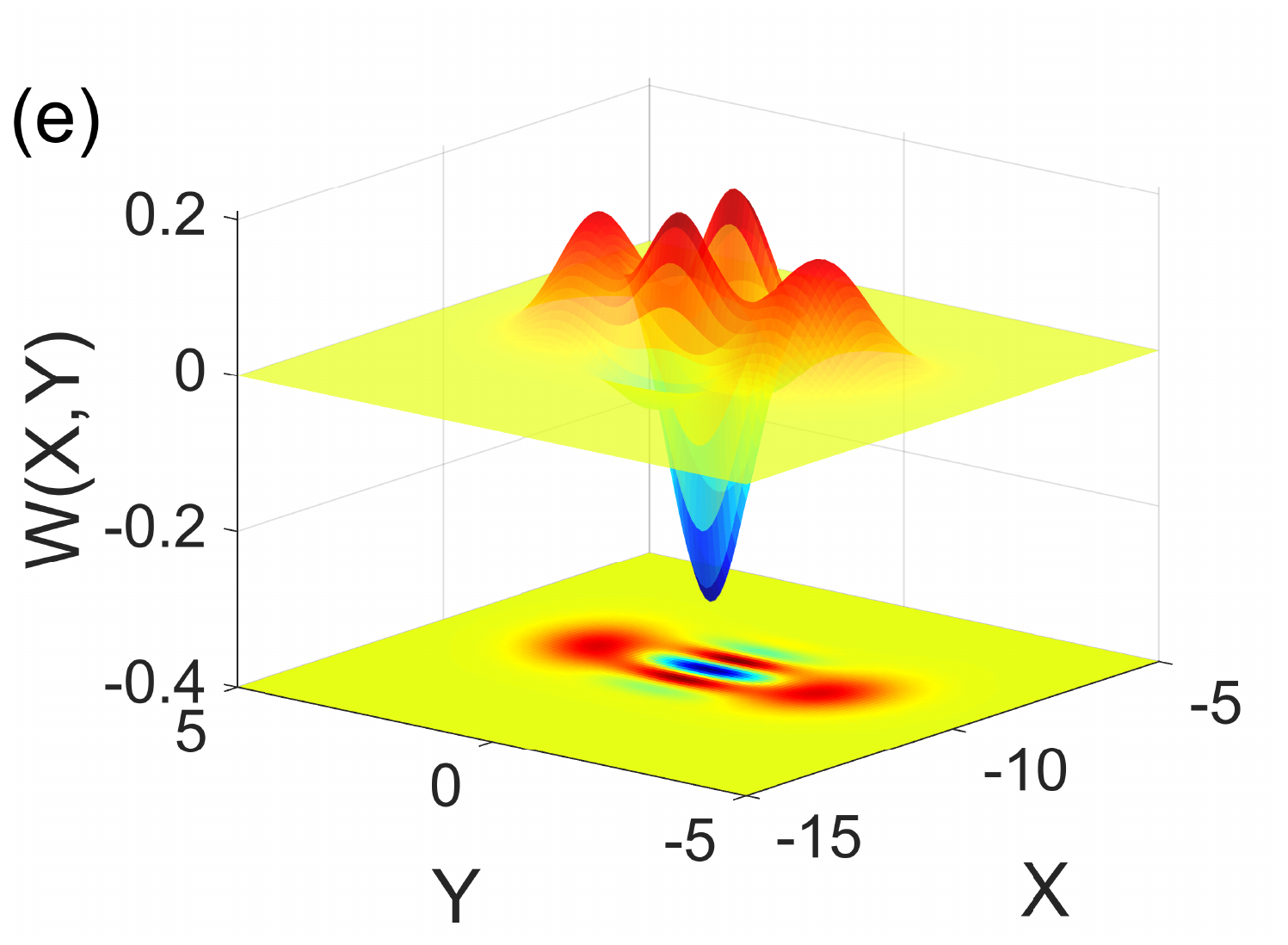}
	\includegraphics[width=0.3\textwidth]{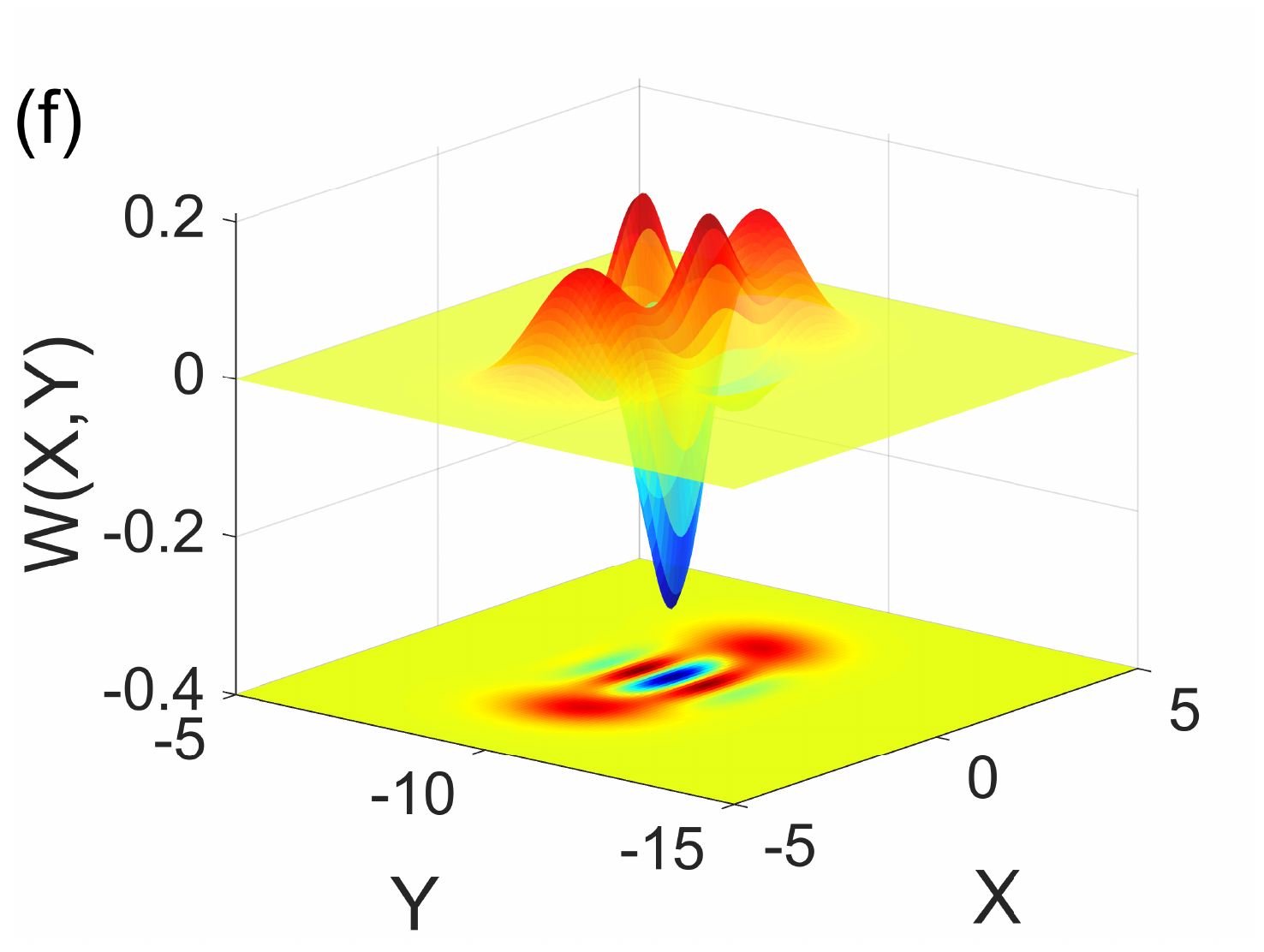}
	\caption{Wigner functions for the optical states $|\phi_j\rangle_p$ with the phase of the initial coherent state (a) $\phi=0$, (b) $\phi=\pi/4$, (c) $\phi=\pi/2$, (d) $\phi=3\pi/4$, (e) $\phi=\pi$ and (f) $\phi=3\pi/2$. The postselection on the electron is set as $k=0$, the coupling strength is $|g|=1.53$, and the initial photon number of optical state is $|\alpha|^2=50$.}
	\label{fig:cat_phase}
\end{figure}

In addition, as we have mentioned in the main text, the coherent amplitude is in general complex, $\alpha=|\alpha| e^{i\phi}$, and the electron energy-ladder states are insensitive to the phase $\phi$. Thus, we will apply the following transition $\tilde{a}=a e^{-i\phi}$ and $\tilde{b}=b e^{i\phi}$, where the scattering matrix becomes
\begin{equation}
	S=\exp(g\tilde{a}^{\dagger}\tilde{b}-g^{\ast}\tilde{a}\tilde{b}^{\dagger})=\exp(-\frac{|g|^2}{2})\sum_{j,l}\frac{g^{j}(-g^{\ast})^{l}}{j!l!}(\tilde{a}^{\dagger}\tilde{b})^{j}(\tilde{a}\tilde{b}^{\dagger})^{l}.
\end{equation}
Therefore, the entangled state takes the form of
\begin{equation}
	|\psi\rangle_{p-e}\propto\sum_{j,l=0}^{\infty}\frac{(g)^j(-g^\ast|\alpha|)^l}{j!l!}(\tilde{a}^{\dagger})^{j}|\alpha\rangle_p|l-j\rangle_e,
\end{equation}
and the conditional optical state is
\begin{equation}
	|\psi^{(k)}\rangle_p=\mathcal{N}_p\sum_{j=\text{max}\{0,-k\}}^{\infty}\frac{(-|\alpha||g|^2)^{j}}{j!(k+j)!}(\tilde{a}^{\dagger})^{j}|\alpha\rangle_p.
\end{equation}
This means that we can only focus on the cases of $\alpha>0$ in the analysis of the quantum channels with superposition coefficients $C_j^{(k)}$. The phase $\phi$ of the initial coherent state just modifies the direction of the generated optical cat state, as indicated in Fig.~\ref{fig:cat_phase}. Without loss of generality, we denote $\tilde{a}$ ($\tilde{b}$) as $a$ ($b$) and take $\alpha>0$ in the main text.

\section{The ideal cat state compared with the created optical state}

To evaluate the quality of the generated optical cat states, we compare them with the ideal odd cat state expressed as follows,
\begin{equation}
	|\varphi_\text{cat}\rangle=\mathcal{N}_c(|\beta\rangle-|\beta^\ast\rangle)=2i\mathcal{N}_c e^{-\frac{|\beta|^2}{2}}\sum_{n=0}^{\infty}\frac{|\beta|^{n}\sin n\phi}{\sqrt{n!}}|n\rangle.
	\label{eq:cat}
\end{equation}
Here, the complex amplitude of the coherent state $\beta=|\beta|\exp(i\phi)$ and the normalized parameter takes the form,
\begin{equation}
\mathcal{N}_c=\left[2-\exp(-|\beta|^2+\beta^2)-\exp(-|\beta|^2+\beta^{\ast 2})\right]^{-\frac{1}{2}}.
\end{equation}
It is more intuitive by rewriting the complex amplitude with its real part $\beta_x=|\beta|\cos\phi$ and imaginary part $\beta_y=|\beta|\sin\phi$, i.e. $\beta=\beta_x+i\beta_y$. Hence, the cat state (\ref{eq:cat}) is equivalent to,
\begin{equation}
	|\varphi_\text{cat}\rangle=\mathcal{N}_c(|\beta_x+i\beta_y\rangle-|\beta_x-i\beta_y\rangle).
\end{equation}
It means that the cat state (\ref{eq:cat}) is a quantum superposition of two coherent states with opposite phase-quadratures $Y=\pm\sqrt{2}\beta_y$ in the phase space. The quadratures are defined as $X=(a+a^\dagger)/\sqrt{2}$ for the amplitude and $Y=(a-a^\dagger)/(\sqrt{2}i)$ for the phase. Compared with the commonly used cat states that are separated in $Y$ direction, $|i\alpha\rangle-|-i\alpha\rangle$, the symmetric center of the cat state (\ref{eq:cat}) is displaced along the amplitude-quadrature with $X_c=\sqrt{2}\beta_x$. 
\begin{figure}[tbp]
	\centering
	\includegraphics[width=0.24\textwidth]{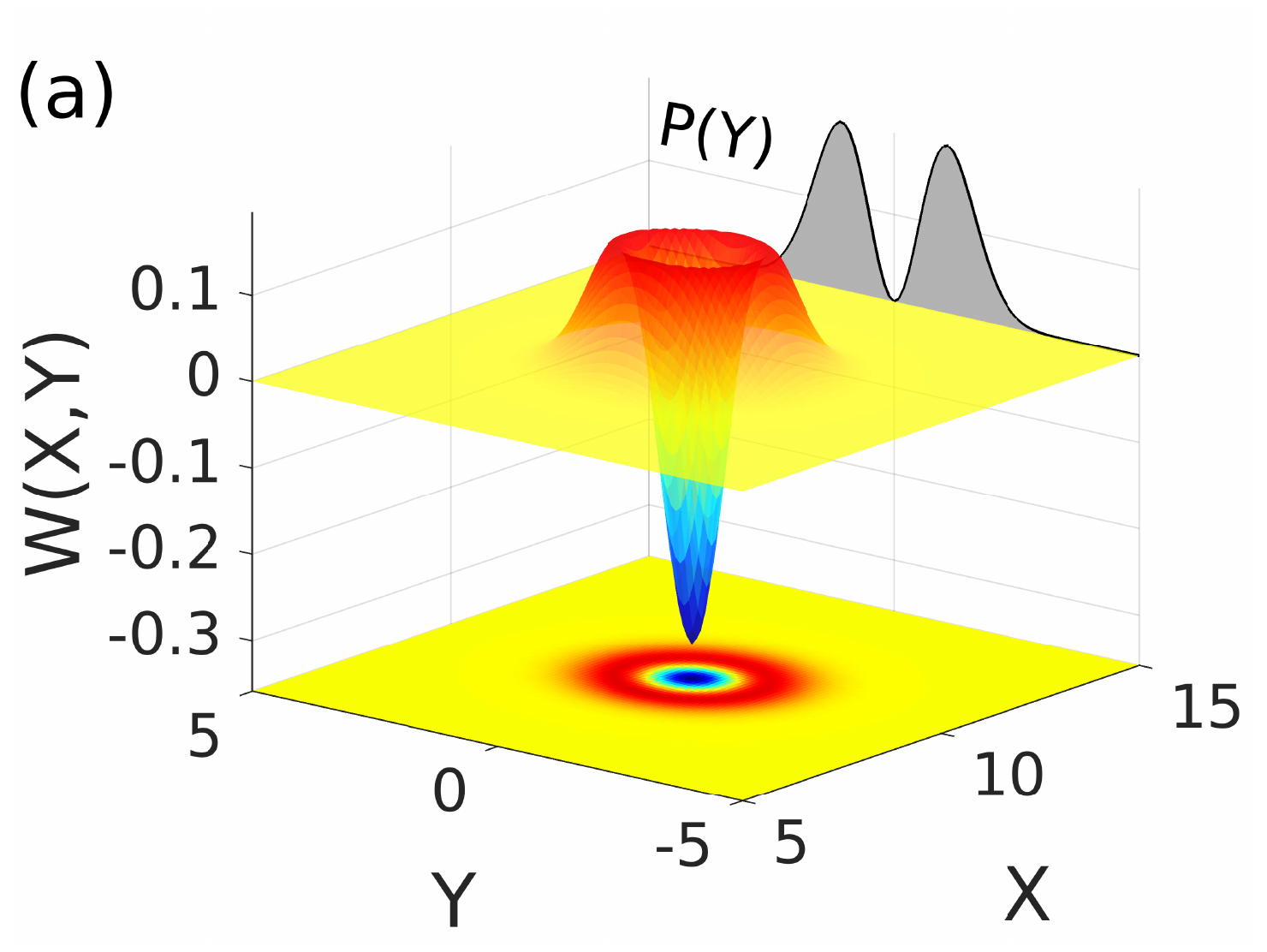}
	\includegraphics[width=0.24\textwidth]{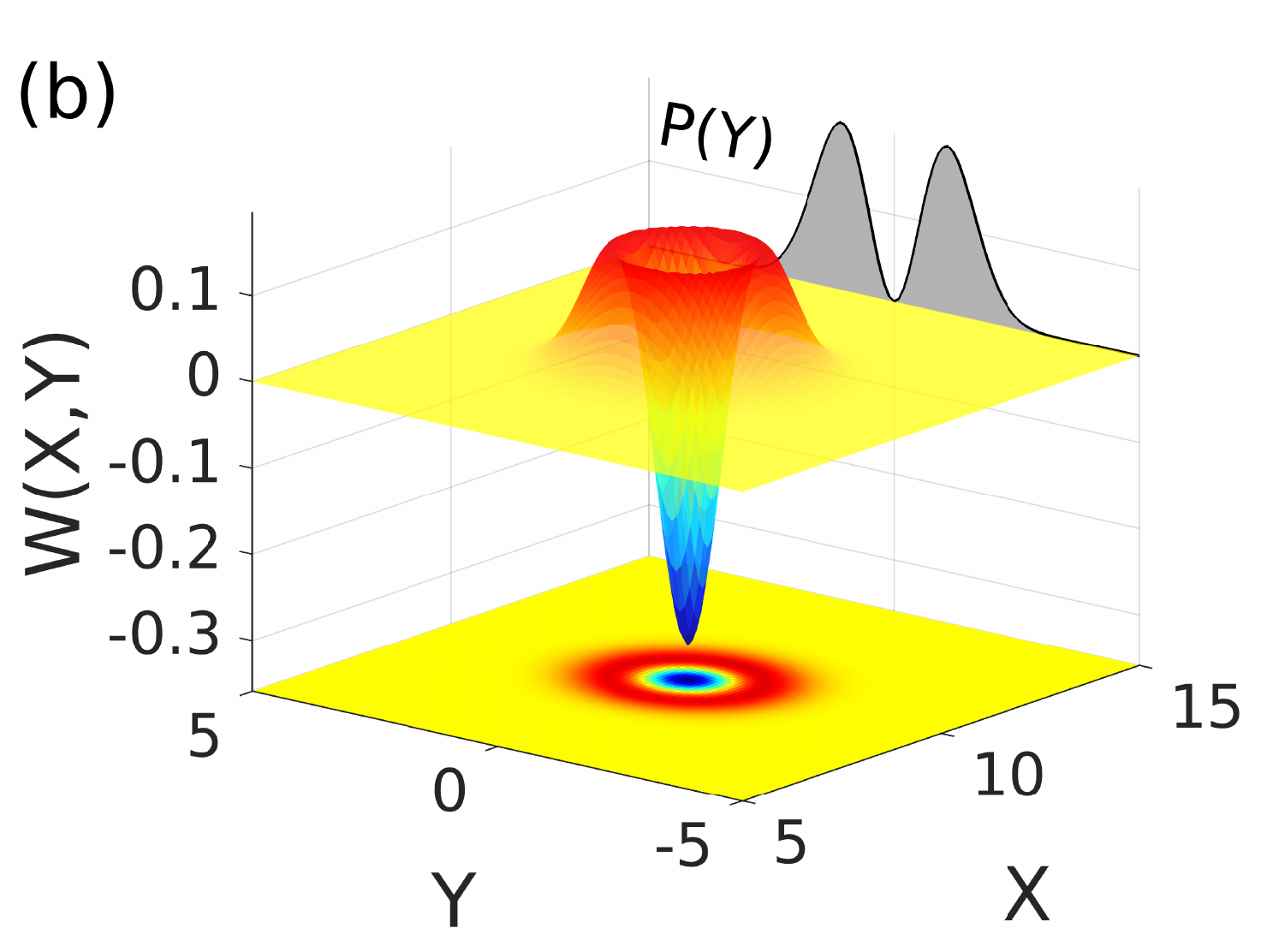}
	\includegraphics[width=0.24\textwidth]{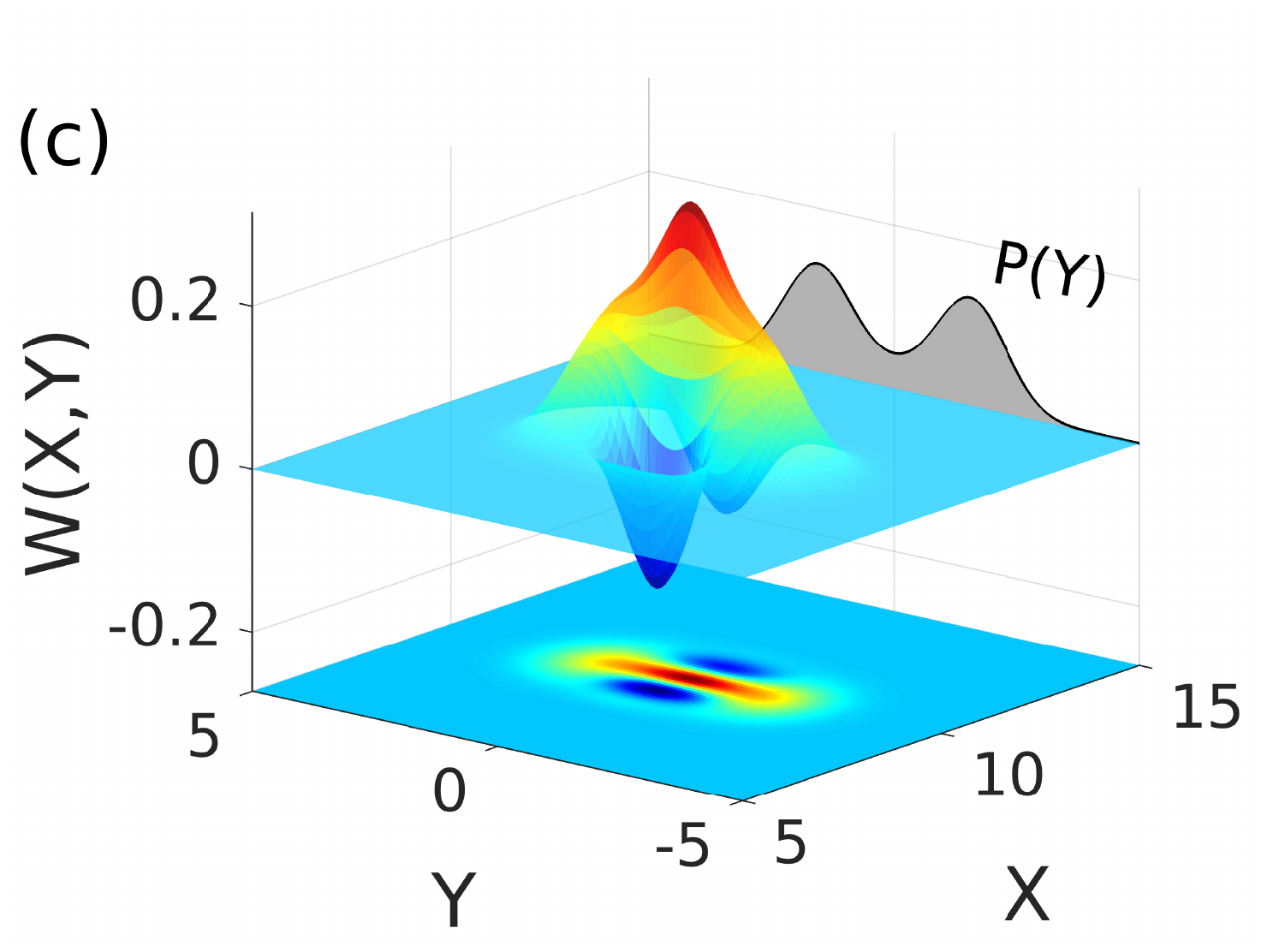}
	\includegraphics[width=0.24\textwidth]{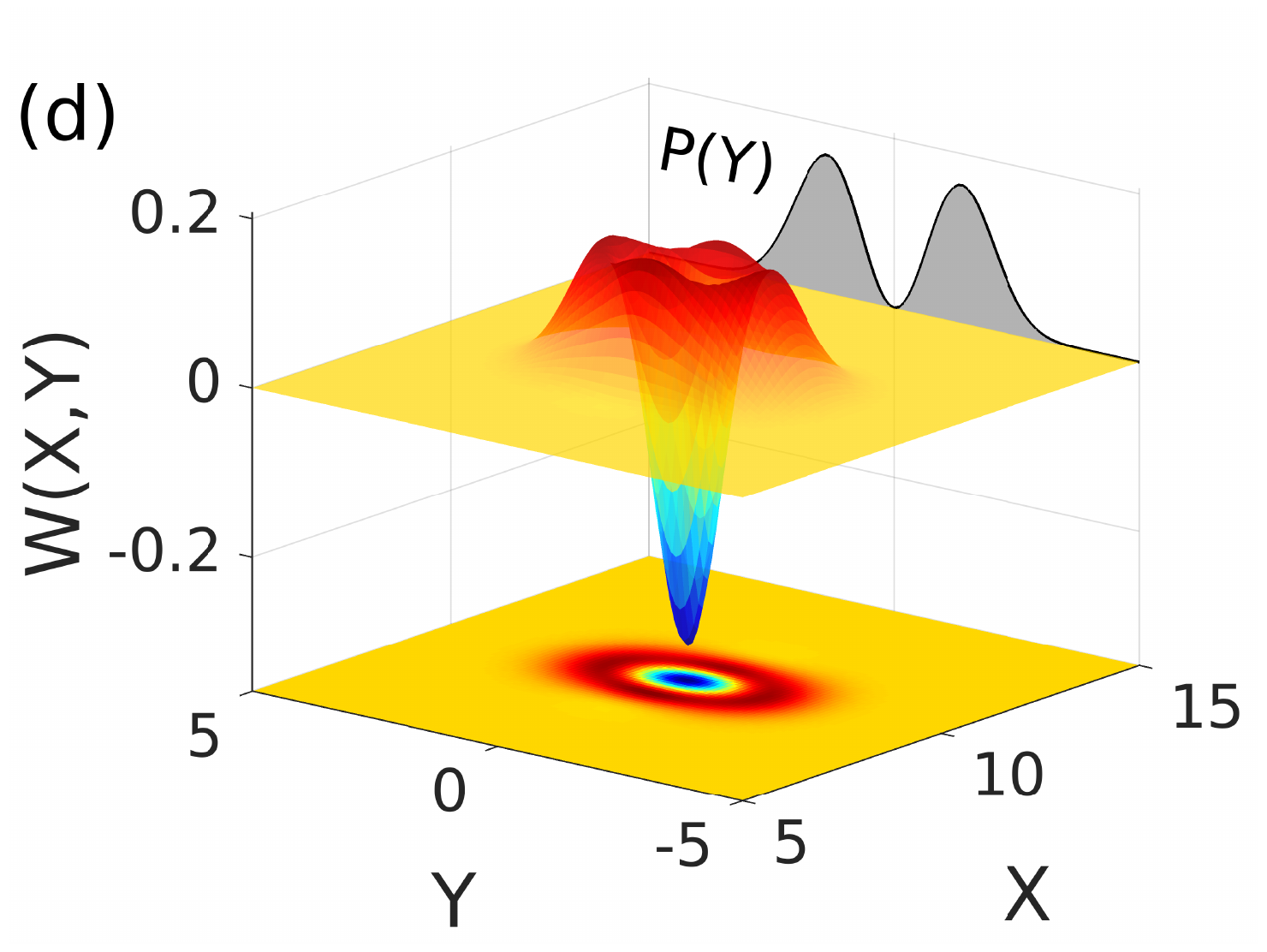}
	\caption{The Wigner functions of the ideal cat states which are taken to compare with the generated optical states displayed in Fig.~2 in the main text. Here, the complex amplitudes of the ideal cat states are (a) $|\beta|=7.021$, $\phi=0.020\pi$, (b) $|\beta|=6.951$, $\phi=0.021\pi$, (c) $|\beta|=6.981$, $\phi=0.051\pi$ and (d) $|\beta|=6.961$, $\phi=0.042\pi$, respectively.}
	\label{fig:cat-WN}
\end{figure}

In order to indicate the reason why the generated optical states (\ref{eq:photonic}) in PINEM possess high fidelity with the cat states (\ref{eq:cat}), we now expand the state (\ref{eq:final}) in the Fock basis, which takes the form of,
\begin{align}
|\psi\rangle_{p-e}=&e^{-\frac{|\alpha|^2+|g|^2}{2}}\sum_{k=-\infty}^{\infty}\sum_{n=\text{max}\{0,-k\}}^{\infty}\frac{\alpha^{k+n}g^{-k}(-|g|^2)^{\text{max}\{0,k\}}\sqrt{n!}}{\text{max}\{0,k\}!(k+n-\text{max}\{0,k\})!(\text{max}\{0,k\}-k)!}\nonumber\\
&\hspace{1em}\times{}_{2}F_{2}(1,\text{max}\{0,k\}-k-n;1+\text{max}\{0,k\},1+\text{max}\{0,k\}-k;|g|^2)|k\rangle_{e}|n\rangle_{p}.
\label{eq:psi_fock}
\end{align}
Here, the function ${}_{2}F_{2}$ is the generalized hypergeometric function. By projecting the electron to its energy-ladder state $|k\rangle_e$ with $k\ge0$, the conditional optical state can be expressed as,
\begin{align}
|\psi^{(k)}\rangle_p&=e^{-\frac{|\alpha|^2+|g|^2}{2}}\sum_{n=0}^{\infty}\frac{\alpha^{k+n}g^{-k}(-|g|^2)^{k}}{k!\sqrt{n!}}{}_{2}F_{2}(1,-n;1+k,1;|g|^2)|n\rangle_{p}\nonumber\\
&=e^{-\frac{|\alpha|^2+|g|^2}{2}}\sum_{n=0}^{\infty}\frac{\alpha^{k+n}g^{-k}(-|g|^2)^{k}}{k!\sqrt{n!}}{}_{1}F_{1}(-n;1+k;|g|^2)|n\rangle_{p}\nonumber\\
&=e^{-\frac{|\alpha|^2+|g|^2}{2}}\sum_{n=0}^{\infty}\frac{\alpha^{k+n}g^{-k}(-|g|^2)^{k}\sqrt{n!}}{(k+n)!}L_{n}^{(k)}(|g|^2)|n\rangle_{p}.
\end{align}
If $k<0$, the conditional optical state reads,
\begin{align}
|\psi^{(k)}\rangle_p&=e^{-\frac{|\alpha|^2+|g|^2}{2}}\sum_{n=-k}^{\infty}\frac{\alpha^{k+n}g^{-k}\sqrt{n!}}{(n+k)!(-k)!}{}_{2}F_{2}(1,-k-n;1,1-k;|g|^2)|n\rangle_{p}\nonumber\\
&=e^{-\frac{|\alpha|^2+|g|^2}{2}}\sum_{n=-k}^{\infty}\frac{\alpha^{k+n}g^{-k}\sqrt{n!}}{(n+k)!(-k)!}{}_{1}F_{1}(-k-n;1-k;|g|^2)|n\rangle_{p}\nonumber\\
&=e^{-\frac{|\alpha|^2+|g|^2}{2}}\sum_{n=-k}^{\infty}\frac{\alpha^{k+n}g^{-k}}{\sqrt{n!}}L_{k+n}^{(-k)}(|g|^2)|n\rangle_{p}.
\end{align}
\begin{figure}[tbp]
	\centering
	\includegraphics[width=0.32\textwidth]{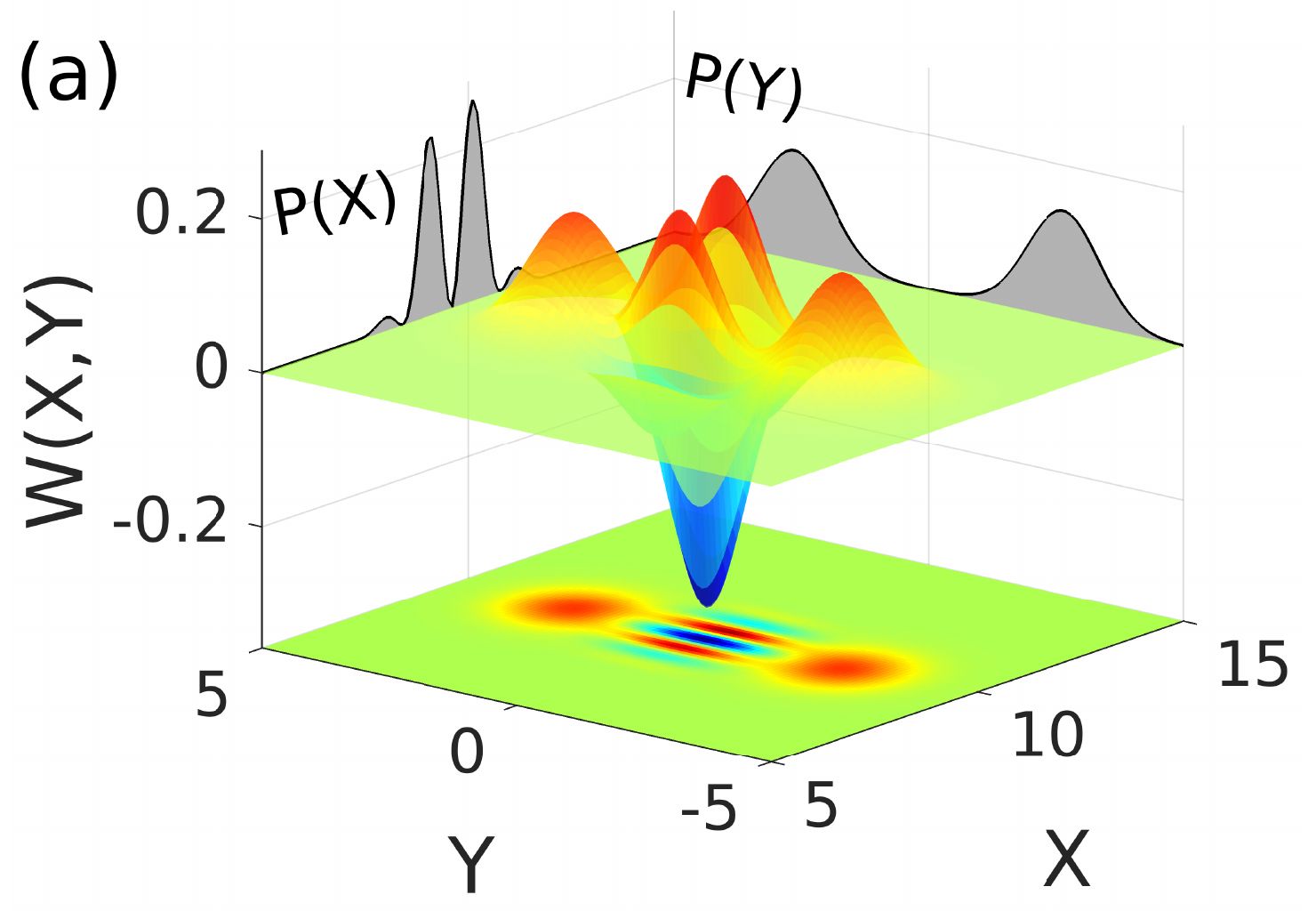}
	\includegraphics[width=0.28\textwidth]{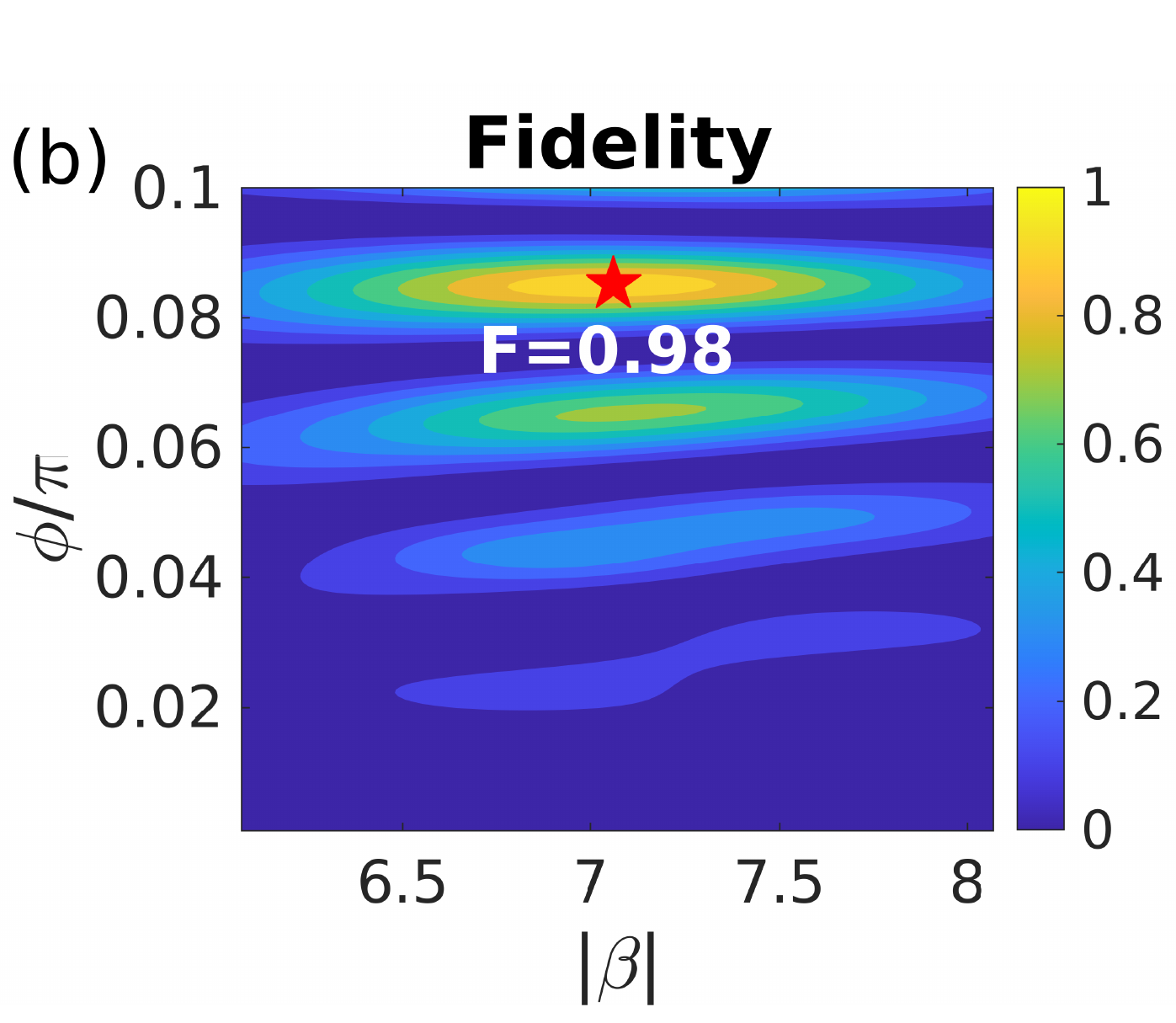}
	\caption{(a) The Wigner function of the ideal odd cat state that is taken to evaluate the fidelity of the generated optical state shown in the insert of Fig.~3(a) of the main text. (b) The corresponding fidelity with amplitude $|\beta|=7.061$ and phase $\phi=0.085\pi$ marked by the red star.}
	\label{fig:photonic-WN}
\end{figure}
Here, $L_{n}^{(a)}(x)$ is the generalized Laguerre function. Considering that the asymptotic behavior of $L_{n}^{(a)}(x)$ for large $n$ with fixed $a$ and positive $x$ is approximate to a sine function as in the ideal cat state (\ref{eq:cat}), the fidelity between the generated optical state $|\psi^{(k)}\rangle_p$ and the ideal odd cat state $|\beta\rangle-|\beta^\ast\rangle$ can be nearly perfect in certain cases. However, due to the complicated formation, we would rather combine the analytical analysis and the numerical calculation, which demonstrates that high-fidelity optical cat states can be successfully generated as shown in Fig.~2 in the main text. Here we present the Wigner functions of the corresponding ideal cat states in Fig.~\ref{fig:cat-WN}. By comparing them, it is obvious that the Wigner functions in two figures are very similar, which indicates high fidelity being achieved. In addition, in Fig.~\ref{fig:photonic-WN}(a) we plot the Wigner function of the ideal cat state that is taken to evaluate the fidelity of the large-size optical cat state created in Fig.~3(a) of the main text. The high fidelity $F=0.98$ is marked by the red star in Fig.~\ref{fig:photonic-WN}(b). Since the two contrary outcomes are separated in $Y$ direction, the size of the cat state can be quantified by the imaginary part of the complex amplitude $\beta_y=|\beta|\sin\phi$. We see that the high-fidelity cat state corresponds to a larger value of $\phi$, which means that large-size optical cat state is achieved with strong coupling.

The generated optical cat states bear excellent resemblance to the ideal cat states (\ref{eq:cat}), which can be understood in the following way. Considering the input optical coherent state $|\alpha\rangle_p$ where a large number of photons are included, it can be characterized by a Gaussian Wigner distribution in the phase space whose symmetric center is located at $X_c=\sqrt{2}\alpha$ and $Y_c=0$ (real $\alpha$ is assumed). Taking a particular postselection $k=0$ as an example, in the weak QPINEM coupling regime ($|g|\ll 1$) only a single interaction channel ($j=0$) is involved where the optical state is unchanged. Then by increasing $|g|$, more channels (e.g. $j=1,2,3,\dots$) are included where $j$ photons are added into the optical state. Thus in the phase space, a non-Gaussian Wigner distribution without negativity is introduced for each channel, and the symmetric center is still located at $X$ axis but with a slightly larger $X_c$ value (as indicated by the plot of `Multiple channels' in Fig.~1 of the main text). Hence, the interference between the interaction channels will result in an interference pattern along $X$ axis and two simultaneous contrary outcomes in $Y$, as indicated by the probability distribution of $Y$ in Fig.~1 in the main text. This is coincident with the ideal cat state described by Eq.~(\ref{eq:cat}). 

\section{Interference between multi-path free-electron--photon interactions}
\begin{figure}[tbp]
	\centering
	\includegraphics[width=0.48\textwidth]{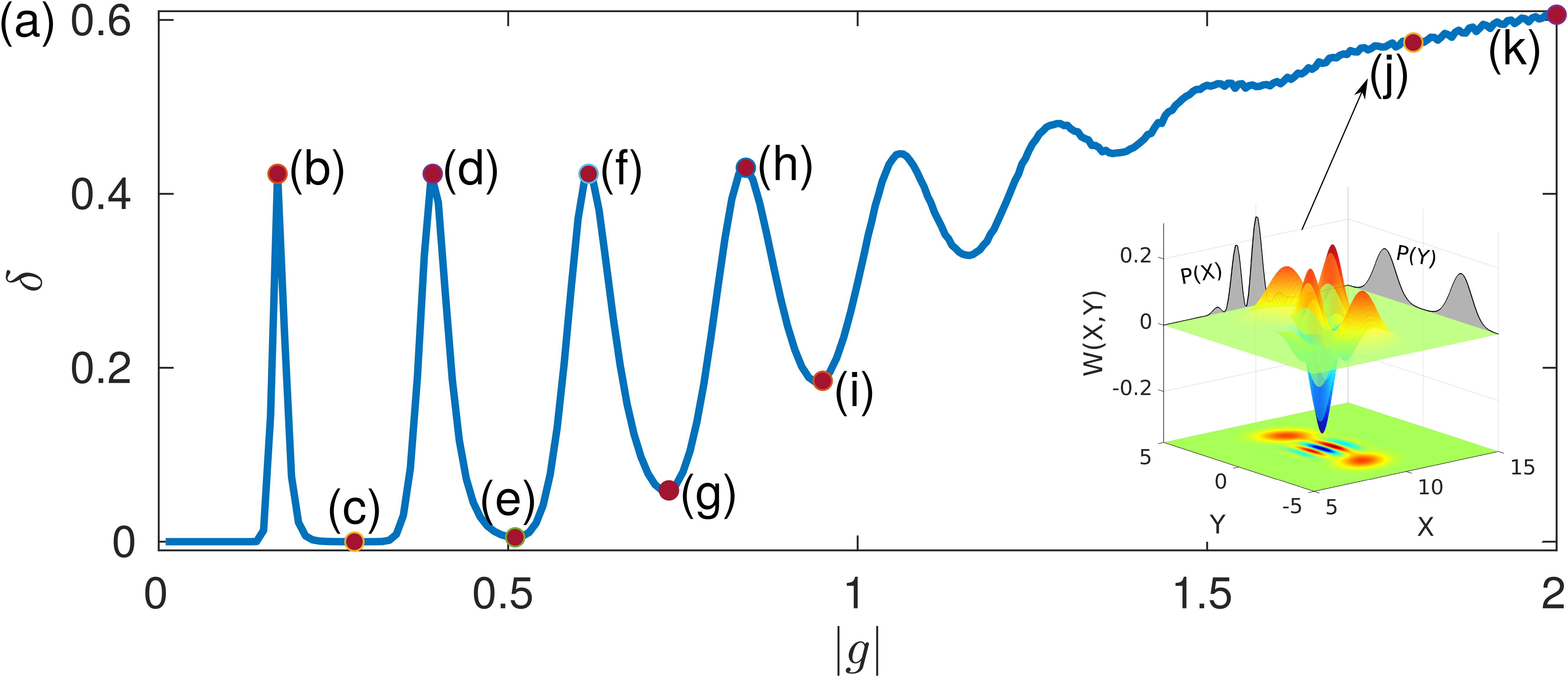}
	\includegraphics[width=0.24\textwidth]{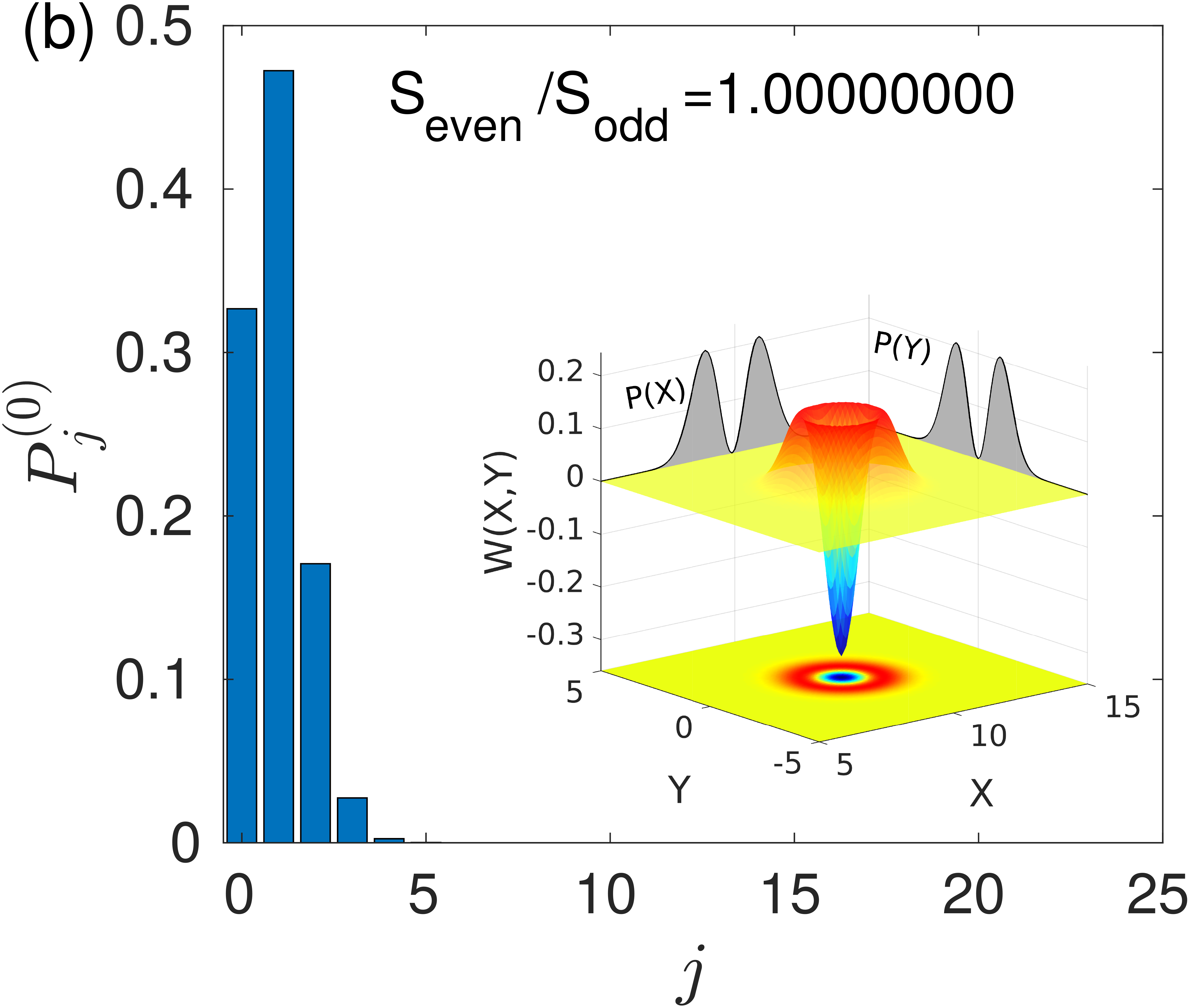}
	\includegraphics[width=0.24\textwidth]{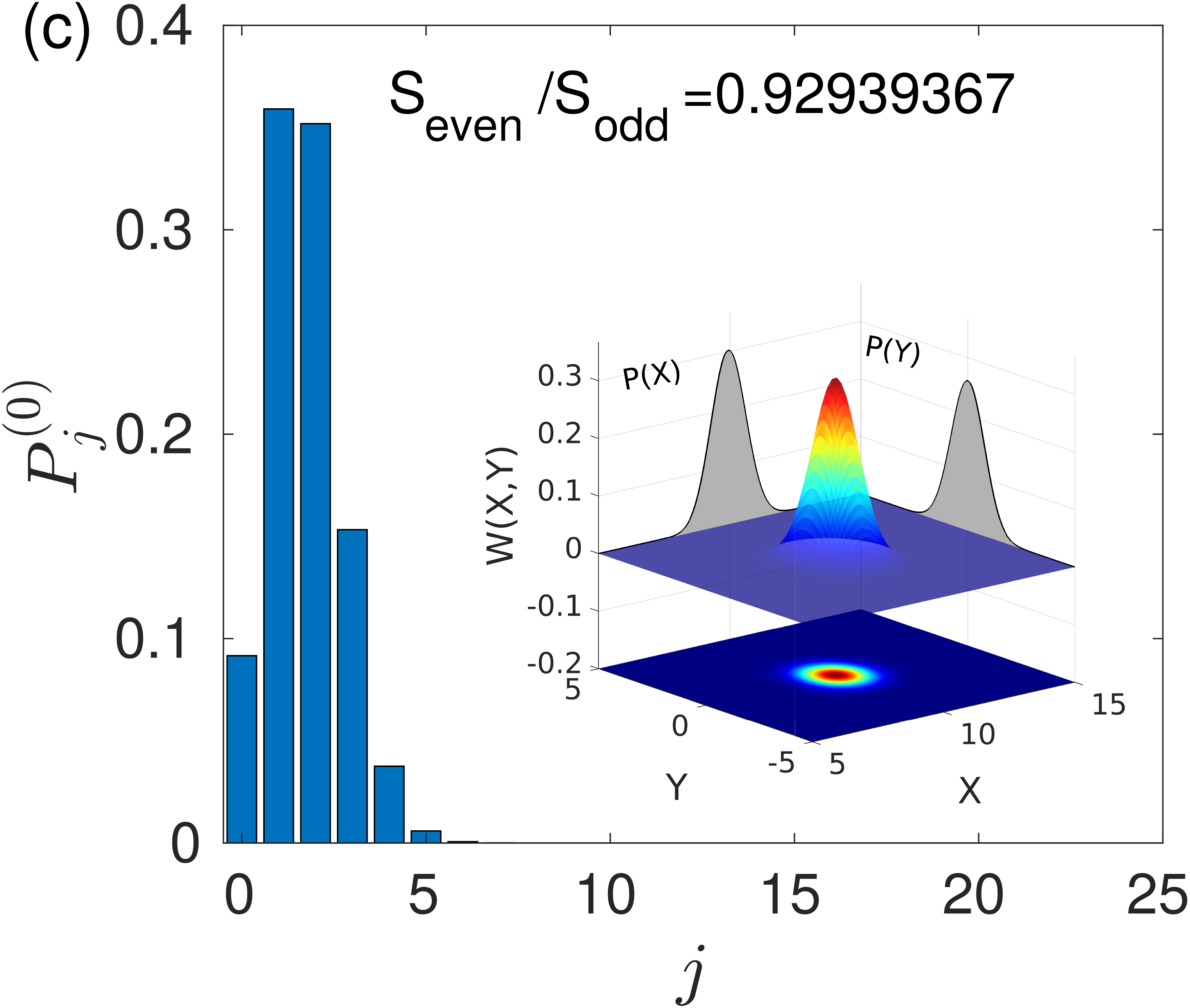}\\
	\includegraphics[width=0.24\textwidth]{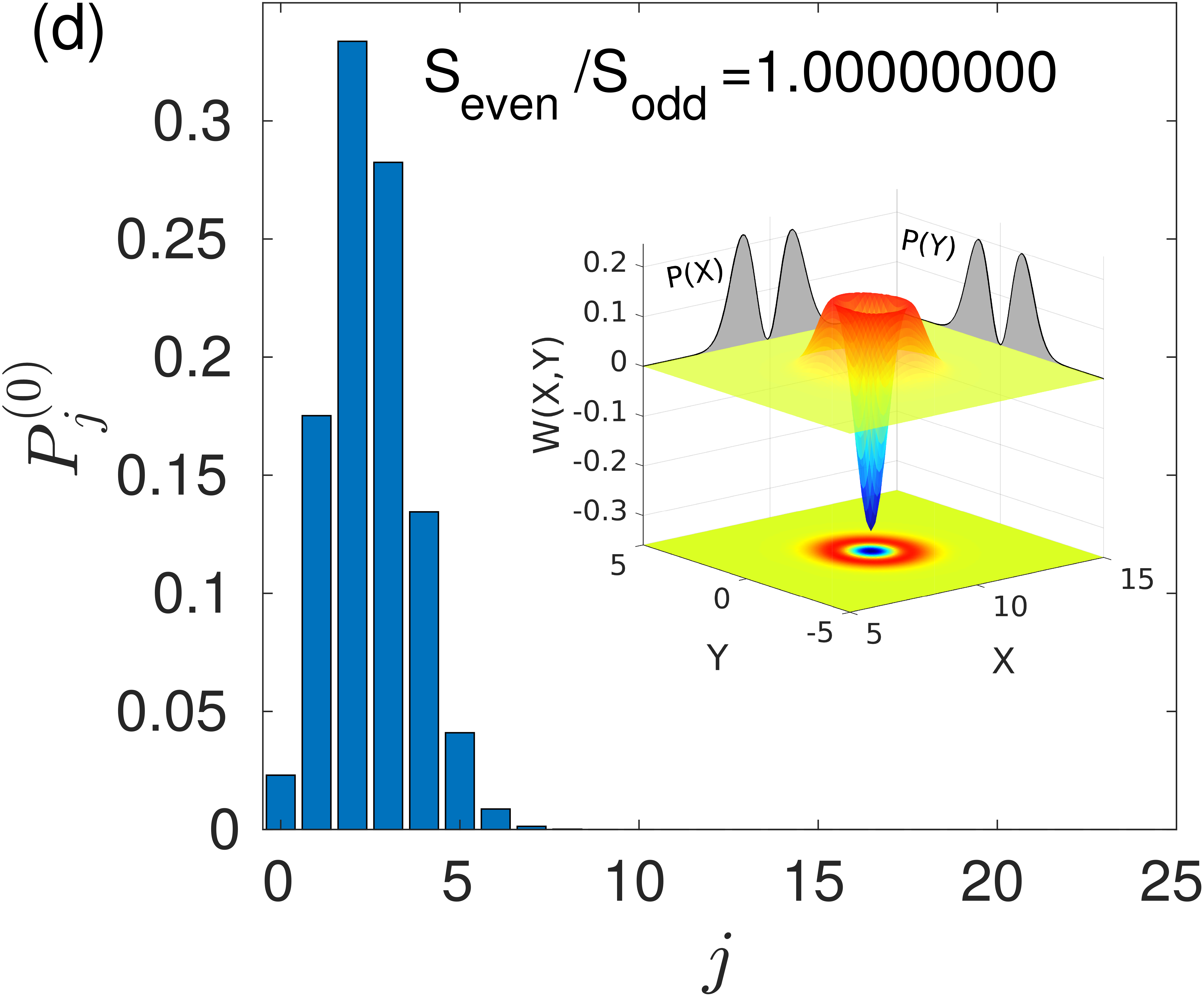}
	\includegraphics[width=0.24\textwidth]{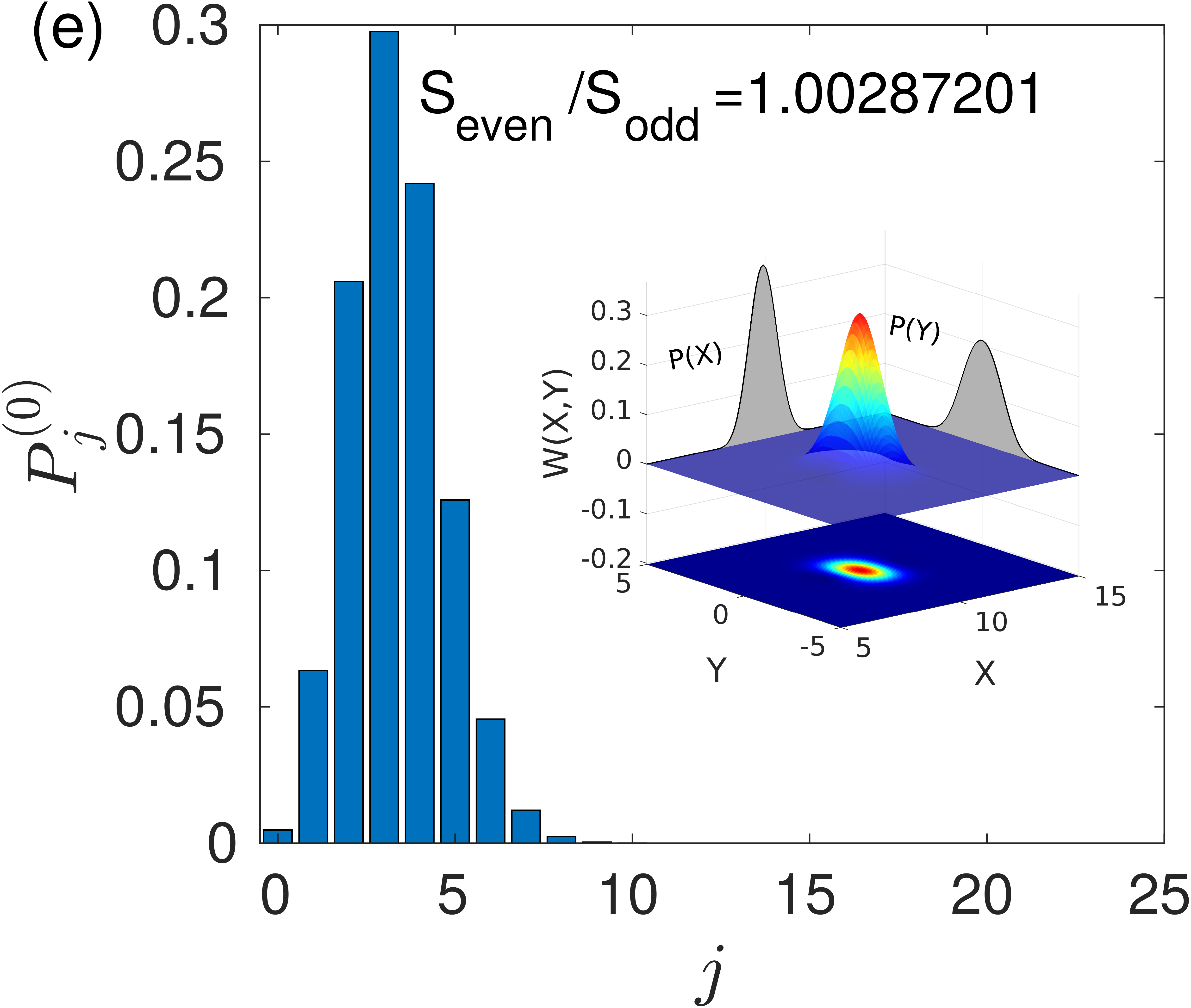}
	\includegraphics[width=0.24\textwidth]{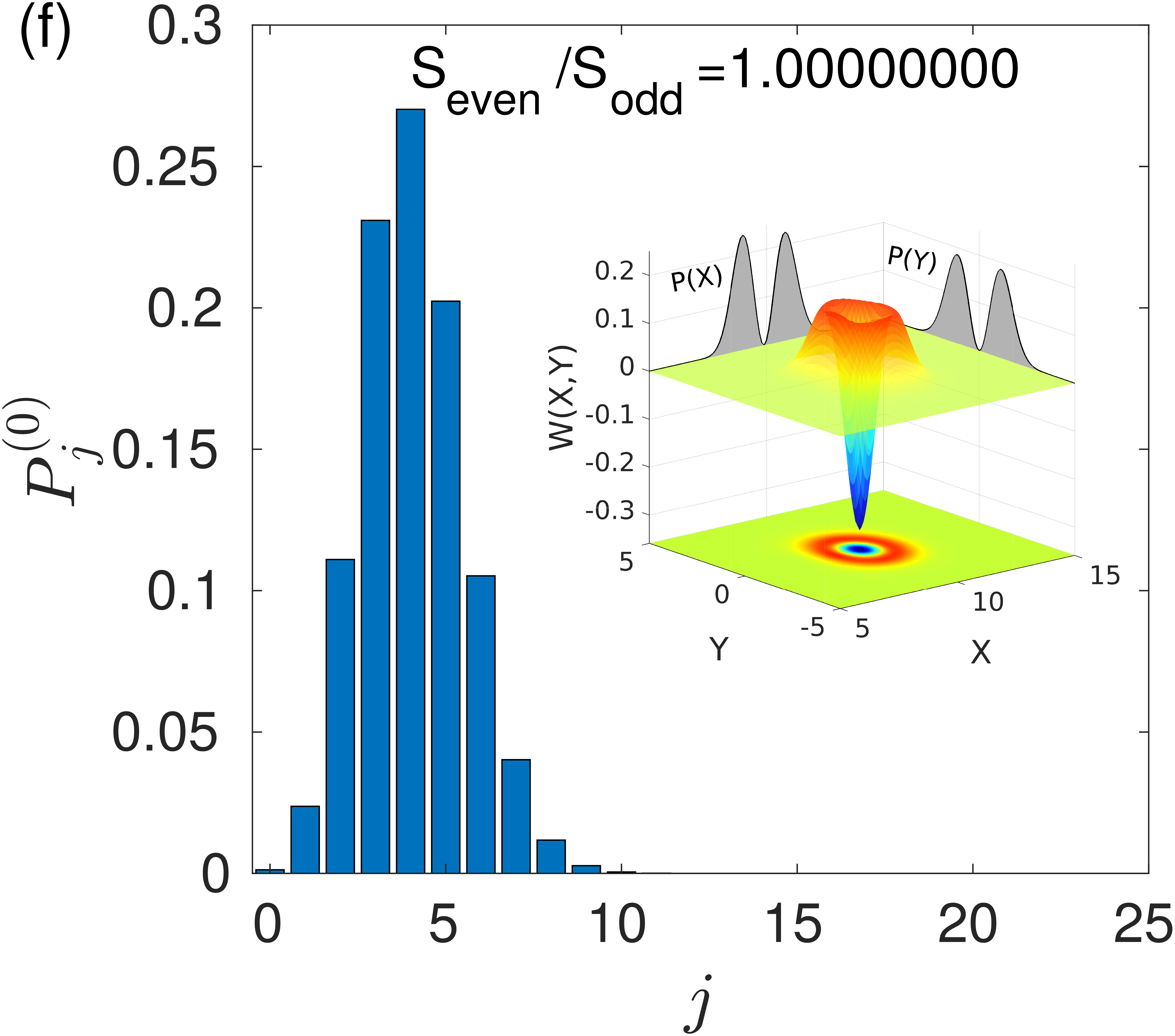}
	\includegraphics[width=0.24\textwidth]{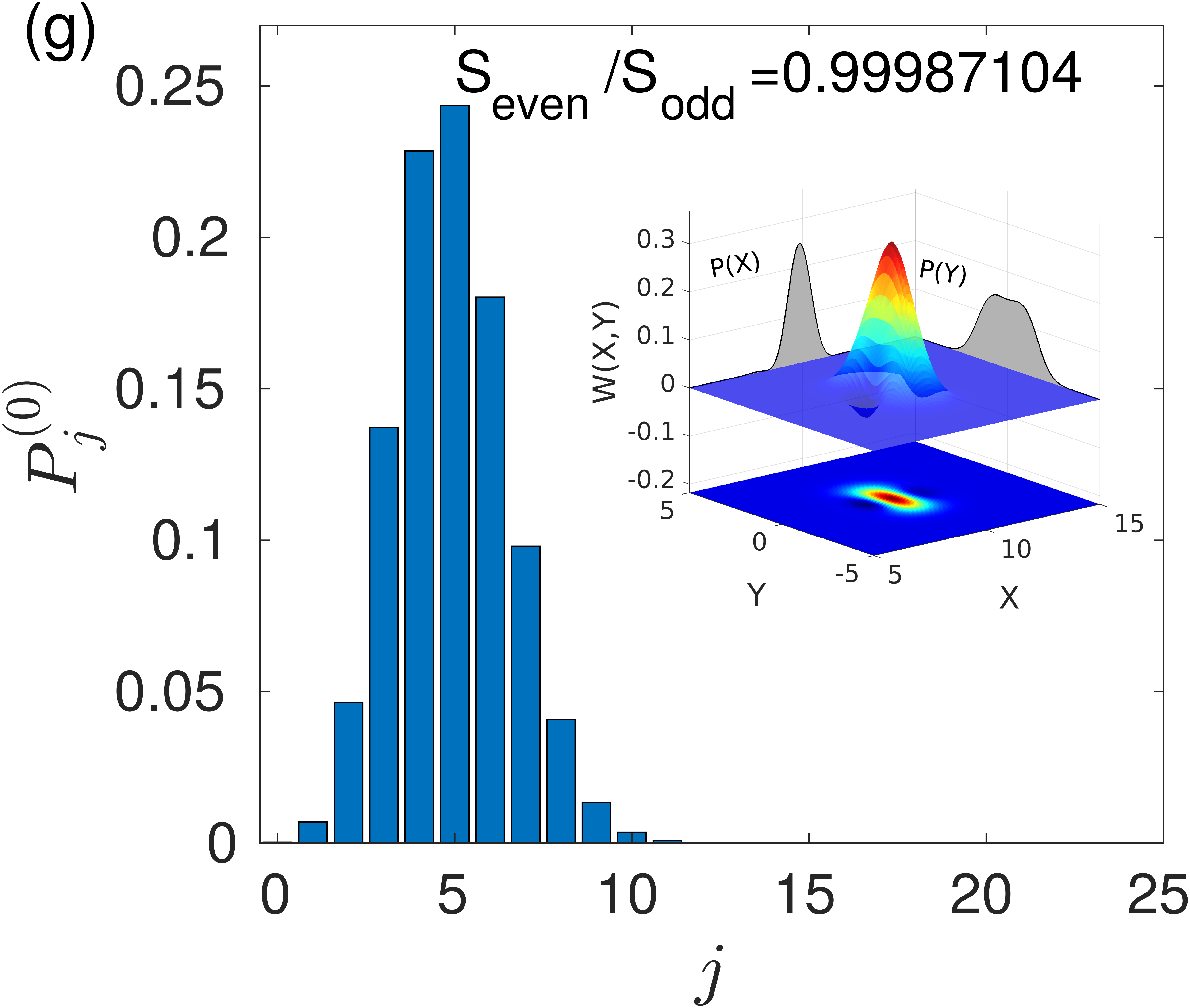}\\
	\includegraphics[width=0.24\textwidth]{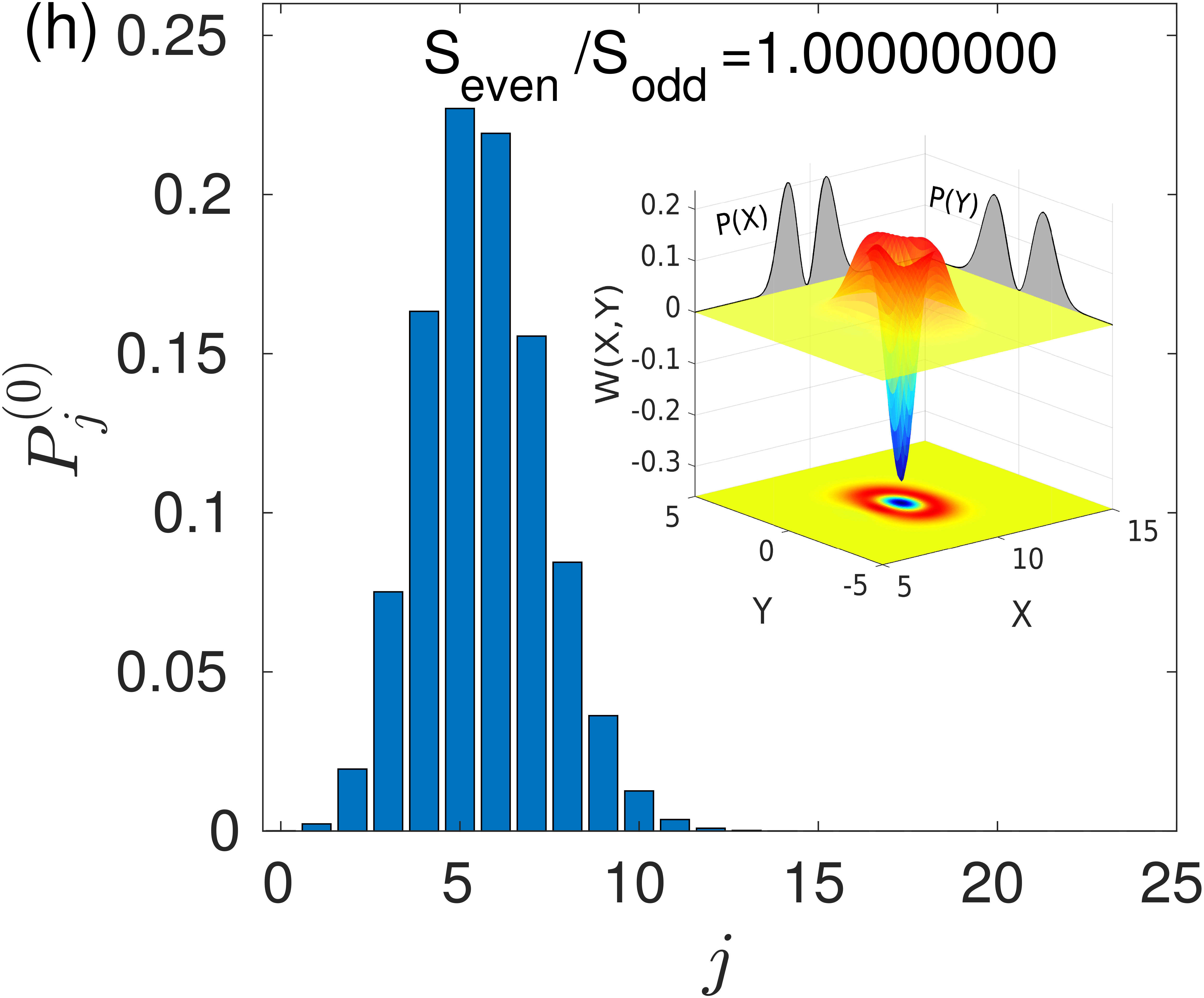}
	\includegraphics[width=0.24\textwidth]{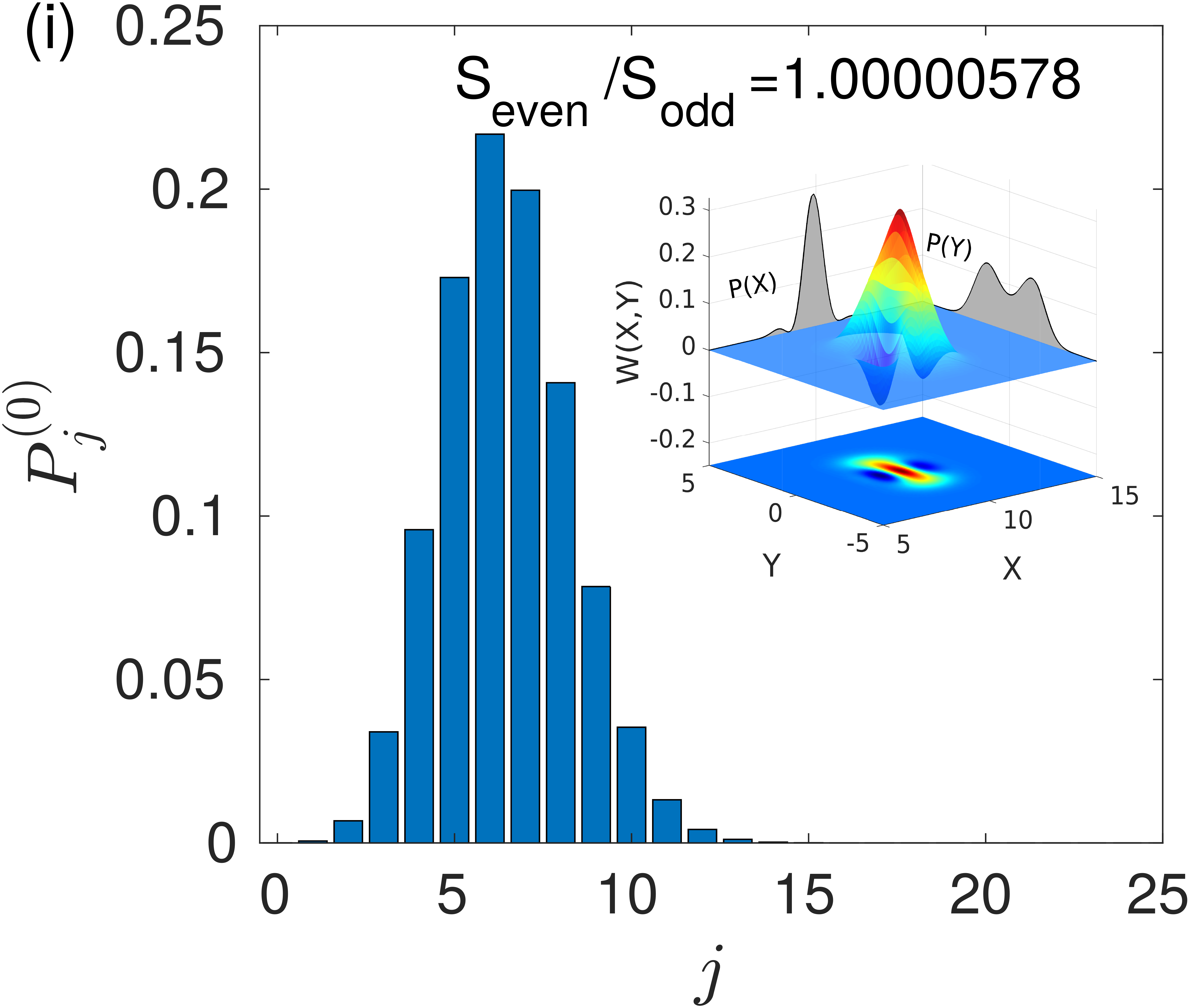}
	\includegraphics[width=0.24\textwidth]{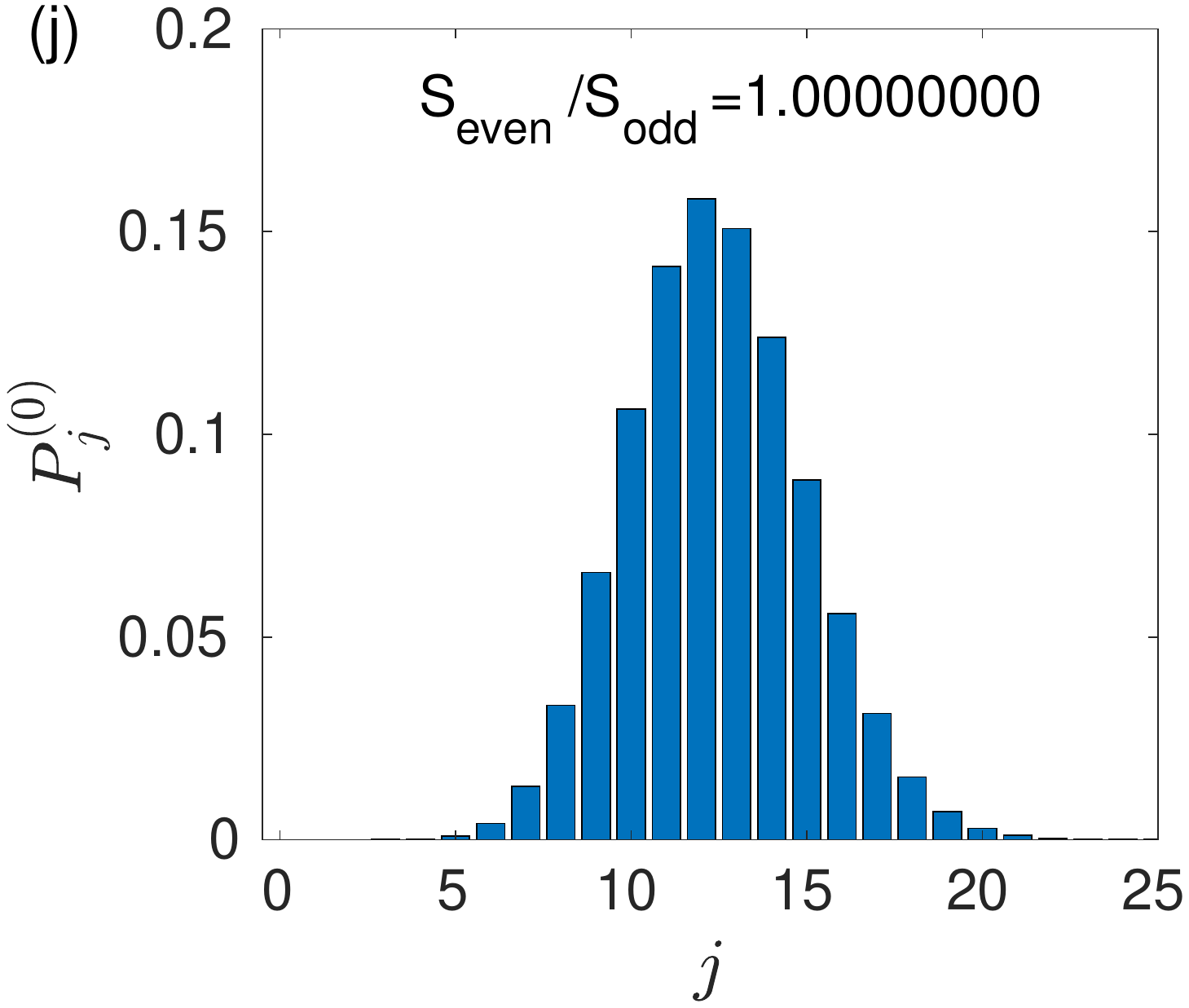}
	\includegraphics[width=0.24\textwidth]{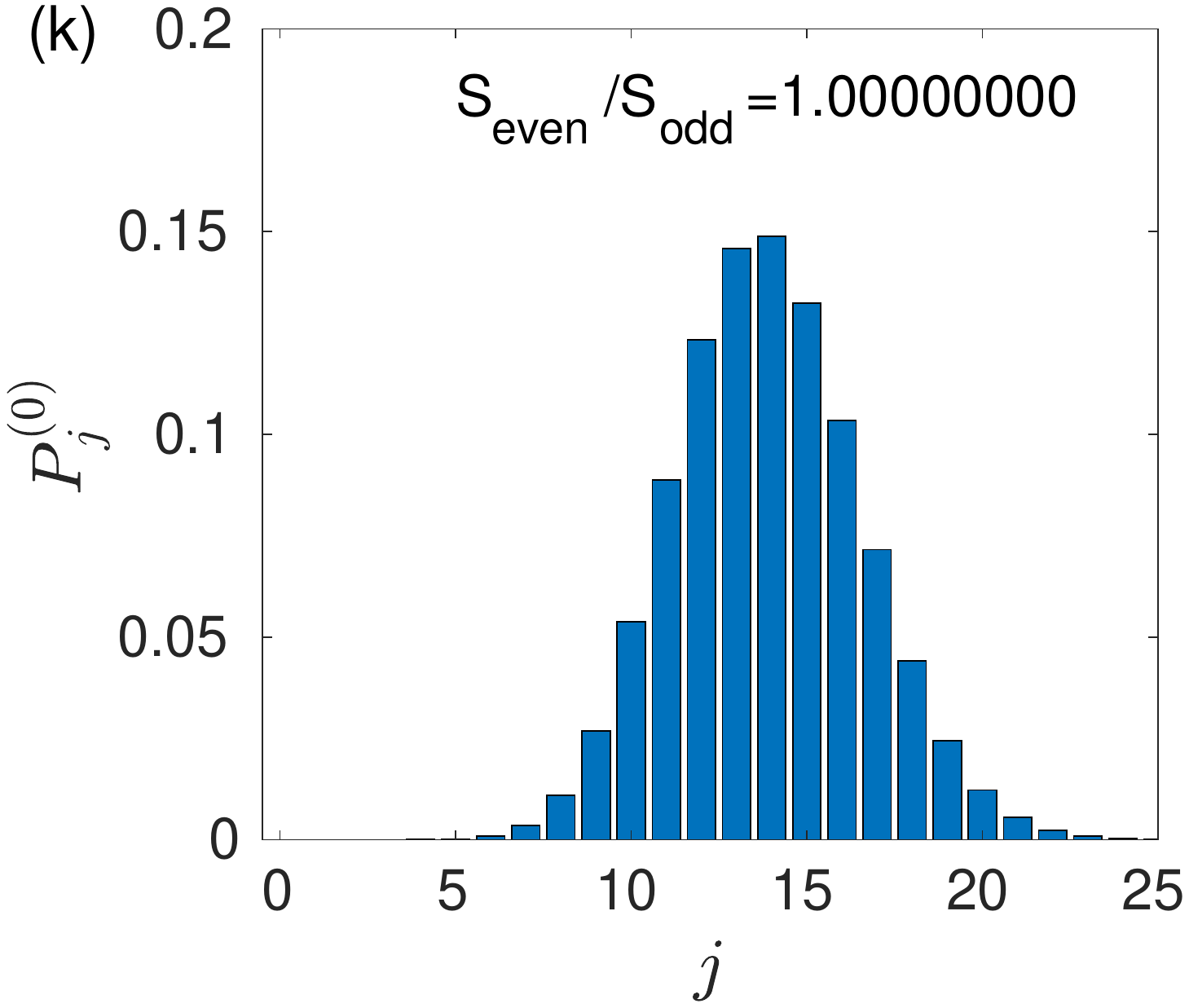}
	\caption{(a) The Wigner negativity of the generated optical states changing with the coupling strength $|g|$. Other parameters are set as $|\alpha|^2=50$ and $k=0$. (b)-(k) display the corresponding weights of the quantum channels for the cases marked in (a). The coupling strengths are (b) $|g|=0.171$, (c) $|g|=0.28$, (d) $|g|=0.391$, (e) $|g|=0.51$, (f) $|g|=0.612$, (g) $|g|=0.73$, (h) $|g|=0.834$, (i) $|g|=0.95$, (j) $|g|=1.8$, and (k) $|g|=2$. The insets of (b-i) are the Wigner functions of the corresponding optical states with all quantum channels considered, respectively. The inset of (a) corresponds to the Wigner function of the optical state labeled as (j). And the Wigner function for the optical state (k) can be found in Fig.~3(a) in the main text.}
	\label{fig:interference-WN}
\end{figure}

In the main text, we have concluded that the generation of optical cat states results from the quantum interference of the multi-path free-electron--photon interactions. In this section, we will further analyze the interference effect and clarify that the high-fidelity odd optical cat states are successfully created because of the constructive interference. To make it clear, we will take the postselection on the electron $k=0$ as an example and illustrate how the interference affects the nonclassicality of the generated optical states. 
 \\

We will make use of the Wigner negativity to quantify the nonclassicality of the generated optical states, which also acts as an important signature of cat states, 
\begin{equation}
 \delta=\int |W(\alpha,\alpha^\ast)|d^2 \alpha-1.
\end{equation}
The oscillation of the Wigner negativity for $k=0$ with the increase of the coupling strength $|g|$ is provided in Fig.~\ref{fig:interference-WN}(a), which is the same with Fig.~3(a) in the main text. In order to understand the relation between the oscillation of Wigner negativity and the interference between multi-channel interactions, it is necessary to study the phase of each channel. According to the generated optical state (\ref{eq:photonic}), it is noticed that the quantum channels can be divided into two sets due to their opposite phases, i.e. channels of even $j$ with positive $C_{j}^{(k)}$ and channels of odd $j$ with negative $C_{j}^{(k)}$. Then the collective weights of the two sets can be defined as $S_{\text{even/odd}}=\sum_{\text{even/odd}\, j}P_{j}^{(k)}$. 
 \\

In the weak coupling limit $|g|\ll 1$, where there is only a single channel $\{j=0: I\}$ to affect the optical state, the ratio of the collective weights of quantum channels $S_{\text{even}}/S_{\text{odd}}\to \infty$. In this case, there is no interference, and thus only Wigner-positive states can be obtained. Then by increasing the coupling strength $|g|$, another channel $\{j=1:a^{\dagger}a\}$ becomes involved, which decrease the ratio of $S_{\text{even}}/S_{\text{odd}}$. With more and more channels included, the interference between the channels becomes more pronounced and nonclassical optical states with Wigner negativity begin to be observed. When the coupling strength is increased to $|g|=0.17$, four quantum channels of $j=0,1,2,3$ are included, where the two sets of channels become equally weighted $S_{\text{even}}/S_{\text{odd}}=1$, as shown in Fig.~\ref{fig:interference-WN}(b). In this case, the constructive interference is achieved where the Wigner negativity reaches its local maximum and an optical odd cat state is prepared. This is exactly the result of Fig.~2(a) in the main text.
 \\
 
By further increasing the QPINEM coupling strength, the ratio $S_{\text{even}}/S_{\text{odd}}$ continues to decrease and the destructive interference takes place. Thus, the Wigner negativity $\delta$ decreases to nearly $0$ and the distinct peaks in $Y$ direction disappear, as shown in Fig.~\ref{fig:interference-WN}(c). As the coupling strength continues to be increased, more channels take part in, which results in the oscillation of the ratio  $S_{\text{even}}/S_{\text{odd}}$. Once the two sets of channels achieve a balance $S_{\text{even}}/S_{\text{odd}}=1$, the constructive interference will occur so that the maximum of the Wigner negativity can be reached where a high-fidelity optical odd cat state is observed, as indicated in Fig.~\ref{fig:interference-WN}(d)(f)(h). And if the two sets of channels deviate from the equally weighted case, $S_{\text{even}}/S_{\text{odd}}\neq 1$, the destructive interference will take place where the Wigner negativity decreases and the two peaks in phase-quadrature are not well separated, as shown in Fig.~\ref{fig:interference-WN}(e)(g)(i). Please note that Fig.~\ref{fig:interference-WN}(i) corresponds to the case of Fig.~2(c) in the main text. 
 \\

It is also observed from Fig.~\ref{fig:interference-WN}(c)(e)(g)(i) that ratio $S_{\text{even}}/S_{\text{odd}}$ at the minimal points of the Wigner negativity becomes closer to $1$ with the increase of the coupling strength. It is because that with stronger and stronger coupling strength, a large number of quantum channels are involved where almost half of them are with even $j$ and the remaining half are with odd $j$. This reduces the difference between the two sets of channels and thus decreases the amplitude of the oscillation of $S_{\text{even}}/S_{\text{odd}}$. When the coupling strength gets strong enough ($|g|>1.5$), the ratio of the weights $S_{\text{even}}/S_{\text{odd}}$ will eventually saturate to $1$, as illustrated in Fig.~\ref{fig:interference-WN}(j) and (k). This means that the constructive interference of the channels can always be achieved in this strong coupling regime. Therefore, high-quality optical cat states with large Wigner negativities are anticipated to be prepared, which is confirmed as shown in Fig.~\ref{fig:interference-WN}(a). It is shown that the oscillation of the Wigner negativity becomes less pronounced for $|g|>1.5$, and large value of Wigner negativity is always achieved. The Wigner functions for Fig.~\ref{fig:interference-WN}(j) (inset of Fig.~\ref{fig:interference-WN}(a)) and for Fig.~\ref{fig:interference-WN}(k) (inset of Fig.~3(a) in the main text) demonstrate the successful creation of large-size optical cat states, where two well-separated peaks in $Y$ quadrature and a distinct interference pattern in $X$ quadrature are clearly observed.
 \\

\begin{figure}[t]
	\centering
	\includegraphics[width=0.24\textwidth]{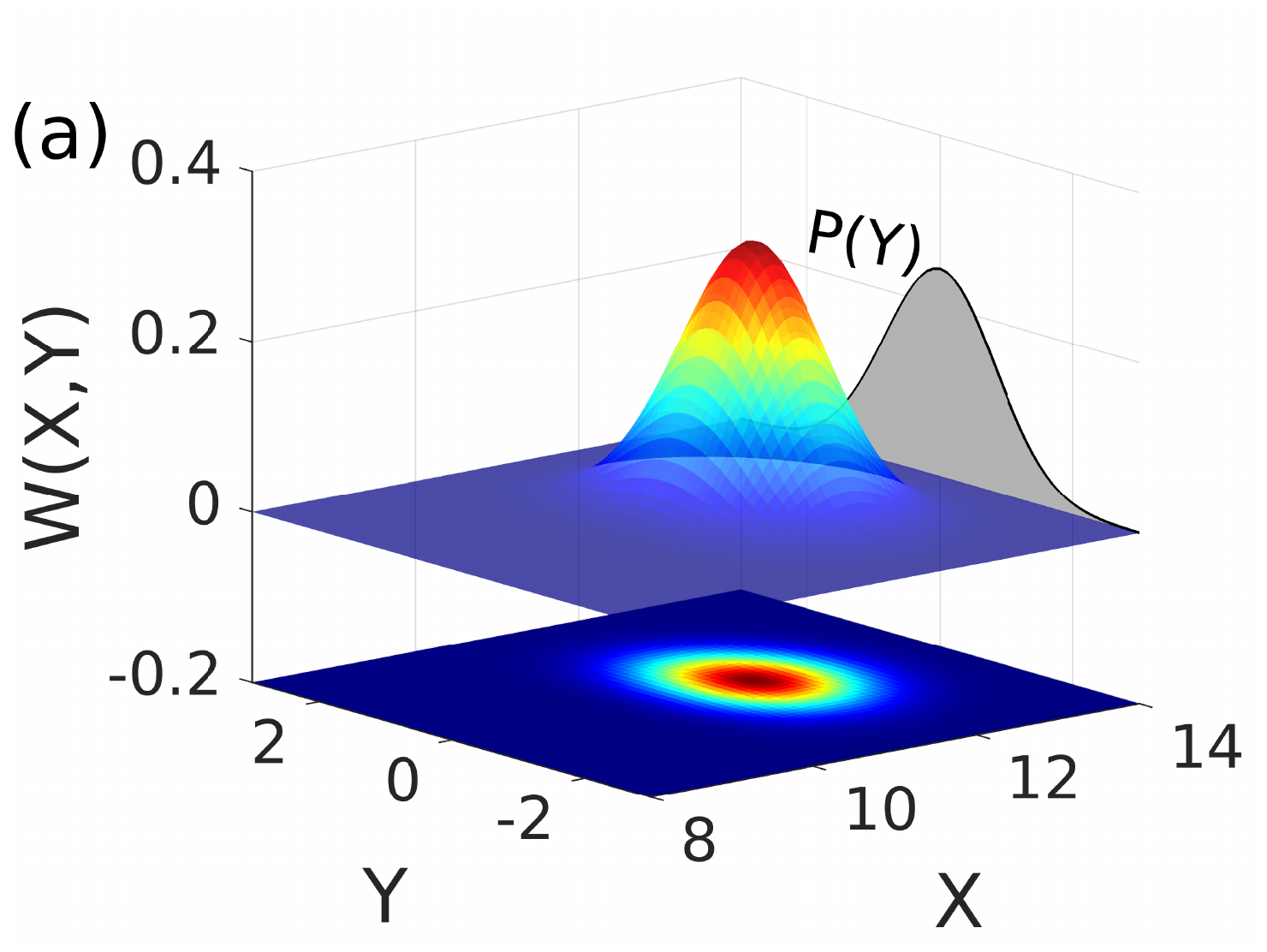}
	\includegraphics[width=0.24\textwidth]{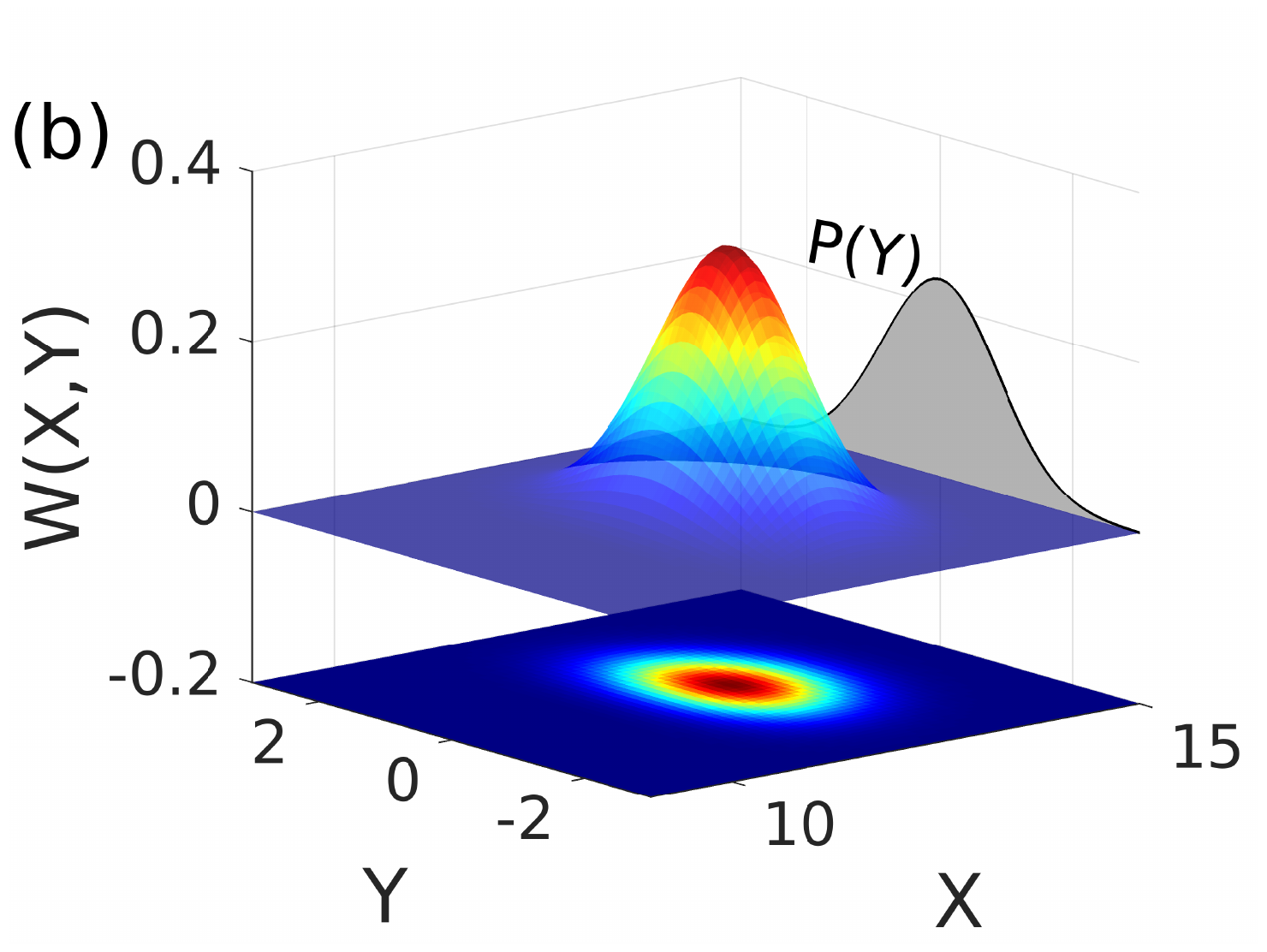}
	\includegraphics[width=0.24\textwidth]{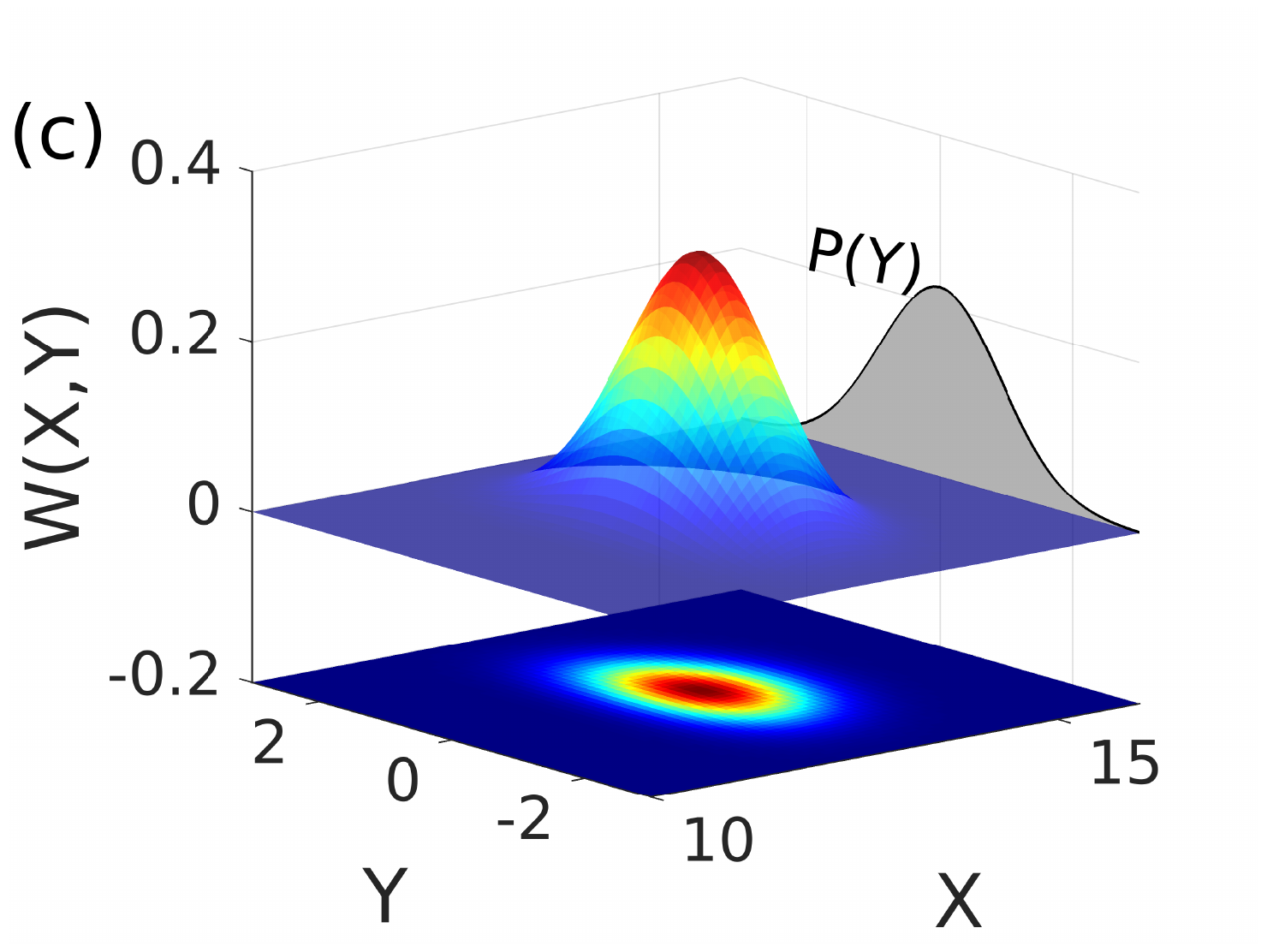}
	\includegraphics[width=0.24\textwidth]{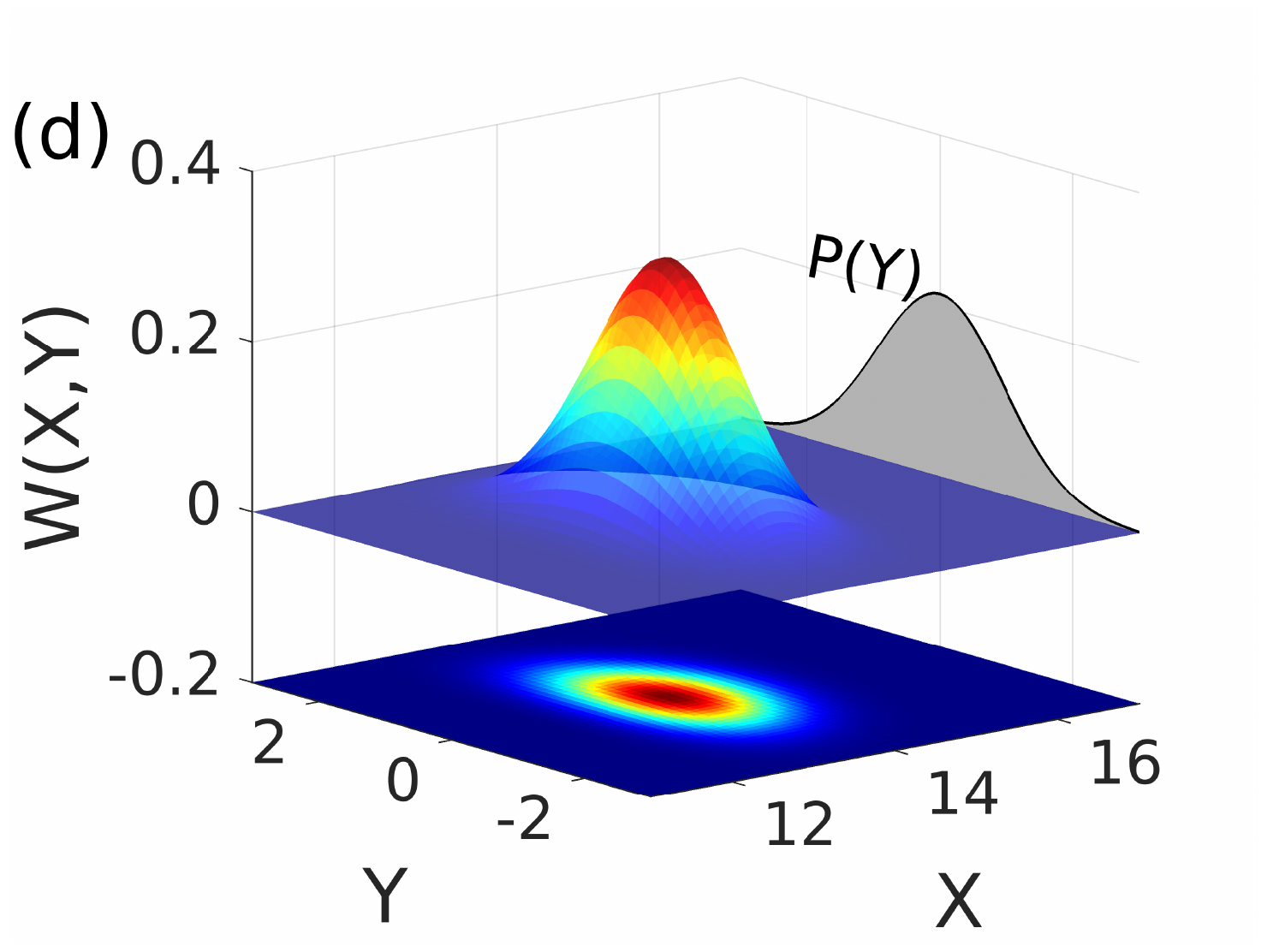}
	\caption{Wigner functions for the states generated by each quantum channel individually. Here the quantum channels are chosen as (a) $j=10$, (b) $j=15$, (c) $j=20$, (d) $j=25$. The postselection on electrons is set as $k=0$, and the initial photon number of optical state is $|\alpha|^2=50$.}
	\label{fig:large_channel}
\end{figure}

Notice that in Fig.~\ref{fig:interference-WN}(k), the channels with $j<25$ are stimulated, which conflicts with the assumption that $|\alpha|^2$ is much larger than the achievable values of $k$ and $j$. We emphasize that the aim of this assumption is to make sure the generated optical state in each quantum channel remaining a positive Wigner function. In this way, we can conclude with confidence that the generation of Wigner negativity only results from the interference of the multi-channel interactions. Here, we verify that the optical states created by individual channel of $j\le25$ are indeed positive in Wigner functions, as shown in Fig.~\ref{fig:large_channel} for quantum channels $j=10,15,20,25$. 
 \\
 
In Fig.~\ref{fig:interference-WN}, we have shown some typical results of the Wigner functions and the weights of channels. More results with the continuous increase of the coupling strength $|g|$ can be found in the Supplementary Video named ``OpticalStates.avi''. In this video, we have fixed the input optical state with $|\alpha|^2=50$ and the postselected electron energy with $k=0$. The upper left figure shows the Wigner functions of the generated optical states, with the probability distributions of the quadrature $Y$ projected on the vertical plane. The upper right plot displays the corresponding Wigner negativity, which is marked by the red dot. The lower left panel is the fidelity between the generated optical state and the ideal cat state, where the amplitude $|\beta|$ and the phase $\phi$ are labeled with a red star. The weights of the quantum channels $P_{j}^{(k)}$ as well as the ratio of the two sets $S_{\text{even}}/S_{\text{odd}}$ can be acquired in the lower right image. Consequently, the constructive interference between the multi-channel free-electron--photon interactions only takes place at certain coupling strength $|g|$ when it is not strong enough. Therefore, it is important to choose proper PINEM coupling strengths $|g|$ for the preparation of the optical cat states. Since the minimal coupling strength required for creating optical cat states, $|g|=0.17$, is only slightly larger than the reported coupling in the recent experiment~\cite{dahan2021imprinting}, our scheme is promising to be realized with the state-of-the-art experimental techniques.

\section{Discussion on the experimental realization}

In order to observe the optical cat states in PINEM with our scheme, strong coupling of $|g|>0.1$ is required between the free electron and the photons. Due to the form of the coupling strength $g=e/\hbar\omega\int_{-\infty}^{\infty}dz'\exp(-i\omega z'/v)E_z(z')$, the most commonly used method to modify $g$ is to change the strength and the distribution of $E_z$ by using designed photonic nanostructures. Theoretically, the strong coupling of $|g|\sim 0.5$ has been proposed based on the electron interaction with microcavities~\cite{kfir2019entanglements}. And in a recent work by Raphael Dahan \textit{et al.}~\cite{dahan2020resonant}, it has been clarified that the coupling strength can be further increased $|g|>1$ through a long phase-matched interaction in extended structures. As the current coupling strength reported in experiments is up to $|g|\sim 1$~\cite{adiv2022observation}, it should be possible to observe the optical cat states and the oscillation of Wigner negativity in PINEM.
\begin{figure}[t]
	\centering
	\includegraphics[width=0.3\textwidth]{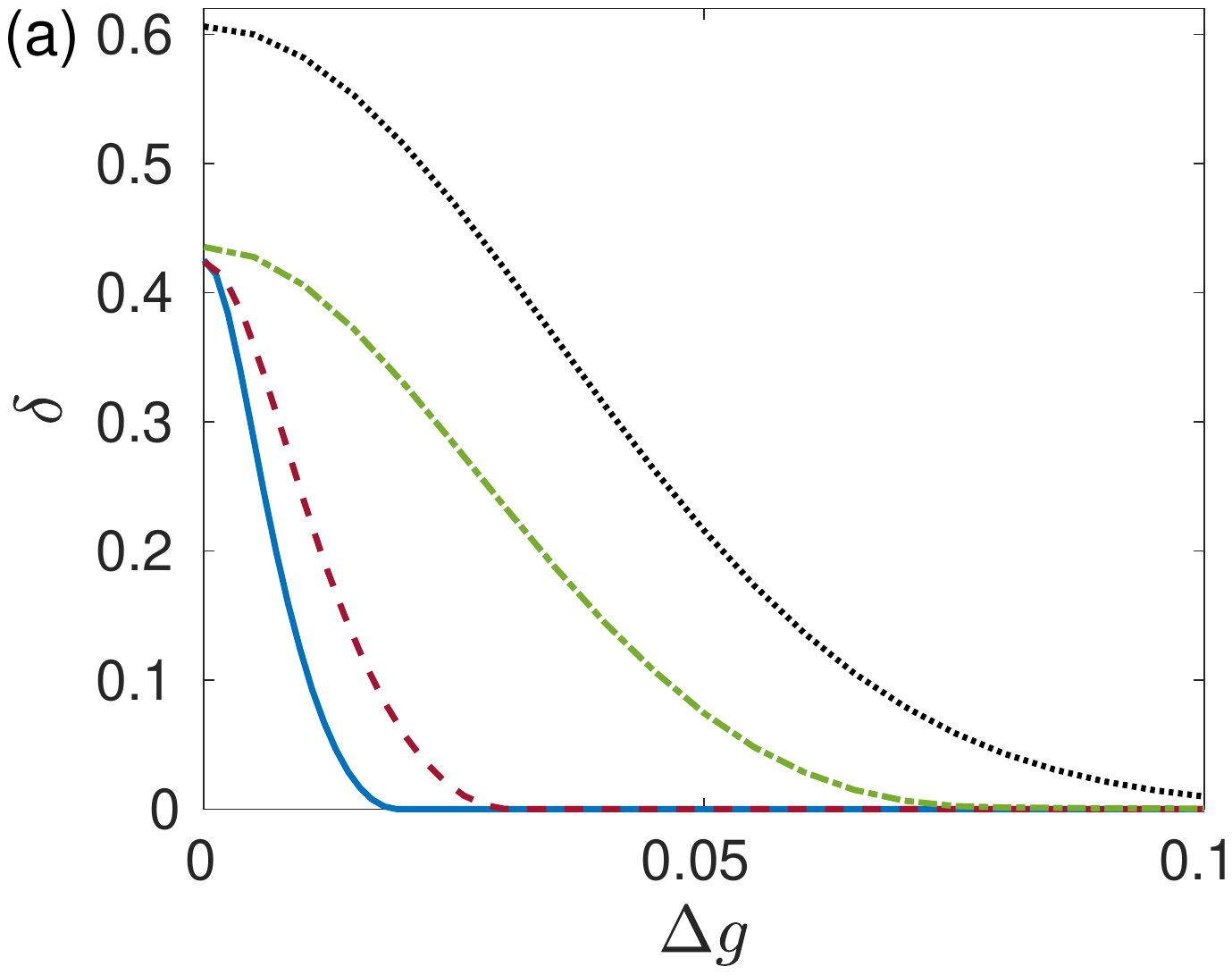}
	\includegraphics[width=0.3\textwidth]{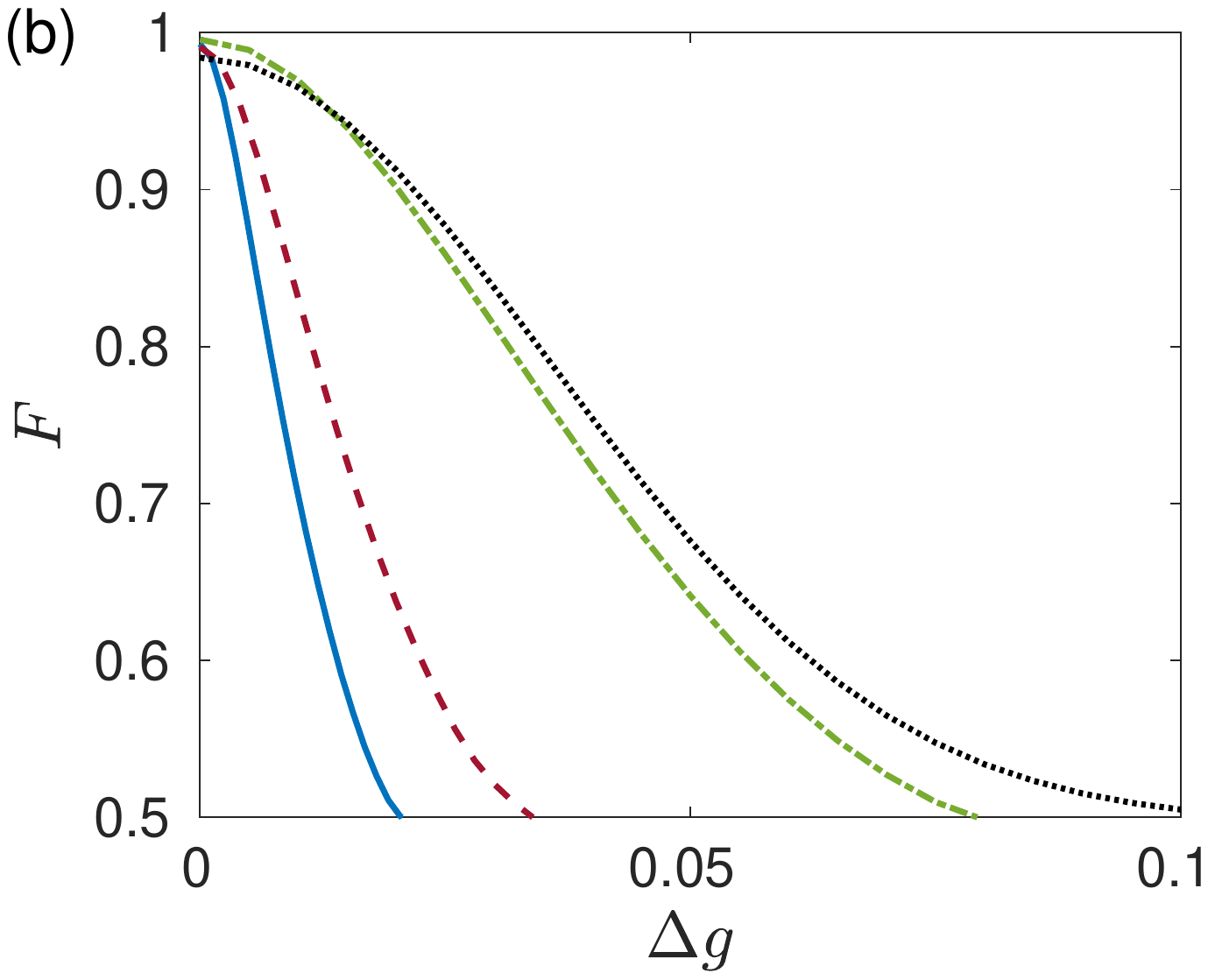}
	\caption{The dependence of (a) the Wigner negativity $\delta$ and (b) the fidelity $F$ on the fluctuation of the coupling strength $\Delta g$. The blue solid line corresponds to the state in Fig.2(a) in the main text ($|g|=0.17$, $k=0$), the red dashed line corresponds to Fig.2(b) ($|g|=0.275$, $k=1$), the green dash-dotted line corresponds to Fig.2(d) ($|g|=0.95$, $k=1$), and the black dotted corresponds to Fig.3(a) ($|g|=2$, $k=0$). Other parameters are set as $|\alpha|^2=50$.}
	\label{fig:dg}
\end{figure}

The property of the produced optical states is determined by the value of $g$. In experiments, the variance of the coupling strength is inevitable due to the bandwidth of the pumping mode, the dispersion of the electrons, etc. So it is necessary to study how the fluctuation of $g$ affects the results. Here we assume that the coupling strength $g$ follows a Gaussian distribution with a standard deviation $\Delta g$. Then by calculating the ensemble average, we obtain the dependence of the Wigner negativity $\delta$ and the fidelity $F$ on the fluctuation of the coupling $\Delta g$ in Fig.~\ref{fig:dg}. It is found that the stronger the coupling strength $|g|$ is set, the more robust the optical cat state is against the fluctuation $\Delta g$. For the optical cat state with the minimal coupling $|g|=0.17$ (blue solid line), the optical cat state with $\delta>0$ and $F>0.5$ can be obtained with $\Delta g<0.02$. And for the case with the largest coupling $|g|=2$ (black dotted line), The fluctuation of the coupling strength can be relaxed to $\Delta g<0.1$.
\begin{figure}[t]
	\centering
	\includegraphics[width=0.35\textwidth]{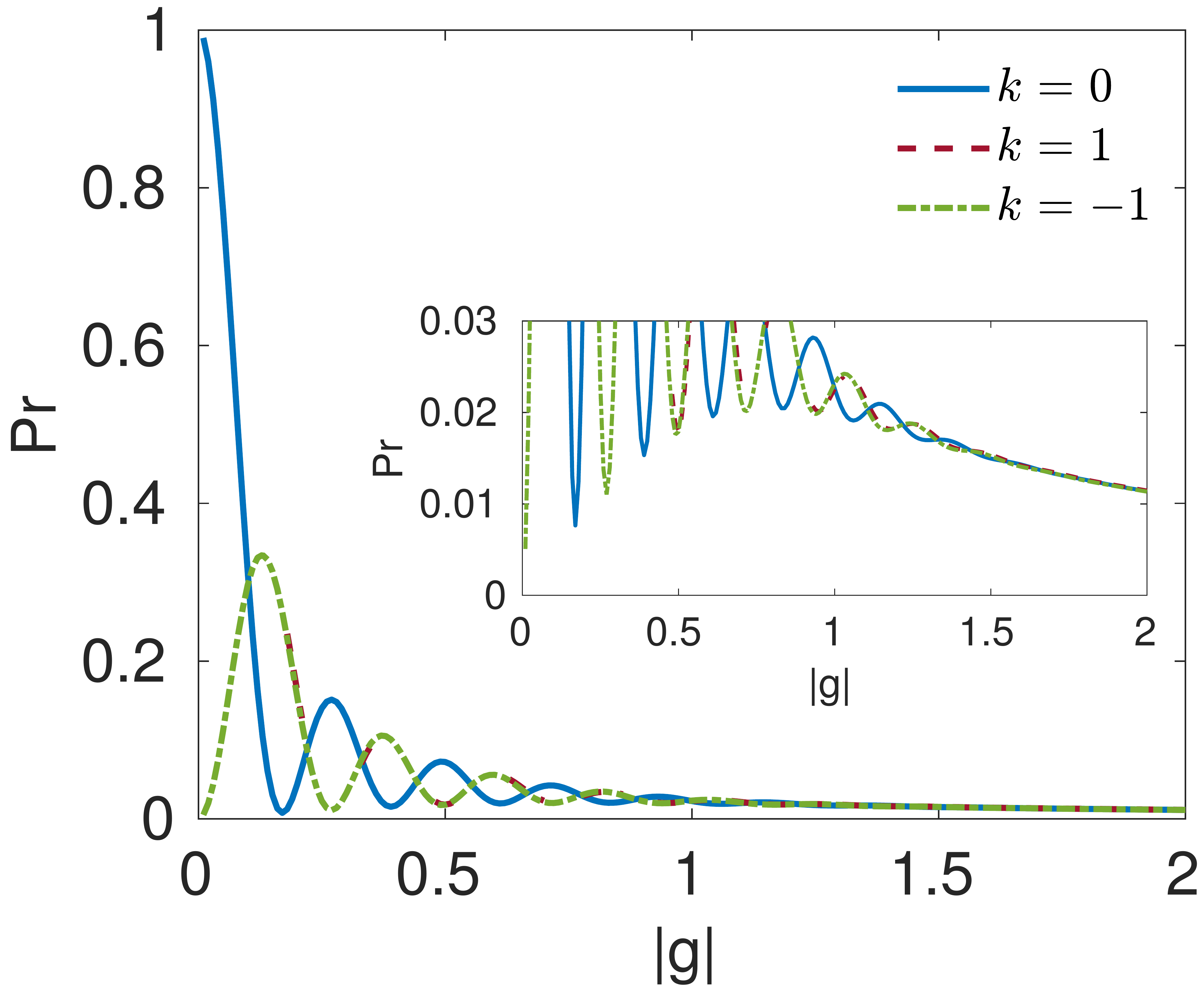}
	\caption{The success probability for projecting the electron to the energy-ladder state with $k=0$ (blue solid), $k=1$ (red dashed) and $k=-1$ (green dash-dotted). Here, the results for $k=1$ and $k=-1$ coincide with each other. The insert indicates that the success probabilities are always positive even when they are close to zero. Other parameters are set as $|\alpha|^2=50$.}
	\label{fig:SucPr}
\end{figure}

After the free-electron-photons interaction, we will then perform the energy measurement on the emitting electron with the electron energy loss spectrometer (EELS), which will project the electron to its energy-ladder state $|k\rangle$. The direct detection of individual electrons is experimentally possible due to the technical advances where the EELS with detective quantum efficiency higher than $0.8$ has been reported~\cite{plotkin2020hybrid}. Recent experiments also reported the coincidence measurements of individual electron energy loss and the single-photon emission \cite{jannis2019spectroscopic,varkentina2022cathodoluminescence}. And the success probability to post-select a particular energy-ladder state with $k$ can be obtained according to Eq.~(\ref{eq:psi_fock}), which takes the form of 
\begin{align}
Pr&=e^{-(|\alpha|^2+|g|^2)}\sum_{n=\text{max}\{0,-k\}}^{\infty}\frac{|\alpha|^{2(k+n)}(|g|^2)^{2\text{max}\{0,k\}-k}n!}{(\text{max}\{0,k\}!(k+n-\text{max}\{0,k\})!(\text{max}\{0,k\}-k)!)^2}\nonumber\\
&\hspace{2em}\times{}_{2}F_{2}^2(1,\text{max}\{0,k\}-k-n;1+\text{max}\{0,k\},1+\text{max}\{0,k\}-k;|g|^2)\nonumber\\
&=e^{-(|\alpha|^2+|g|^2)}\left\{
\begin{array}{ll}
\sum_{n=0}^{\infty}\frac{|\alpha|^{2(k+n)}|g|^{2k}n!}{((k+n)!)^2}L_{n}^{(k)2}(|g|^2), &k\ge0,\\
\sum_{n=-k}^{\infty}\frac{|\alpha|^{2(k+n)}|g|^{-2k}}{n!}L_{n+k}^{(-k)2}(|g|^2), &k<0.
\end{array}
\right.
\end{align}
The results for $k=0$ (blue solid), $k=1$ (red dashed) and $k=-1$ (green dash-dotted) are shown in Fig.~\ref{fig:SucPr}. Especially, to observe the optical odd cat states in the main text, the success probabilities are $Pr=0.76$\% for $|g|=0.17$, $k=0$ [Fig.~2(a) in the main text], $Pr=1.2$\% for $|g|=0.275$, $k=1$ [Fig.~2(b) in the main text], $Pr=2.0$\% for $|g|=0.95$, $k=1$ [Fig.~2(d) in the main text] and $Pr=1.1$\% for $|g|=2$, $k=0$ [Fig.~3(a) in the main text], respectively. Considering these small probabilities, the free-electron-photons interaction should be repeated thousands of times to successfully postselect the required outcome of the electron energy, where multiple electrons are needed. Since the free-electron-photons interaction is implemented by the ultrafast optics technique, where the typical length of the free-electron--photons interaction regime is $L\lesssim500\mu$m~\cite{dahan2020resonant}, and the timescale of a single experiment is usually $t\lesssim1000$fs, our scheme should be experimentally achievable.

\begin{figure}[t]
	\centering
	\includegraphics[width=0.3\textwidth]{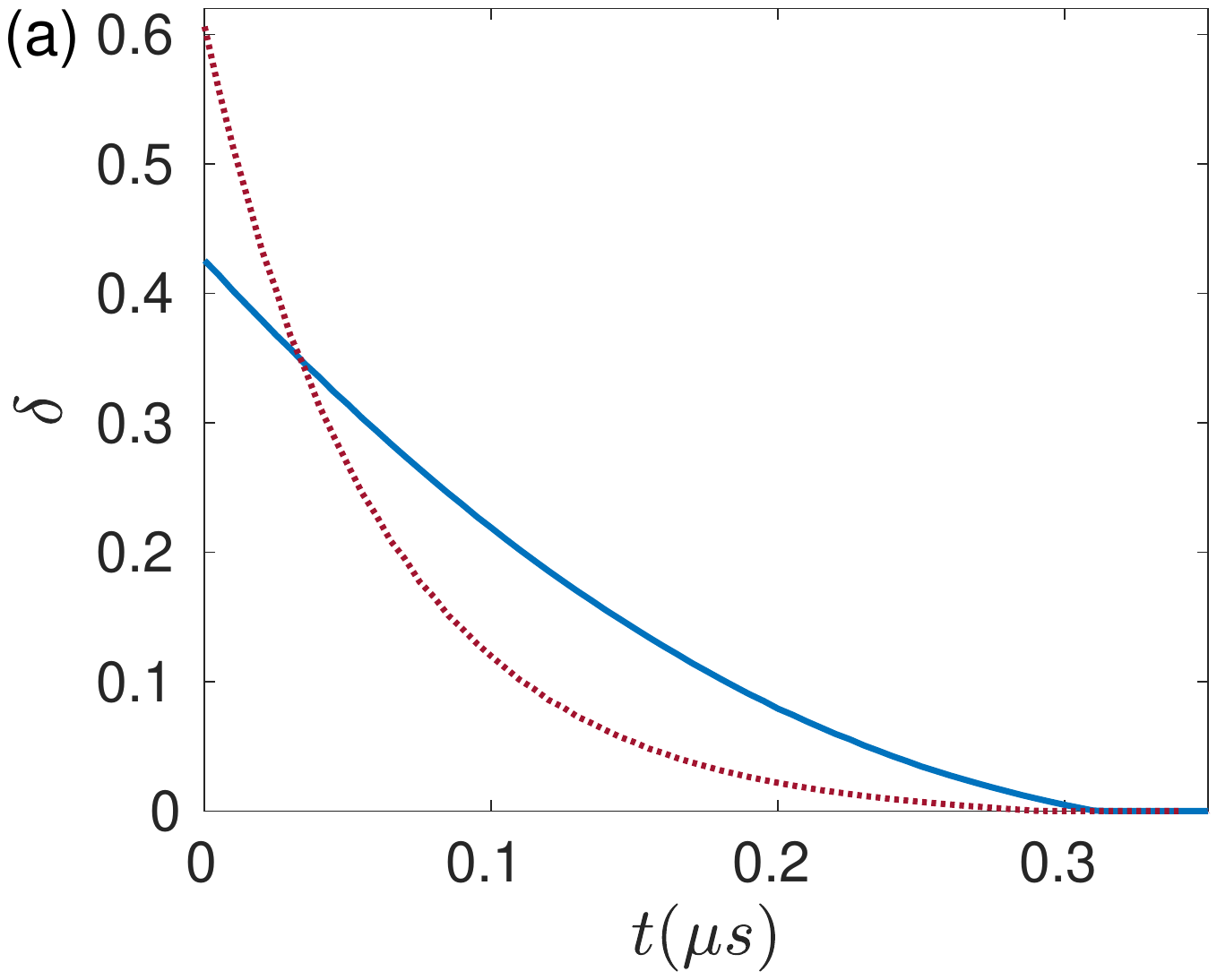}
	\includegraphics[width=0.3\textwidth]{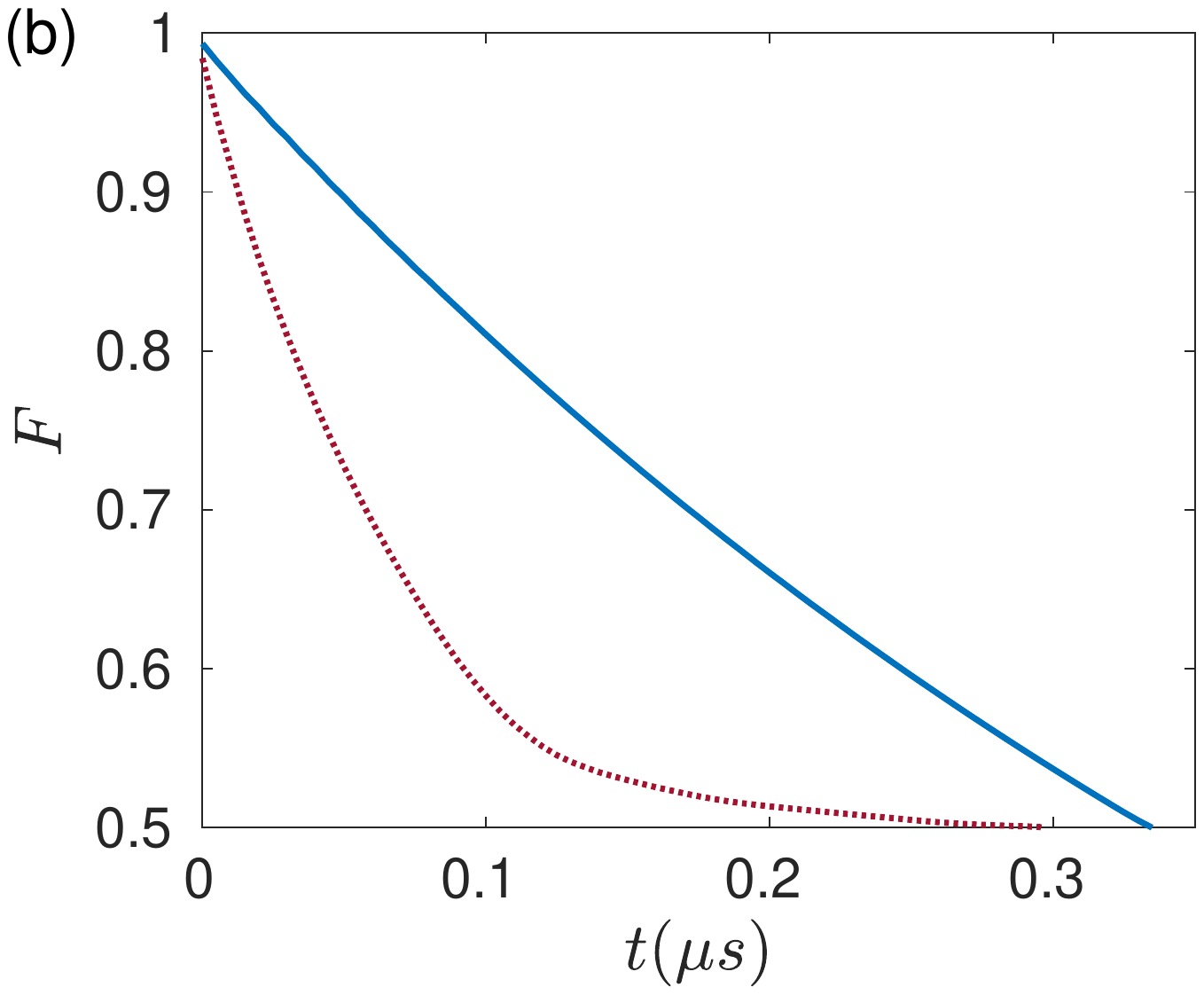}
	\caption{The time evolution of (a) the Wigner negativity $\delta$ and (b) the fidelity $F$ of the generated optical states. The blue solid line corresponds to the state in Fig.2(a) in the main text ($|g|=0.17$), and the red dotted corresponds to Fig.3(a) ($|g|=2$). Other parameters are set as $|\alpha|^2=50$, $k=0$ and the optical loss $\kappa=1$MHz.}
	\label{fig:lifetime}
\end{figure}

Besides, to validate the potential application of the produced cat states, we also study the lifetime of the generated optical cat state. The optical loss during the free-electron-photons interaction can be negligible, because the timescale $t\lesssim1000$fs is extremely short. After the interaction, the emitting photons will be collected in a high-Q cavity. In the current experiments, the typical decay rate of a high-Q cavity is about $\kappa\sim1$MHz~\cite{purdy2013observation,peterson2016laser,zupancic2019p}. Here, we take the optical loss as $\kappa=1$MHz and refer to the master equation of
\begin{equation}
\dot{\rho}=\kappa(2a\rho a^\dagger -a^\dagger a\rho-\rho a^\dagger a),
\end{equation}
which leads to the evolution of the Wigner negativity $\delta$ and the fidelity $F$ of the generated optical states as shown in Fig.~\ref{fig:lifetime}. It is found that the lifetimes will be $\tau\sim0.3\mu$s for both the optical cat states with $|g|=0.17$ and $|g|=2$. With the increase of the coupling strength, the scale of the optical cat states has been enlarged while the decrease of the lifetime is insignificantly small. This means that the lifetime of the optical cat state is dominantly affected by the decay rate of the high-Q cavity. Considering that the Q factor of the cavity can be further improved with the state-of-the-art technology, the lifetime of the optical cat state is promising to be further enhanced.
